\documentclass[
aps
,prx
,twocolumn,10pt
,superscriptaddress
,longbibliography
,amsmath
,amssymb
,amsfonts
]{revtex4-1}
\usepackage{epsfig}
\usepackage{makecell}
\usepackage{dcolumn}
\usepackage{float}
\usepackage{commath}
\usepackage{graphicx}
\usepackage{verbatim}
\usepackage{color}
\usepackage{braket}
\usepackage{lipsum}
\usepackage[dvipsnames,svgnames,x11names,hyperref]{xcolor}
\usepackage{times}
\usepackage[compat=1.1.0]{tikz-feynman}
\usepackage[sf,tight]{subfigure}
\usepackage[toc,page]{appendix}
\usepackage{siunitx}
\DeclareSIUnit\gauss{G}

\definecolor{linkcolor}{RGB}{6,69,173} 
\definecolor{diffcolor}{RGB}{175,31,36} 

\usepackage[colorlinks=true,
            linkcolor=linkcolor,
            urlcolor=linkcolor,
            citecolor=linkcolor,
            unicode,
            pdfencoding=auto]{hyperref}

\usepackage[
]{physpack}

\newcommand{\COMMENT}[1]{}

\begin{document}
\title{Topological quantum phase transitions driven by a displacement field in the twisted MoTe$_2$ bilayers}

\author{Prakash Sharma}
\email{sharmaprakash078@gmail.com}
\affiliation{Department of Physics and Astronomy, California State University, Northridge, California 91330, USA}

\author{Yang Peng}
\email{yang.peng@csun.edu}
\affiliation{Department of Physics and Astronomy, California State University, Northridge, California 91330, USA}
\affiliation{Institute of Quantum Information and Matter and Department of Physics, California Institute of Technology, Pasadena, CA 91125, USA}

\author{D.N. Sheng}
\email{donna.sheng@csun.edu}
\affiliation{Department of Physics and Astronomy, California State University, Northridge, California 91330, USA}

\begin{abstract}
We study twisted bilayer MoTe$_2$ systems at fractional fillings of the lowest hole band under an applied out-of-plane displacement field. By employing exact diagonalization in finite-size systems, 
we systematically map out the ground state quantum phase diagram for two filling fractions, $\nu=1/3$ and $2/3$, and provide a detailed characterization of each phase.
We identify the phase transition between a fractional Chern insulator (FCI) and a layer-polarized charge density wave (CDW) at a filling fraction of $\nu=1/3$, denoted as CDW-$1$. Additionally, we demonstrate that the competition between the displacement field and twist angle leads to another phase transition from a layer-polarized CDW-$1$ to a layer-hybridized CDW-$2$, identified as a first-order phase transition. Furthermore, at $\nu=2/3$ filling of the lowest hole band, we observe that the FCI remains stable against the displacement field until it approaches proximity to a transition in single-particle band topology at a smaller twist angle. 
\end{abstract}

\maketitle

\section{Introduction}
Twisted bilayer systems offer a unique platform for exploring the interplay of topology and interactions, facilitated by the emergence of topological, remote, and nearly flat bands. 
These systems provide a rich landscape for investigating diverse quantum phenomena, which can be finely tuned by controlling the relative twist angle between the layers.
This synergy has led to diverse many-body quantum phenomena that includes the experimental observations of Mott insulators~\cite{cao2018correlated,tang2020simulation}, generalized Wigner crystals~\cite{regan2020mott,li2021imaging,huang2021correlated,zhou2021bilayer}, fractional quantum Hall effects ~\cite{ki2014observation,dean2013hofstadter,kou2014electron,maher2014tunable}, and unconventional superconductivity~\cite{cao2018unconventional,yankowitz2019tuning,stepanov2020untying,oh2021evidence}. 
These experimental discoveries of many-body correlated states in bilayers have stimulated intense theoretical studies~\cite{po2018origin,chen2021realization,ledwith2020fractional,abouelkomsan2020particle, matty2022melting,ledwith2020fractional,abouelkomsan2020particle,tao2024valley, balents2020superconductivity, xu2018topological, kezilebieke2022moire, crepel2023topological,lee2023triangular}. 

A particular class of bilayer semiconductors based on transition metal dichalcogenides (TMDs) has recently garnered significant research attention due to their diverse topological and correlated phases observed both experimentally and theoretically. Notable examples include the observation of integer and fractional quantum anomalous Hall (FQAH) effects in twisted bilayer MoTe$_2$ ($t$MoTe$_2$)~\cite{cai2023signatures, park2023observation, xu2023observation, zeng2023thermodynamic}, and experimental evidence of fractional quantum spin Hall effects at filling factor $\nu=3$~\cite{kang2024evidence}.
 A similar observation has also been reported more recently in experiments with pentalayer rhombohedral graphene~\cite{lu2024fractional}.

FQAH effects in the absence of Landau levels (LLs), which are also called fractional Chern insulators (FCIs), 
have been predicted~\cite{wu2019topological,devakul2021magic, li2021spontaneous, crepel2023anomalous,abouelkomsan2020particle}
to emerge in the partially filled Chern bands of TMD bilayers.
Such FCIs are lattice analogy of fractional quantum Hall effect
first established in artificial interacting  models~\cite{neupert2011fractional,sheng2011fractional, regnault2011fractional,tang2011high, sun2011nearly}.
Their material realization without ambiguity was only recently achieved in $t$MoTe$_2$. These experimental breakthroughs have further spurred extensive theoretical investigations into the origin and characteristics of many-body correlated states in the partially filled Chern bands~\cite{reddy2023fractional,reddy2023toward,dong2023theory,abouelkomsan_band_2023,wang2024fractional, yu2024fractional, xu2024maximally,song2024intertwined, reddy2024non, xu2024multiple, ahn2024landau, wang2024higher, ChaomingJian2024}. Studies reveal that the FQAH states in $t$MoTe$_2$ originate as a Jain sequence from their parent anomalous composite Fermi liquid (CFL) at half-filling, similar to those in the lowest Landau level physics~\cite{goldman2023zero, dong2023composite}. However, deviations from the ideal quantum geometry of moir\'e bands~\cite{parameswaran2013fractional, liu2022recent, roy2014band}, specifically away from magic angles~\cite{devakul2021magic,bistritzer2011moire}, can lead to exotic many-body effects not anticipated within lowest Landau physics.
For example, recent numerical findings suggest the emergence of a quantum anomalous Hall crystal at large twist angles in $t$MoTe$_2$~\cite{sheng2024quantum,song2024intertwined}. 
Apart from twist angles, externally applied out-of-plane electric fields in bilayers can induce dramatic changes in the Chern bands, leading to new correlated quantum phenomena. 
Additionally, electric fields can serve as a controlling parameter to manipulate magnetism in bilayers. 
A recent theoretical study demonstrated that electric fields can be used to manipulate both spin and valley degrees in TMD-encapsulated chirally stacked graphene multilayers~\cite{wang2024electrical}.
The role of electric displacement fields in driving quantum phase transitions and layer polarization was further experimentally demonstrated in Hofstadter bands of twisted double bilayer graphene~\cite{adak2022perpendicular,cao2020tunable}.

Given that a perpendicular displacement field can significantly change the single-particle Bloch states and band topology, a systematic analysis is crucial to understand their effects on partially filled Chern bands. For instance, a recent theoretical study observed the transition from a composite Fermi liquid (CFL) to a Fermi liquid driven by externally applied displacement fields at half-filling~\cite{dong2023composite}. However, systematic exploration of the effects of displacement fields away from half-filling remains lacking, while there is experimental evidence emphasizing their significance~\cite{cai2023signatures}.

In this study, we focus on the topological quantum phase diagram and transitions driven by externally applied perpendicular electric fields in $t$MoTe$_2$. The out-of-plane displacement field creates a layer potential difference that can modify the single-particle band topology, potentially rendering the valence band topologically trivial. The suppression of the FQAH state at $\nu = -2/3$ \textcolor{blue}{($+2/3$ in the convention of hole filling number)} at high displacement fields, observed in a recent optical measurements~\cite{cai2023signatures}, is consistent with this effect. However, interestingly, a theoretical study has shown that the FCI phase ceases well before the transition in single-particle band topology~\cite{wang2024fractional}. Our study focuses on these moderate values of displacement fields where the single-particle band topology remains unchanged. We investigate whether the survival of the FCI is tied to ferromagnetism (FM) and, if the FQAH state disappears before the vanishing of FM and the band Chern number, what other competing phases emerge.

Employing the microscopic Hamiltonian for the semiconductor $t$MoTe$_2$, we perform exact diagonalization calculations at \textcolor{blue} {hole} filling factors $\nu=\frac{1}{3}$ and $\frac{2}{3}$ of the lowest moir\'e hole band and map out the topological phase diagram by varying the out-of-plane displacement fields and twist angles. At $\nu=\frac{1}{3}$ \textcolor{blue}{hole filling}, we observe clear evidence of a topological phase transition from the FCI to a layer-polarized  Charge Density Wave (CDW) driven by the displacement field, which we term CDW-$1$. This phase transition occurs well before the suppression of FM and the change in single-particle band topology.
We further verify this transition under band mixing at large displacement field, where the band gap is small. 
Additionally, we observe another phase transition from a layer-polarized CDW-$1$ to a layer-hybridized CDW-$2$, arising from the competition between twist angle tuned kinetic energy and displacement field (favoring strong layer polarization). We further identify this as a first-order phase transition through finite-size numerical analysis. On the other hand, at $\nu=\frac{2}{3}$, we find that the FCI remains resilient against displacement fields, only exhibiting a phase transition in the proximity of the transition in single-particle band topology at the smaller twist angle side. 
This is different from the results of a previous theoretical study for larger twist angles, where the FCI is suppressed well before the single-particle topological phase transition~\cite{wang2024fractional}.

\section{Band topology at finite displacement field}
$\textit{Continum model---}$  The two time-reversal partner Brillouin Zone corners $\mathbf{K}$ and $\mathbf{K^\prime}$ of MoTe$_2$ monolayer, where the valence band edges reside, are intrinsically locked to the opposite hole spins due to strong spin-orbit coupling~\cite{wu2019topological,xiao2012coupled}, forming a single two-component spin/valley degree. 
This allows for the independent construction of microscopic Hamiltonians for valley/spin degrees of the TMD bilayer, which are simply related to each other by time-reversal symmetry. Therefore, low energy electronic bands at $\mathbf{K}$ and $\mathbf{K^\prime}$
for opposite spin flavors have opposite Chern numbers.
We consider the MoTe$_2$ bilayer, where the two layers are relatively twisted by a small angle $\theta$ measured from the AA-stacking configuration. This introduces a long wavelength moir\'e period with periodicity $a_M=\frac{a}{2\sin(\theta/2)}$ (where $a$ is monolayer lattice constant), enabling for modeling the low-energy electronic structure with a continuum model in the sub-layer space~\cite{wu2019topological,yu2020giant, reddy2023fractional}.
In the presence of an out-of-plane displacement field $V_D$, the Hamiltonian of this model for $\mathbf{K}$ valley reads,
\begin{figure}
\includegraphics[width=0.49\linewidth,height=4.1cm]{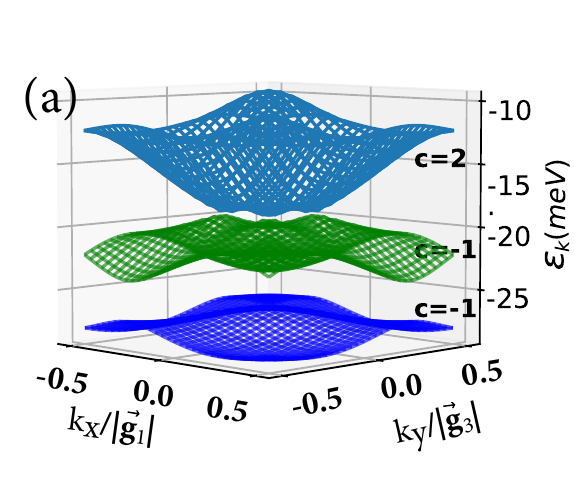}
\includegraphics[width=0.5\linewidth]{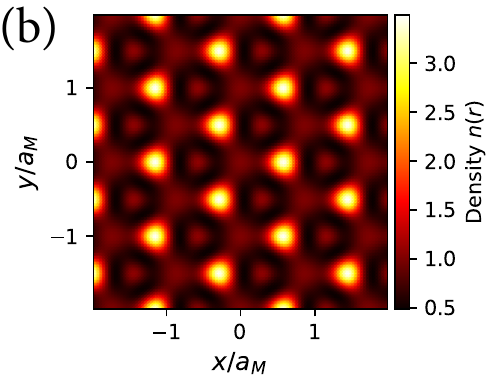}
\\
\includegraphics[width=0.49\linewidth,height=3.3cm]{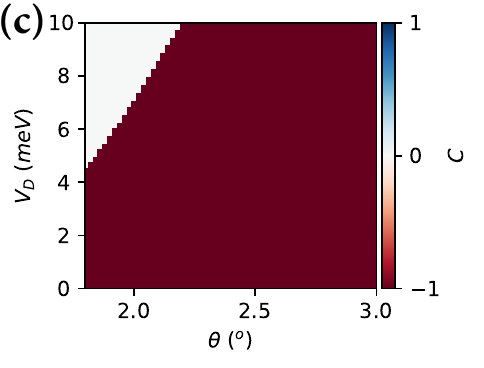}
\includegraphics[width=0.49\linewidth]{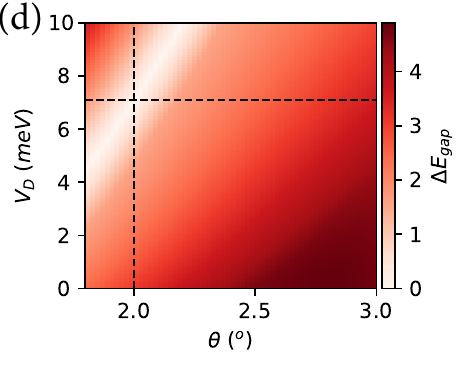}
\caption{(a) The band dispersion of continuum model Hamiltonian at magic angle $\theta=2^\circ$ with a finite displacement field $V_D=4 \text{ meV}$ for the lowest three topological bands with their corresponding Chern number shown. We partially fill the lowest moir\'e hole band shown in blue, assuming higher hole bands are all empty. 
(b) The charge density $n(\mathbf{r})=\frac{1}{N_{uc}}\sum_\mathbf{k}|\psi_\mathbf{k}(\mathbf{r})|^2$ corresponding to the lowest band highlighted in blue, where $N_{uc}$ is the number of moi\'re unit cells. 
(c) Chern number, $C$, of the lowest band, 
and (d) band gap, $\Delta E_{gap}$ (in meV), between the lowest two bands (highlighted in blue and green in (a)) at various twist angles and displacement fields.  
The vertical and horizontal dashed lines crossing represent the band gap closing at twist angle $\theta=2^\circ$, which occurs at $V_D \approx 7.1$ meV.
} 

\label{fig:band_topology}
\vspace{-0.15cm}
\end{figure}
\begin{equation}
h_K = 
\begin{pmatrix}
 h_t + V_D/2 & T(\textbf{r}) \\
T^\dagger(\textbf{r}) &  h_b - V_D/2
\end{pmatrix}
\label{eqn:continuum_model}
\end{equation}
where $h_{t/b} = \frac{\hbar^2 (\mathbf{k}-\mathbf{k}_{t/b})^2}{2m^*} +V_{t/b}(\textbf{r})$ are the top ($t$) and bottom ($b$) layer Hamiltonians with quadratic dispersion subjected to the intra-layer moir\'e potential $V_{t/b}(\mathbf{r}) = -2v\sum_{i=1,3,5} \cos(\mathbf{g_i}\cdot{\mathbf{r}} + \phi_{t/b})$, where $\mathbf{g_i}$ is the reciprocal moi\'re lattice vectors defined as $\mathbf{g}_i = \frac{4\pi}{\sqrt{3}a_M} \left\{ \cos(\frac{\pi(i-1)}{3}), \sin(\frac{\pi(i-1)}{3}) \right\}$. For $t$MoTe$_2$, the effective mass is  $m^*=0.62m_e$, where $m_e$ is the bare electron mass.
The inter-layer tunneling obeying $C_{3z}$ symmetry can be written as $T(\mathbf{r})=w (1+e^{i\mathbf{g_2}\cdot\mathbf{r}} + e^{i\mathbf{g_3}\cdot\mathbf{r}})$. 
Due to relative displacement introduced by twist, the $\mathbf{K}$ points of the top and bottom layers fold into the two corners of moir\'e Brillouin Zone (mBZ), given by $\mathbf{k}_t = \frac{2\mathbf{g}_1+\mathbf{g}_3}{3}$ and  $\mathbf{k}_b=\frac{\mathbf{g}_1-\mathbf{g}_3}{3}$ respectively.  
\begin{figure*}
\includegraphics[width=0.25\linewidth]{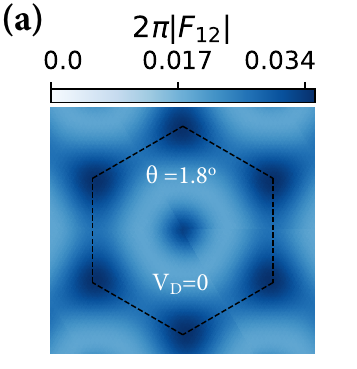}
\includegraphics[width=0.24\linewidth]{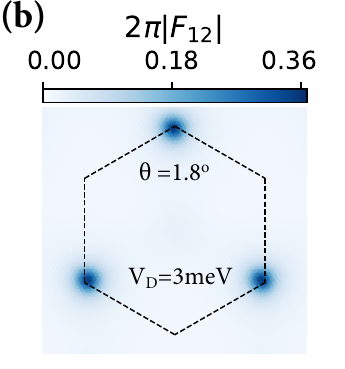}
\includegraphics[width=0.23\linewidth]{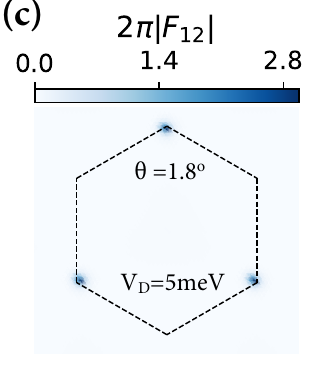}
\includegraphics[width=0.25\linewidth]{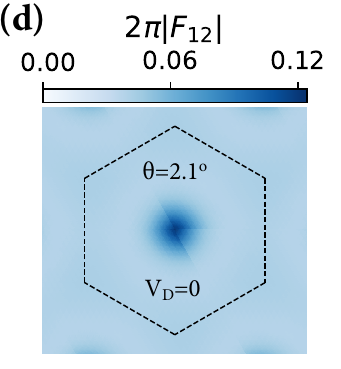}
\\
\includegraphics[width=0.25\linewidth]{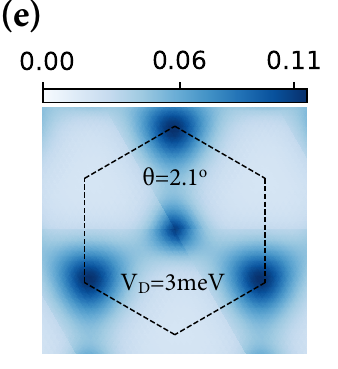}
\includegraphics[width=0.25\linewidth]{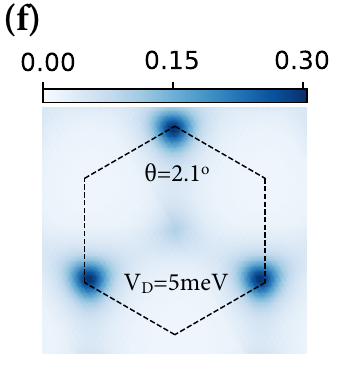}
\includegraphics[width=0.24\linewidth]{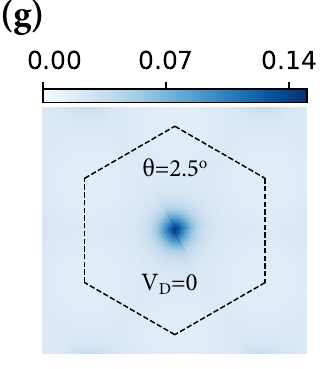}
\includegraphics[width=0.24\linewidth]{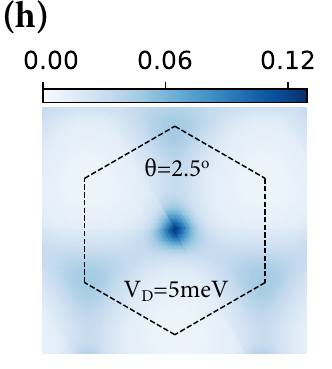}
\caption{The Berry curvature evolution of the lowest moi\'re band at representative twist angles $\theta=1.8^\circ,2.1^\circ, \text{ and } 2.5^\circ$ as a function of displacement fields $V_D$. The hexagonal dotted line on each plot represents the moi\'re Brillouin zone.  
} 
\label{fig:berry_curvature}
\vspace{-0.15cm}
\end{figure*}
 
Note that in equation~\eqref{eqn:continuum_model}, we use the convention previously considered in Ref.~\cite{reddy2023fractional}, where the single-particle bands are flipped so that the energy spectrum of $h_\mathbf{K}$ is bounded from below and has its vacuum at charge neutrality. 
Therefore, we have a positive sign for band filling (the number of holes per moir\'e unit cell), opposite to the convention used in experiments~\cite{cai2023signatures}. 
The free parameters in the model ($v,w,\phi$) are adapted from the ref.~\cite{reddy2023fractional}, which for $text{MoTe}_2$ bilayer are $v=11.2 \text{ meV}$, $\phi_{t/b}=\mp 91^\circ$ (negative for top and positive for bottom), and $w=-13.3 \text{ meV}$. 
We note that there is some disagreement on these single particle parameters given by refs.~\cite{wang2024fractional,jia2024moire} and ref.~\cite{reddy2023fractional}, which require future density functional theory studies. However, the physics of FCIs and composite Fermi liquid of  the lowest hole band is robust insensitive to detailed parameters.
Our main conclusion holds across different parameter sets, demonstrating the  broad applicability of our results in twisted TMD bilayers. Appendix~\ref{APP:wang}
 discusses the main results using parameters from ref.~\cite{wang2024fractional}.

$\textit{Band topology with displacement field---}$ 
The presence of the displacement field, $V_D$, breaks the layer symmetry that reduces the six-fold rotational symmetry to three-fold rotation $(C_{3z})$, while the translation symmetry remains preserved. This allows block diagonalization of the continuum model at each crystal momentum $\mathbf{k}$.
At $V_D=0$ and for twist angles $\theta\leq 3^\circ$, the lowest hole band 
is nearly flat, well separated from the second band by a reasonable band gap, and has Chern number $C=\pm 1$ for each spin/valley. Since the application of a perpendicular displacement field can change the band dispersion and band topology significantly, we first examine the single-particle properties by varying the displacement field and twist angle.

Fig.~\ref{fig:band_topology}(a) represents the lowest three bands with their corresponding Chern numbers at magic angle $\theta=2^\circ$ and displacement field $V_D=4 \text{ meV}$ for spin-up. Notably, the lowest Chern band (represented in blue) is nearly flat with a bandwidth of $\Delta E_w \sim 2.53\text{ meV}$  and separated from the second band (in green) minima by a band gap of $\Delta E_{gap}\sim 1.76 \text{ meV}$. 
The Chern number of the lowest band vanishes at critical $V_D$ that depends on the twist angle as shown in Fig.~\ref{fig:band_topology}(c). 
For example, the Chern number at $\theta=2^\circ$ changes at around $V_D=7.1 \text{ meV}$ following a band gap closing as shown in Fig.~\ref{fig:band_topology}(d). 
We shall see later that the displacement field plays a crucial role in the many-body phase diagram even before the transition in single particle band topology.

At intermediate twist angles ($2.3<\theta\leq 3^\circ$), despite the increased dispersion of the lowest Chern band (see Appendix~\ref{sec:single_particle}), displacement fields exhibit modest impact on the band gap. 
Only a marginal variation of about $1.4 \text{ meV}$ is observed across the range of $V_D$ from $0$ to $10 \text{meV}$. 
This allows us to investigate the Coulomb interaction, albeit approximately, by projecting onto the lowest Chern band as performed in previous studies both with and without displacement fields~\cite{reddy2023fractional,dong2023composite,goldman2023zero}. 

Another important consequence of the displacement field is the change in lattice geometry due to layer potential. 
At small twist angles, the charge density $n(\mathbf{r})$ (see definition in the Appendix~\ref{sec:local density}) equally localized around the MX and XM stacking sites of moir\'e cells forming a honeycomb lattice at $V_D=0$~\cite{reddy2023fractional}, which continuously deforms into a triangular lattice when $V_D$ is turned on. 
This happens due to the relocation of charges from the  MX sites of the top layer to the XM sites of the bottom layer, as shown in the density plot in Fig.~\ref{fig:band_topology}(b). 
In this particular example for $\theta=2^\circ$ and $V_D=4 \text{ meV}$, hole filling density on the top layer reduces from $0.5$ to $0.33$ while that of the bottom layer increases to $0.67$, indicating a partially layer polarized integer QAH state at $\nu=1$, 
which is different from the spontaneous layer polarized state induced by strong Coulomb interaction found in previous studies~\cite{wang2023topological,dong2023composite}. The latter is a topologically trivial insulator.

At large twist angles, where kinetic energy is high, layer polarization tendency
is much weakened for the same given value of $V_D$.
However, significant changes occur in lattice geometry. 
For instance, we observe a kagome geometry with almost uniform charge distribution between top and bottom layers at $\theta=3^\circ$ (see Appendix~\ref{sec:single_particle} for detail). 
This alteration in lattice geometry, combined with interactions, can significantly influence the underlying physics of correlated phases in bilayer systems.

A major role of the displacement field in the topology can be understood more qualitatively by observing its influence on local geometric characteristics such as the Berry curvature distribution. 
The single-particle Berry curvature distribution plays a crucial role in elucidating the physics of 2D systems and 1D pumps, irrespective of whether the global topological invariant (Chern number) is non-trivial~\cite{thouless1983quantization,xiao2010berry,wimmer2017experimental}. 
Fig.~\ref{fig:berry_curvature} illustrates the Berry curvature fluctuation across the mBZ at representative $\theta$ and $V_D$, highlighting the substantial impact of $V_D$ on its distribution. 
Notably, its deviation from a nearly uniform distribution due to the out-of-plane electric field, particularly at small twist angles, sets the stage for competition between FQAH states and other possible correlated phases. 
However, at large twist angles, the distribution of Berry curvature remains nearly unaffected by displacement fields, as depicted in Fig.~\ref{fig:berry_curvature}(g)-(h). 
This suggests the possible existence of a robust many-body state that is resilient against the displacement field. 
Later, we will explore how Berry curvature variation with twist angles and $V_D$ contributes to understanding the many-body phase diagram and competition among various correlated phases within the lowest moi\'re band.

\section{Coulomb interaction}
The continuum model Hamiltonian with projected Coulomb interaction can be conveniently written as,
\begin{equation}
    \mathcal{H} = \sum_{\mathbf{k},\sigma} \epsilon_{\mathbf{k},\sigma} \gamma^\dagger_{\mathbf{k},\sigma} \gamma_{\mathbf{k},\sigma}  
    + \frac{1}{2A} \sum_{\Tilde{\mathbf{q}}} V(\Tilde{q}) :\Bar{\rho}(\Tilde{\mathbf{q}}) \Bar{\rho}(-\Tilde{\mathbf{q}}):
    \label{eq:coulomb interaction}
\end{equation}
where $\gamma^\dagger_{\mathbf{k},\sigma}$ ($\gamma_{\mathbf{k},\sigma}$) creates (annihilates) a Bloch state $\ket{u_{\mathbf{k},\sigma}}$ in the lowest band at crystal momenta $\mathbf{k}$ and spin/valley $\sigma$, $\epsilon_{\mathbf{k},\sigma}$ is the corresponding Bloch band energy, $V(\Tilde{q})=\frac{2\pi e^2}{\epsilon_r |\Tilde{\mathbf{q}}|}$ is the Fourier transform of the Coulomb interaction in 2D, and $A=\frac{\sqrt{3}}{2}a^2_M N_1\times N_2$ is the area of the system, where $N_1$ and $N_2$ denote the number of  moir\'e unit cells along the two lattice vectors $\mathbf{L_1}$ and $\mathbf{L_2}$. 
The density operator $\Bar{\rho}(\pm\Tilde{\mathbf{ q}})=\sum_{\mathbf{k},\sigma}\langle u_{\mathbf{k\pm q + g_{k\pm \Tilde{q}}},\sigma}\ket{u_{\mathbf{k},\sigma}} \gamma_{\mathbf{k\pm q},\sigma}^\dagger \gamma_{\mathbf{k},\sigma}$ is projected to the lowest moir\'e band.
Notice that we have explicitly denoted the momentum belonging to the mBZ without a tilde sign, otherwise with a tilde sign. 
For example, in the moir\'e band basis, $\mathbf{\Tilde{q}}$ transforms to $\mathbf{q}$ by a reciprocal vector $\mathbf{g_q}$, such that $\mathbf{\Tilde{q}}\xrightarrow{}\mathbf{q}+\mathbf{g_q}$.
See Appendix~\ref{sec:coulomb} for more details on the projected Coulomb interaction. 
%
\begin{figure}
\includegraphics[width=0.49\linewidth]{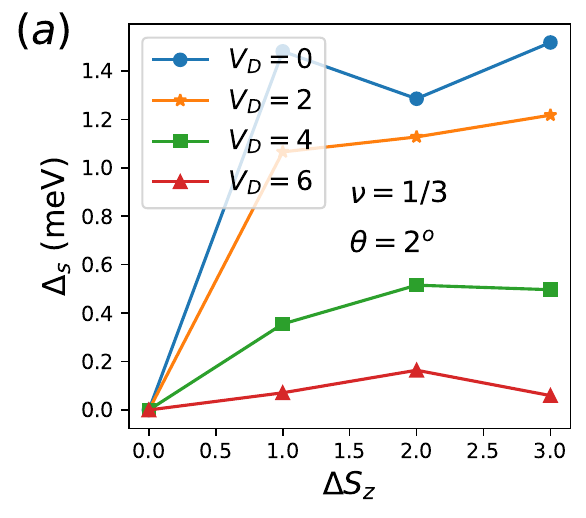}
\includegraphics[width=0.5\linewidth]{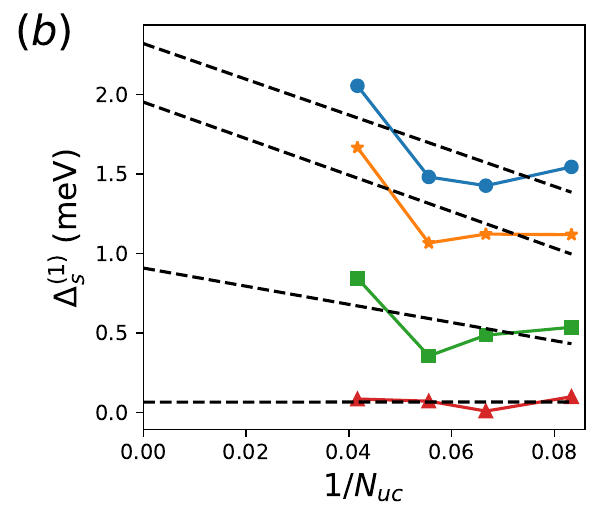}
\\
\includegraphics[width=0.48\linewidth]{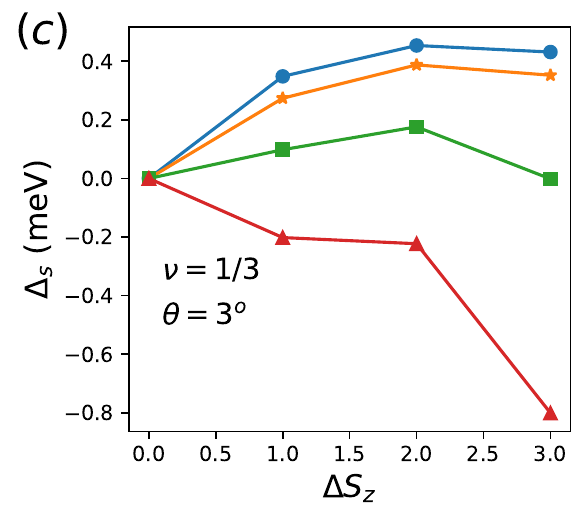}
\includegraphics[width=0.51\linewidth]{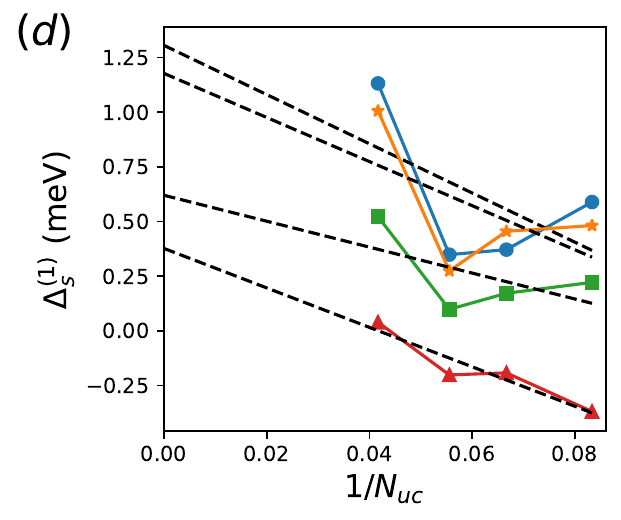}
\\
\includegraphics[width=0.49\linewidth]{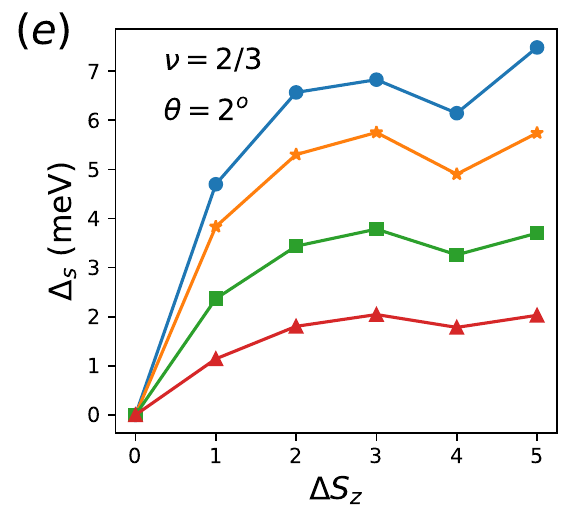}
\includegraphics[width=0.49\linewidth]{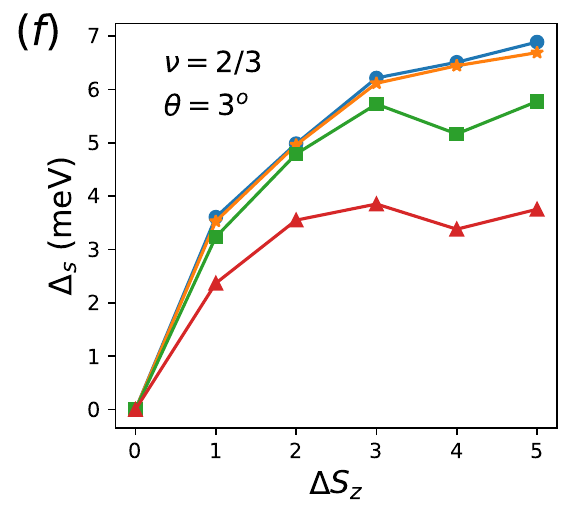}
\caption{Spin gap as a function of spin flips, defined in the main text, for representative values of $V_D$ (in meV). At $\nu=\frac{1}{3}$, (a) the full spin gap, $\Delta_S$, across all the spin sectors in $N_{uc}=18$ unit cell clusters, and (b) finite size scaling of spin-$1$ gap, $\Delta_S^{(1)}$, for $\Delta S_Z=1$ at $\theta=2^\circ$. (c) The full spin gap on $N_{uc}=18$, and (d) spin-$1$ gap at $\theta=3^\circ$. Finite-size analysis was conducted with cluster sizes $N_{uc}=12, 15, 18, \text{ and } 24$, with the black dashed line representing extrapolation to the thermodynamic limit.
(e)-(f) Spin gap for $\nu=\frac{2}{3}$ at $\theta=2^\circ$ and $3^\circ$, respectively, calculated on $N_{uc}=15$ unit cell clusters. 
} 
\label{fig:FM_onethird}
\vspace{-0.15cm}
\end{figure}

Typically, the lowest band projection is valid when characteristic Coulomb energy $\frac{e^2}{\epsilon_ra_M}$ is small compared to the band-gap ($\Delta E_{gap}$) between the lowest two moir\'e bands. 
However, there are evidences that the single band projection provides reasonable results for FCIs even when the characteristic energy far exceeds the band gap~\cite{kourtis2014fractional,grushin2015characterization}. 
Therefore, FCI states are robust against appreciable band mixing, and the candidate systems do not need to have a very large gap to host them.
For this qualitative region. we fixed the Coulomb interaction strength to a practically feasible value by setting $\epsilon_r=10$ throughout our calculations.

\textit{Ising ferromagnetism}---
It has been shown in the literature that the ground state of \textit{t}MoTe$_2$ remains ferromagnetic across a wide range of twist angles and filling fractions~\cite{reddy2023fractional,sheng2024quantum}. 
These ferromagnetic ground states exhibit a finite spin gap due to the absence of SU$(2)$ symmetry, broken by spin-valley locking~\cite{crepel2023anomalous,sheng2024quantum,reddy2023fractional}.
However, particularly at large $V_D$, the kinetic antiferromagnetic exchange can dominate the interactions~\cite{anderson2023programming}.
Therefore, we revisit these scenarios, with a specific focus on the filling fractions $\nu=\frac{1}{3} \text{ and } \frac{2}{3}$, considering the effects of the displacement fields.

We first consider full spin Hamiltonian. Since spin/valley degrees are conserved, we perform exact diagonalization within each $S_z=\frac{(N_\uparrow-N_\downarrow)}{2}$ sectors separately, where $N_\uparrow$ and $N_\downarrow$ are numbers of spin-$\uparrow$ and spin-$\downarrow$ holes such that total holes occupying the lowest moir\'e band is $N_h=N_\uparrow+N_\downarrow$. 
Using different cluster sizes, we analyze the ground state energies, $E_{GS}$, in different $S_z$ sectors and determine the polarization by calculating full spin gap $\Delta_S=E_{GS}(S_z)-E_{GS}(S_{zmax})$ and spin-$1$ gap $\Delta_S^{(1)}=E_{GS}(S_{zmax}-1)-E_{GS}(S_{zmax})$ as a function of $\Delta S_z=S_{zmax}-S_z$, as shown in Fig~\ref{fig:FM_onethird} for representative values of $V_D$ and $\theta$. 
Clearly, the spin gap decreases with an increase in $V_D$ as expected.  
%
\begin{figure}
\includegraphics[width=\linewidth,height=4.4cm]{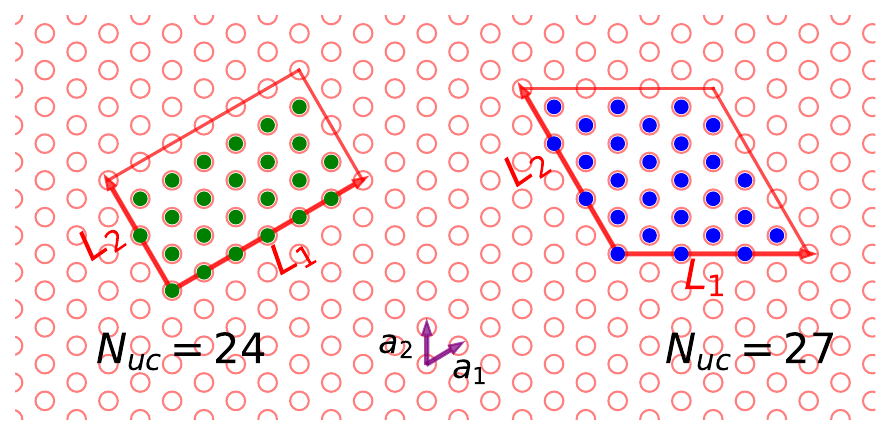}
\caption{ The real space cluster geometries with $N_{uc}=24$ (green) and $N_{uc}=27$ (blue) used in our exact diagonalization calculations. $\mathbf{L_1}$ and $\mathbf{L_2}$ determine the periodicity. These clusters contain the symmetric points $\gamma$ and $\kappa_\pm$ in the momentum space.
} 
\label{fig:clusters}
\vspace{-0.15cm}
\end{figure}

At $\nu=\frac{1}{3}$, the spin gap is much weaker, especially for larger values of $V_D$, as shown in Fig.~\ref{fig:FM_onethird}(a)-(c). 
Finite size analysis up to $N_{uc}=24$ moi\'re unit cells shown in Fig.~\ref{fig:FM_onethird}(b) and (d) for $\theta=2^\circ \text{ and } 3^\circ$, respectively, indicate that spin-$1$ gap, $\Delta_S^{(1)}$, increases with the system size and remains positive for larger system size (although it is negative at $V_D=6$ meV for small cluster sizes $N_{uc}\leq 18$ at $\theta=3^\circ$). 
Importantly, at large displacement fields, a thorough analysis of the full spin gap is crucial, as the $S_z=0$ sector may potentially dominate the low-energy magnetic phase. 
Therefore, we perform finite-size scaling for large $V_D = 6$ meV, shown in Appendix~\ref{APP:Ising_FM}, which illustrates that the full spin gap at $\theta=2^\circ$ increases towards positive value with the system size and extrapolates to $\Delta_S = +0.85$ meV in the thermodynamic limit. 
For a larger twist angle, $\theta=3^\circ$, finite size effects are more severe and pose challenges in predicting underlying magnetism in the thermodynamic limit, underscoring the necessity for larger cluster size analysis.
On a larger cluster with $24$ unit cells, although the full spin gap is not practically feasible to simulate, we found the spin-1 gap to be $\sim 0.0393$ meV and the spin-2 gap (for $N_\uparrow=6$ and $N_\downarrow=2$) to be $\sim-0.0032$ meV. Although both of which are larger when compared with corresponding gaps in $18$-site clusters shown in Fig.~\ref{fig:FM_onethird}(c), these results
may also indicate that the ground state can be in smaller or zero spin sector.
Given the small or vanishing spin gap, the application of small magnetic fields in  experiments can enhance the gap, facilitating measurements at finite $V_D$ within the fully polarized sectors at experimentally accessible low temperatures.  
%
\begin{figure}
\includegraphics[width=0.495\linewidth]{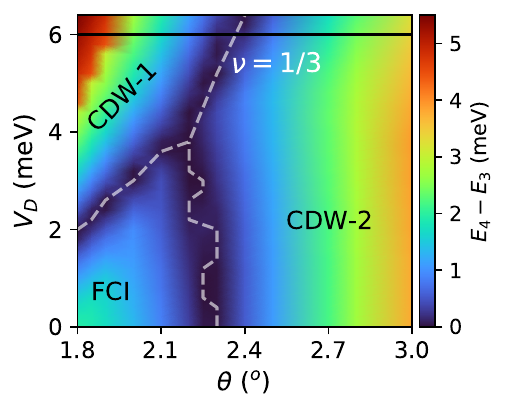}
\includegraphics[width=0.495\linewidth]{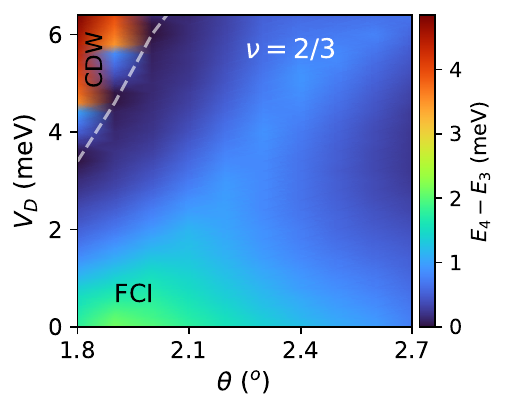}
\caption{Phase diagrams for fillings $\nu=\frac{1}{3}$ (left) and $\nu=\frac{2}{3}$ (right) as functions of twist angle $\theta$ and out-of-plane displacement field $V_D$, obtained from ED calculations on a $N_{uc}=27$ unit cell cluster. 
Color reflects the intensity of the many-body spectrum gap $E_4-E_3$. For a given finite-size cluster, the phase boundaries shown in a white dashed line in each panel are determined by connecting the points of many-body gap closing and other information in the main text. The horizontal solid line in left panel represents the region above which Ising FM is uncertain. The dark red region on each plot is included to complete the phase diagram; these regions correspond to topologically trivial band in Fig.~\ref{fig:band_topology}(c). Our focus is below these regions.}
\label{fig:phase_diagram}
\vspace{-0.15cm}
\end{figure}

Unlike $\nu=\frac{1}{3}$, we observe robust spin gaps at $\nu=\frac{2}{3}$ across all the spin sectors up to $V_D=6$ meV, as shown in Fig~\ref{fig:FM_onethird}(e)-(f), which is consistent with previous studies, where an increased spin gap is reported as hole density approaches from $\nu=1/3$ to $\nu=1$~\cite{reddy2023fractional,crepel2023anomalous}.
It is worth noticing that experimental results for $\nu=\frac{2}{3}$ indicate the vanishing of the Ising ferromagnet at a displacement field of approximately $\frac{D}{\epsilon_o}\sim 140 \text{ mV nm}^{-1}$~\cite{cai2023signatures}, which in terms of our unit $V_D=e\left( \frac{D}{\epsilon_o} \right)\frac{d}{\epsilon_r}$, corresponds to approximately $V_D\sim 14$ meV, assuming separation between top and bottom layer is $d\sim 1$ nm. 
Since our investigation is focused well below this experimental value, the observed large positive spin gap $(\Delta_S)$ within $V_D=6$ meV is consistent with the experiment results.

In the rest of our work, we will perform ED calculations within the momentum space of the fully polarized sector of $\nu=\frac{1}{3}$ and $\frac{2}{3}$ filling of the lowest moir\'e hole bands. 
While we will mainly use single-band Coulomb projection, which is adequate for capturing the phase transition, we will also present two-band results for completeness, especially at larger values of $V_D$ where the band gap is small.

%
\begin{figure*}
\includegraphics[width=0.25\linewidth]{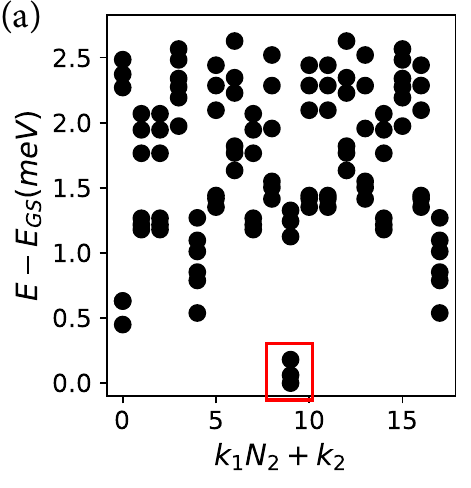}
\includegraphics[width=0.25\linewidth]{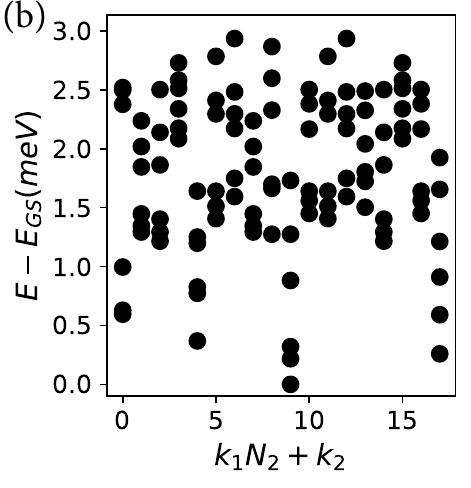}
\includegraphics[width=0.24\linewidth]{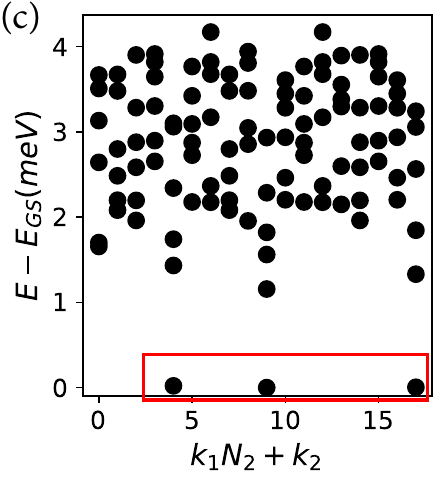}
\includegraphics[width=0.24\linewidth]{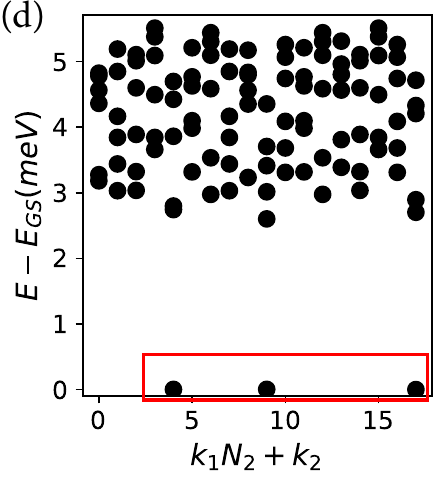}
\caption{The many-body energy spectrum at $\nu=\frac{1}{3}$ for a system size $6\times3$ with $\theta=2^\circ$, considering two bands per valley, is shown for displacement fields (a) $V_D=0$ meV, (b) $V_D=1$ meV, (c) $V_D=3$ meV, and (d) $V_D=6$ meV. The red rectangular box highlights the nearly degenerate ground states in the FCI and CDW-1 phases, respectively.} 
\label{fig:ED_2bands}
\vspace{-0.15cm}
\end{figure*}

\section{Phase diagram}
We track the evolution of the many-body ground state at various twist angles (up to $\theta\leq 3^\circ$), as a function of $V_D$.
We observe the competition between FCI and CDW phases, driven by the displacement fields and twist angles. 
To address this competition, it is essential to use clusters that accommodate the highly symmetric points in the mBZ, such as $\gamma$, and the two corners $\kappa_+$ and $\kappa_-$. 
Therefore, we utilize two symmetric clusters—one with $N_{uc}=24$ and the other with $N_{uc}=27$ moir\'e unit cells—for our single band ED calculations. Additionally, we utilized $6\times 3$ cluster size for two band ED calculations. 

These clusters with periodic boundary conditions are constructed  by selecting $(\mathbf{L_1,L_2})=((6,0),(-2,4))$ for $N_{uc}=24$ and $((6,-3),(-3,6))$ for $N_{uc}=27$, as shown in Fig.~\ref{fig:clusters}, where numbers in the parenthesis denote the scalar components $(m,n)$ such that $\mathbf{L_i}=m\mathbf{a_1}+n\mathbf{a_2}$ with $\mathbf{a_1}=a_M(\frac{\sqrt{3}}{2},\frac{1}{2})$ and $\mathbf{a_2}=a_M(0,1)$.

The precise location of topological degeneracy of the ground states in the $k$-space of the center of mass momenta (COM), $\mathbf{k}_{COM}=\sum_{j=1}^{N_h} \mathbf{k_i}$, provides a key hallmark for identifying the low-energy phases. 
However, it's important to note that this ground state degeneracy, while exact in the thermodynamic limit, may exhibit some splitting in finite cluster sizes. 
Specifically, FCI at filling fractions $\nu=\frac{p}{q}$ manifests $q$-fold degenerate ground states, with their momentum-orbital dependence governed by the generalized Pauli principle~\cite{haldane1991fractional,bernevig2008model,regnault2011fractional}.
This governing rule is sensitive to the geometric characteristics of the clusters utilized in ED simulations.
For instance, both $\nu=\frac{1}{3}$ and $\frac{2}{3}$ FCIs display a three-fold ground state degeneracy at the same COM momentum $\mathbf{k}_{COM}=\mathbf{\gamma}$ when employing tripled unit cell cluster with $N_{uc}=27$, while this degeneracy for $N_{uc}=24$ distributes among three momenta at $\mathbf{k}_{COM}=\mathbf{\gamma}$, $\mathbf{\kappa}_+$, and $\mathbf{\kappa}_-$.
On the other hand, the structure of the ground state degeneracy in the CDW phase depends on their order or the manner in which the spatial translation symmetry is broken. 
For example,  CDW at $\nu=\frac{1}{\mathbf{3}} \text{ and } \frac{2}{3}$ exhibits three-fold degenerate ground states with order wave vector  $\mathbf{q^*}=\mathbf{k_\pm}$. 
For further details on more exotic CDW orders, refer to Ref.~\cite{wilhelm2021interplay}.
It is important to note that the three-fold ground state degeneracy in the $N_{uc}=24$ unit cell cluster occurs at the same COM momenta ($\gamma \text{ and } \kappa_\pm$) for both FCI and CDW phases. Hence, employing a $N_{uc}=27$ unit cell cluster in our simulations is imperative for distinguishing between these two competing phases.

We first map out the phase diagram, as illustrated in Fig.~\ref{fig:phase_diagram}, for $\nu=\frac{1}{3}$ and $\frac{2}{3}$ filling of the lowest moir\'e band using single band ED calculation on $27$ unit cell clusters.  
The phase diagram is constructed based on the momenta at which nearly degenerate three-fold ground states occur~\cite{haldane1991fractional,bernevig2008model,sheng2011fractional,regnault2011fractional}, and the many-body spectrum gap $E_4-E_3$ as indicated through color variations. Here, $E_1, E_2, E_3, \text{ and } E_4$ represent the energies of the lowest four eigenstates. 
The identified phases are further verified through additional analysis of the many-body Chern number, charge structure factor, and Bloch state occupation number in the nearly three-fold degenerate ground states, which we will discuss later. 
The phase boundaries quantitatively match for $24$ unit cell clusters as well.

%
\begin{figure*}
\includegraphics[width=0.245\linewidth]{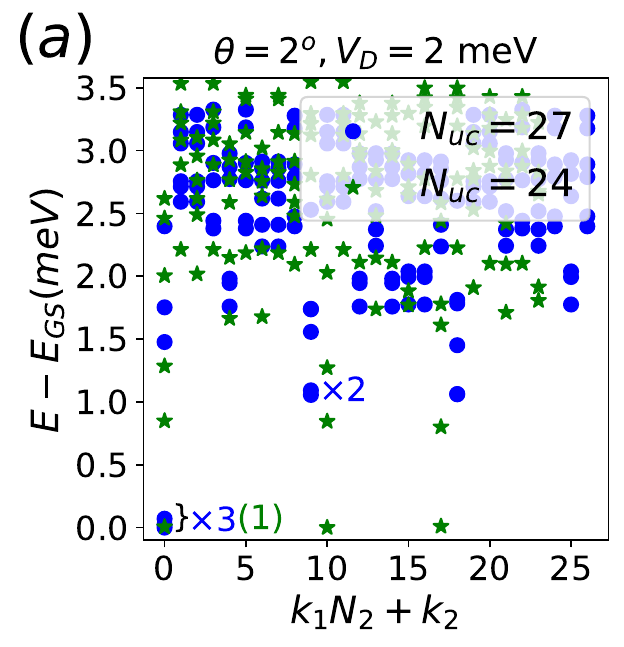}
\includegraphics[width=0.245\linewidth]{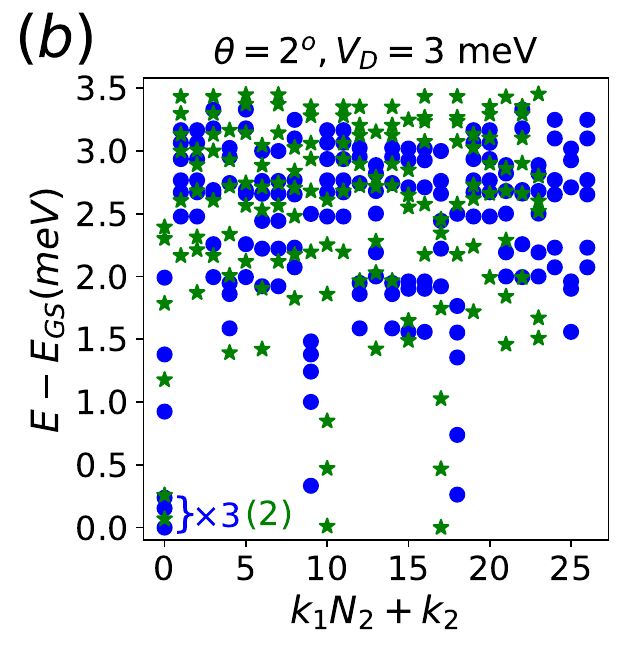}
\includegraphics[width=0.245\linewidth]{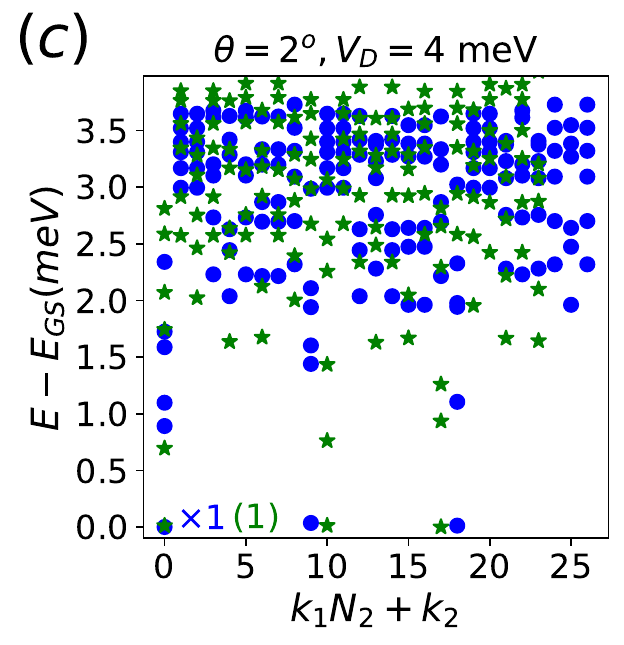}
\includegraphics[width=0.245\linewidth,,height=4.55cm]{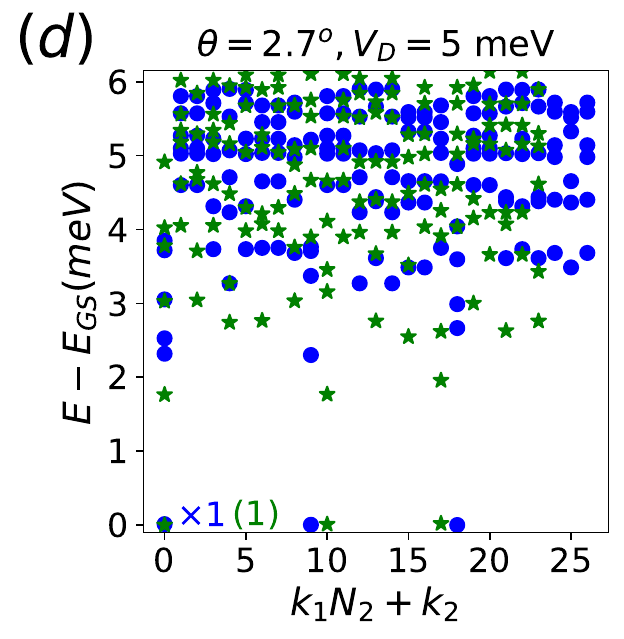}
\\
\includegraphics[width=0.245\linewidth]{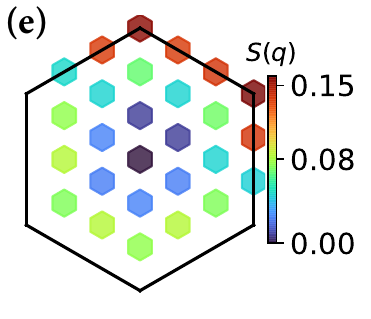}
\includegraphics[width=0.245\linewidth]{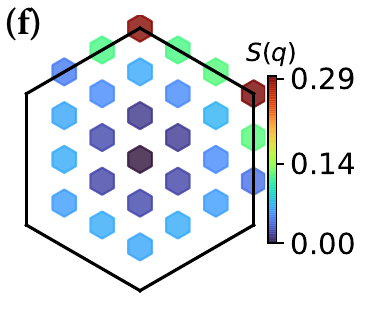}
\includegraphics[width=0.245\linewidth]{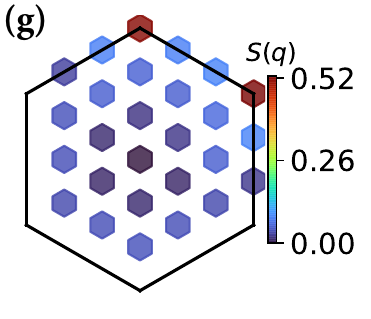}
\includegraphics[width=0.245\linewidth]{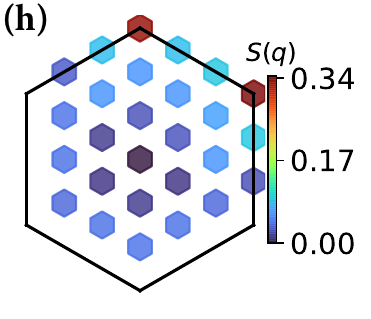}
\\
\includegraphics[width=0.2456\linewidth]{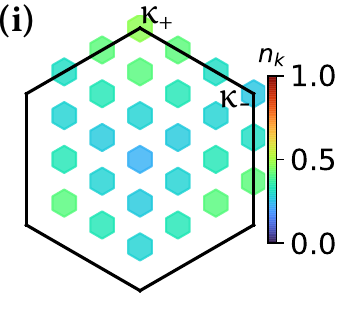}
\includegraphics[width=0.2456\linewidth]{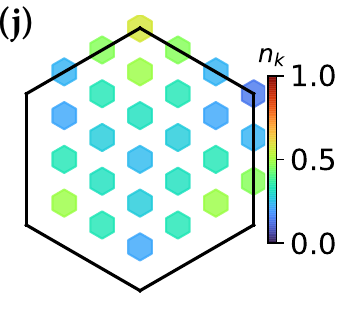}
\includegraphics[width=0.2456\linewidth]{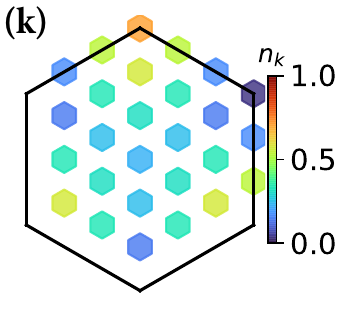}
\includegraphics[width=0.2456\linewidth]{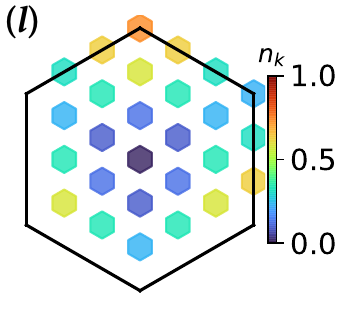}
\caption{(a)-(d) The momentum resolved low-lying many-body energy spectra indicating competition between FCI and CDWs at $\nu=\frac{1}{3}$. ED results are obtained on two different clusters with $N_{uc}=24$ (green stars) and $N_{uc}=27$ (blue dots) moir\'e unit cells containing high symmetric points ($\gamma$ and $\kappa_\pm$). At $\theta=2^\circ$, (a) $V_D=2\text{ meV}$ and (b) $V_D=3\text{ meV}$, and (c) $V_D=4\text{ meV}$, demonstrates a phase transition from FCI to CDW-$1$ driven by displacement field. 
(d) ED spectrum at $\theta=2.7$ and $V_D=5\text{ meV}$  represents a robust CDW-$2$ phase at large twist angles characterized by a large many-body gap. 
The blue (green) labels in (a)-(d) represent the number of states closely clustered together in $N_{uc}=27$ ($24$) site systems.
(e)-(h) The charge structure factors, $S(\mathbf{q})$, and (i)-(l) Bloch state occupations, $n_k$, averaged over the (nearly) degenerate ground states in the panels shown directly above them.
} 
\label{fig:ED_spectra}
\vspace{-0.15cm}
\end{figure*}
%

\subsection{Competing FCI and CDWs at $\nu=\frac{1}{3}$}
At $\nu=\frac{1}{3}$, we observe three distinct correlated phases--FCI, CDW-$1$, and CDW-$2$--separated by spectrum gap closings, as indicated by the white dashed lines in the left panel of Fig.~\ref{fig:phase_diagram}.
At $V_D=0$, the quantum phase transition from FCI to CDW-$2$ at around $\theta=2.3^\circ$ is consistent with that in the ref.~\cite{reddy2023fractional}. 
Notably, this transition is primarily influenced by the Berry curvature distribution and Bloch state wavefunctions rather than the Coulomb interaction.

In the FCI region (small $\theta$ and $V_D$), the many-body gap is large, i.e., $E_1-E_3<E_4-E_3$, until it approaches the phase boundary. 
Remarkably, upon increasing the twist angle from $\theta=1.8^\circ$ towards larger values, the FCI region extends to higher values of the displacement field   (until a phase transition to CDW-$2$ occurs).
This enhancement in the FCI region can be understood from the Berry curvature distribution within the mBZ, presented in Fig.~\ref{fig:berry_curvature}, where the Berry curvature fluctuation is highly suppressed and more uniformly distributed for $\theta=2.1^\circ$ compared to $\theta=1.8^\circ$ for the same given value of $V_D=3\text{meV}$ (compare between Fig.~\ref{fig:berry_curvature}(b) and (e)). 
This relative uniformness in the Berry curvature distribution at higher twist angles enables the holes in the lowest band to experience a smooth magnetic flux, providing a suitable playground for the FQAH states.
This finding is further corroborated by the trace condition violation illustrated in Fig.~\ref{fig:band_width} of Appendix~\ref{sec:single_particle}. 

Since Berry curvature gradually becomes non-uniform and effective kinetic energy of the lowest band increases (only mildly) with increasing $V_D$, 
other competing states may emerge.
Consequently, we observe a phase transition from FCI to a CDW-$1$, which happens well before the disappearance of ferromagnetism and the transition in single particle band topology. 
For example, the FCI to CDW-$1$ transition for $\theta=2^\circ$ takes place at $V_D\approx 3.0 \text{ meV}$, whereas the Chern number of the lowest single-particle band vanishes at a relatively large value, $V_D\approx 7.1\text{ meV}$, as discussed earlier. 
This transition in the single-particle band topology results in a sudden jump in the many-body spectrum gap, as indicated by the dark red region in Fig.~\ref{fig:phase_diagram}. This also indicates the projection fails in the transition regime while we focus our study in smaller $V_D \leq 6.0 \text{ meV}$ (and below the red areas in the phase diagram for other twist angles). 
Since band mixing inevitably occurs near band closing, the stability of CDW-$1$ under such conditions warrants further investigation. 
Therefore, we perform two band projection within a fully polarized sector to further validate our phase diagram, and the resulting many-body spectrum on $6\times 3$ cluster for representative angle $\theta=2^\circ$ is presented in Fig.~\ref{fig:ED_2bands}.
For  $V_D=0, 1$ meV, we see three fold near degenerating ground states consistent with FCI phase. 
The migration of nearly degenerate threefold ground states from the COM $(3,0)$ at $V_D=0$ meV to the three different momentum sectors $(1,1),(3,0),$ and $(5,2)$ (which have relative momentum difference $\kappa_\pm$) for $V_D=3$ meV, clearly illustrates phase transition from FCI to CDW-$1$.
The many-body gap in the CDW-$1$ phase increases further with increase in $V_D$, mirroring the trend observed in the single-band approximation, albeit with a relatively larger gap in the two-band approximation. This qualitative similarity between the one-band and two-band results confirms that the single-band projection provides reasonable approximation even when the band gap is smaller compared to the Coulomb interaction strength. 
Therefore, in the remainder of our discussion, we will consider only the single-band projection, which allows us to access larger cluster sizes.


At large twist angles, where the Berry curvature shows significant fluctuations with most of its weight centered at $\gamma$ and remains qualitatively unaffected by $V_D$ (see Fig.~\ref{fig:berry_curvature}(g)-(h)), holes tend to distribute towards the border of the mBZ, which helps them avoiding substantial exposure to the effective magnetic field (see discussion below). 
Consequently, a stable CDW-$2$ phase forms with a robust many-body gap that remains unperturbed by the displacement field (at least up to the range of $V_D$ we considered). 
This robustness of $\text{CDW-}2$ against the $V_D$ can be attributed to the fact that the band dispersion and Berry curvature distribution are roughly insensitive to the change in the displacement field.

%
\begin{figure}
\includegraphics[width=\linewidth]{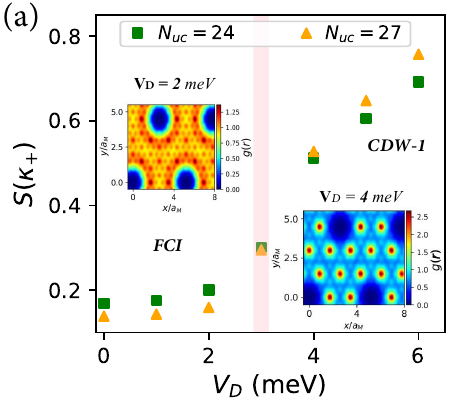} 
\\
\includegraphics[width=\linewidth]{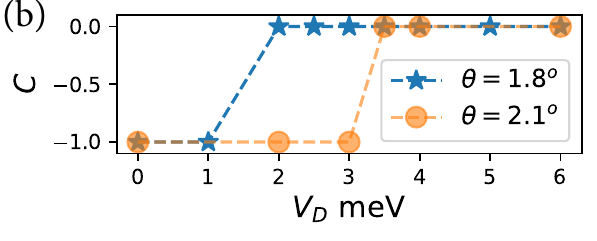} 
\caption{(a) The charge structure factor peak at $\mathbf{q}=\kappa_+$, i.e., $S(\mathbf{\kappa_+})$ for two different cluster sizes as a function of $V_D$ at fixed value of twist angle, $\theta=2^\circ$. 
The insets show pair correlation functions, $g(\mathbf{r})$, at $V_D=2$ meV and $V_D=4$ meV for $\theta=2^\circ$, indicating incompressible fluid and symmetry broken crystalline phase, respectively.
(b) The sum of many-body Chern numbers over the degenerate ground state manifolds of $N_{uc}=27$ unit cell cluster plotted as a function of $V_D$ for representative twist angles. The calculations are performed at $\nu=\frac{1}{3}$ filling. 
}
\label{fig:FCI_CDW}
\vspace{-0.15cm}
\end{figure}

\textit{FCI to CDW-1 transition}---
In Fig.~\ref{fig:ED_spectra}, representative plots for the lowest many-body eigen-energies at each phase are presented as a function of momentum $\mathbf{k}=k_1 \mathbf{b_1} + k_2 \mathbf{b_2}$, where $\mathbf{b_1}$ and $\mathbf{b_2}$ denote the translation vectors that generate a discrete momentum grid in reciprocal space.
Each momentum is assigned a scalar integer pair ($k_1, k_2$), which can be uniquely mapped into a 1D lattice of orbital momentum $\lambda=k_1N_2+k_2$. 
For fixed $\theta=2^\circ$, Fig.~\ref{fig:ED_spectra}(a)-(c) shows the evolution of the energy spectrum as a function of $V_D$. 
At $V_D=2 \text{ meV}$, the three-fold degenerate ground states lie at the COM momenta allowed by the counting rule, exhibiting a relatively enhanced many-body gap for larger cluster size $N_{uc}=27$ compared to $N_{uc}=24$, suggests a FCI state. 
However, this gap closes at around $V_D=3$ meV and subsequently reopens upon a further increase in $V_D$ as shown in Fig.~\ref{fig:ED_spectra}(b)-(c), indicating a phase transition. 
Notably, the degenerate ground states now have the COM momenta separated by $\kappa_\pm$ on both the clusters, strongly suggesting the emergence of the CDW state. 
Moreover, a relatively large gap in the spectrum for the larger cluster (Fig.~\ref{fig:ED_spectra}(c)) points towards the robustness of the CDW-$1$ phase in the thermodynamic limit. This gap further increases with an increase in $V_D$ (see Appendix~\ref{sec:extended13}). 

We further confirmed this quantum phase transition by calculating the charge structure factor within the projected band, defined as
\begin{equation}
     S(\mathbf{q}) = \frac{1}{N_{uc}} \left[  \langle \Bar{\rho}(\mathbf{q}) \Bar{\rho}(\mathbf{-q}) \rangle - \langle \Bar{\rho}(\textbf{q}) \rangle \langle \Bar{\rho}({-\mathbf{q}}) \rangle  \right]
     \label{eq:str_fac}
\end{equation}
and Bloch state occupation,
\begin{equation}
    n_k = \langle \gamma_\mathbf{k}^\dagger \gamma_\mathbf{k} \rangle
    \label{eq:nk}
\end{equation}
where expectations are taken over the (nearly) degenerate ground states and averaged out.
$S(\mathbf{q})$ has no sharp feature in the FCI phase as shown in Fig.~\ref{fig:ED_spectra}(e) (relatively larger weight at $\kappa_{\pm}$ indicates the tendency towards CDW for given cluster size), while the sharpness of the peaks at $\kappa_\pm$ gradually increase with $V_D$ as we approach CDW-$1$ phase, as illustrated in Fig.~\ref{fig:ED_spectra}(f)-(g). 
Additionally, the deviation from nearly uniform distribution of the average Bloch state occupation, $n_k$, over the mBZ between $V_D=2-4$ meV, shown in Fig~\ref{fig:ED_spectra}(i)-(k), further strengthens our assertion regarding the FCI to CDW-$1$ phase transition. 
%
\begin{figure*}
\includegraphics[width=0.3\linewidth]{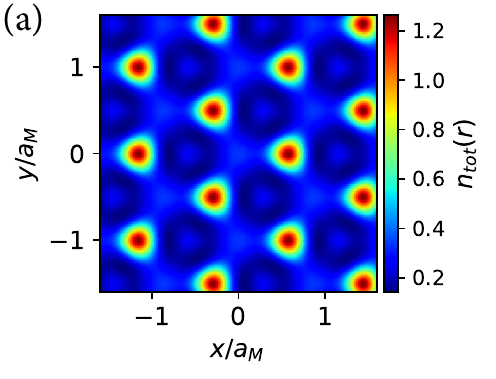} 
\includegraphics[width=0.3\linewidth]{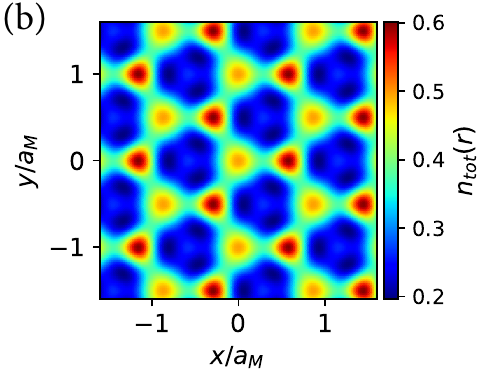} 
\includegraphics[width=0.32\linewidth]{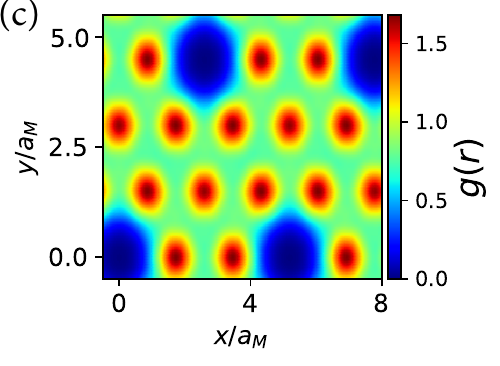}
\\
\includegraphics[width=0.46\linewidth]{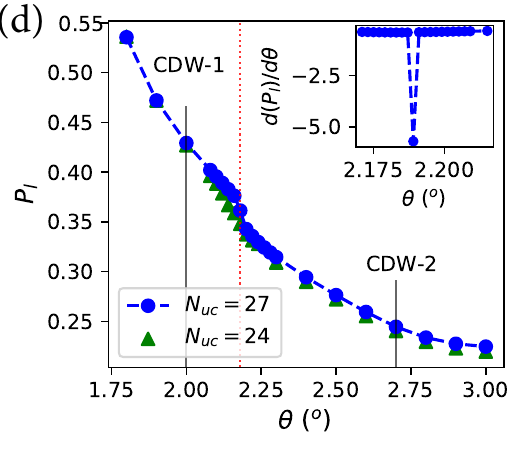}
\includegraphics[width=0.42\linewidth]{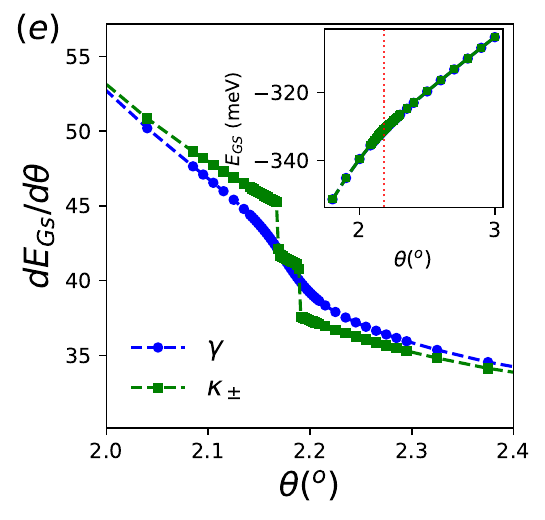}
\caption{For a fixed value of $V_D=4$ meV (a) the real space distribution of the total hole density across the top and bottom layers for twist angles (a) $\theta=2^\circ$, and (b) $\theta=2.7^\circ$, indicating strong layer polarization at small angles.
(c) The pair correlation function, $g(\mathbf{r})$, for $\theta=2.7^\circ$ and $V_D=4$ meV. The calculations are performed on the $27$ unit cell cluster and averaged over the nearly degenerate ground states.  
(d) The average layer polarization, $P_l=\frac{n_{bot}-n_{top}}{n_{bot}+n_{top}}$, calculated on the ground states as a function of twist angle, $\theta$, for a fixed value of $V_D=4$ meV. The dotted vertical line represents the phase boundary between the strongly layer polarized CDW-$1$ and the weakly layer polarized CDW-$2$ phase. 
The two vertical grey lines correspond to the parameters shown in (a) and (b).
The inset is the first derivative of $P_l$ w.r.t. $\theta$ calculated on $27$ unit cell geometry in the vicinity of phase transition.
(e) The first derivative of the lowest energy at $\gamma$, and $\kappa_\pm$ (averaged) with respect to $\theta$ in $27$ unit cell clusters, still for fixed $V_D=4$ meV.
The inset represents the corresponding ground state energies as a function of $\theta$, which shows a small kink at the transition point in the $\kappa_\pm$ sector, indicating the first-order phase transition.
}
\label{fig:LP}
\vspace{-0.15cm}
\end{figure*}
%

In Fig.~\ref{fig:FCI_CDW}(a), we present the structure factor peak at $\mathbf{q}=\kappa_+$ as a function of $V_D$ at fixed twist angle $\theta=2^\circ$, and contrast its intensity across two distinct cluster sizes utilized in the simulation.
Clearly, in the FCI regime ($V_D<3$ meV), the structure factor peak is small and diminishes further with the system size, whereas its magnitude notably increases beyond $V_D=3$ meV and scales with the system size, suggesting a distinct Bragg peak at the ordering wave vector $\mathbf{q}=\kappa_\pm$. 
The insets in the same figure show a pair correlation function, defined as
\begin{equation}
    g(\mathbf{r-r^\prime}) =\frac{1}{N_h^2} \sum_{\mathbf{\Tilde{q}}\in R^2}\langle:\Bar{\rho}({\mathbf{\Tilde{q}}})\Bar{\rho}({-\mathbf{\Tilde{q}}}):\rangle e^{i(\mathbf{r}-\mathbf{r^\prime})\cdot \mathbf{\Tilde{q}}} 
\end{equation}
with $\mathbf{r^\prime}$ 
conveniently taken to be at the origin.
The pair correlation is nearly uniformly distributed at longer distances in the FCI region for $V_D=2$ meV, displaying characteristics of a fluid with short-range correlations. In contrast, deep in the CDW-$1$ phase (at $V_D=4$ meV), the pair correlation clearly reflects spontaneously symmetry-broken crystalline structure with $\sqrt{3}\times \sqrt{3}$ order, consistent with the Bragg peaks at $\kappa_\pm$ observed in the structure factor.

In light of the recent theoretical study of a quantum Hall crystal at $\nu=\frac{1}{2}$, characterized by a topological charge density wave with a non-zero many-body Chern number and an unconventional Hall conductance of $\sigma_{xy}=\frac{e^2}{h}$~\cite{sheng2024quantum}, we investigate the topological nature of different quantum phases by computing the many-body Chern number, defined as
\begin{equation}
    \mathcal{C}=\frac{i}{2\pi}\int_0^{2\pi} d\phi_1 d\phi_2 \left\{ \Braket {\frac{\partial \Phi(k_1,k_2)}{\partial \phi_1}| \frac{\partial\Phi(k_1,k_2)}{\partial\phi_2}}   - c.c\right\},
    \label{eq:mchernnum}
\end{equation}
where $\Phi(k_1,k_2)$ is the many-body ground state at COM momentum $(k_1,k_2)$ and $\phi_j$ ($j=1,2$) is the phase obtained by introducing twist boundary condition $\psi(\mathbf{r}+N_j \mathbf{a_j})=e^{i\phi_j}\psi(\mathbf{r})$, in the single particle state $\psi(\mathbf{r})$. 
This is equivalent to introducing a magnetic flux that shifts the kinetic momentum of each particle by $\mathbf{\delta k}= \frac{\phi_1}{2\pi}\mathbf{b_1}+\frac{\phi_2}{2\pi}\mathbf{b_2}$.
The imaginary part of the integrand in equation~\eqref{eq:mchernnum}  defines the many-body Berry curvature $F(\phi_1,\phi_2)$.
For numerical computation, we discretize the boundary phase into a $12 \times 12$ Brillouin zone mesh and calculate the overlaps $\braket{\Phi(\phi_1,\phi_2)|\Phi(\phi_1^\prime,\phi_2^\prime)}$ around each square plaquettes. This yields the Berry phase contributions from both the single-particle Bloch states and the many-body wavefunction, as detailed in the supplementary material of ref.~\cite{okamoto2022topological}.

In the FCI phase, when the ground state degeneracy occurs at the same COM momentum, the Chern number quantization condition applies to the total Chern number, which may not be evenly distributed among the ground states.
Thus, only the sum of Chern numbers within the degenerate manifold yields the correct quantization.
Therefore, in Fig.~\ref{fig:FCI_CDW}(b), we show the sum of Chern number of the nearly degenerate ground states on the $27$ unit cell clusters as a function of $V_D$ for two representative twist angles $\theta=1.8^\circ$ and $2.1^\circ$. 
For both values of angle, the total Chern number is one for $\nu=\frac{1}{3}$ FQAH states as expected. 
However, the total Chern number vanishes as soon as CDW-$1$ emerges, concluding that the CDW-$1$ is a topologically trivial Wigner crystal~\cite{zhou2021bilayer}, unlike the QAH crystal. In Appendix~\ref{sec:FCI_CDWs}, we provide extended numerical data and demonstrate the transition between FCI and CDW-1 is presumably continuous. 

\textit{CDW-1 to CDW-2 transition}---
Now we turn our focus to understanding the phase transition between two competing CDW phases at $\nu=\frac{1}{3}$, primarily driven by the twist angle at finite displacement field. 
This transition arises from the interplay between $V_D$, which increases the layer polarization, and $\theta$, which increases the bandwidth.

Notably, the CDW-$2$, prevalent at large twist angles, is resilient against the displacement field exhibiting a robust many-body gap, as illustrated in Fig.~\ref{fig:ED_spectra}(d) for representative angle $\theta=2.7^\circ$ and large displacement field, $V_D=5$ meV.
While the Berry curvature distribution is centered around the $\gamma$ point (at large twist angles), the holes are strongly distributed around the corners of the mBZ, as shown in Fig.~\ref{fig:ED_spectra}(l), thereby avoiding their exposure to magnetic flux, which would otherwise favor the FQAH state rather than CDW-$2$ state.
The structure factor, shown in Fig.~\ref{fig:ED_spectra}(h), reveals that CDW-$2$ shares the same wavevector ordering, $\mathbf{q=\kappa_\pm}$, as CDW-$1$. 
The real space patterns of the pair correlation function, $g(\mathbf{r})$, shown in Fig.~\ref{fig:LP}(c), also resemble those of CDW-$1$ (shown in the inset of Fig.~\ref{fig:FCI_CDW}(a)), albeit with less pronounced peaks (due to kinetic energy), strongly suggesting that both exhibit commensurate $\sqrt{3}\times\sqrt{3}$ CDW order.

To understand the differences between these two CDWs, we analyze their layer polarization characteristics.
Particularly, we investigate the real space charge density across the top and bottom layers, $n_{tot}(\mathbf{r})=\sum_l n_l(\mathbf{r})$, where $n_l(\mathbf{r}) = \frac{1}{N_{uc}} \sum_\mathbf{k} \psi^*_{\mathbf{k},l}(\mathbf{r})  \psi_{\mathbf{k},l}(\mathbf{r}) \langle\gamma^\dagger_\mathbf{k} \gamma_\mathbf{k} \rangle$, with $l$ being the layer index (see Appendix~\ref{sec:local density} for details).
For the same given value of $V_D=4$ meV, we compare $n_{tot}(\mathbf{r})$ at $\theta=2^\circ$ (corresponding to the CDW-$1$ phase) and $\theta=2.7^\circ$ (deep in the CDW-$2$ phase), as shown in Fig.~\ref{fig:LP}(a) and (b) respectively.
For $\theta=2^\circ$, the pattern resembles that for $\nu=1$ discussed previously (see Fig.~\ref{fig:band_topology}(b)), forming a triangular lattice due to strong charge localization on the bottom layer of XM moir\'e stacking sites. 
However, for $\theta=2.7^\circ$, the pattern is notably different, forming a kagome geometry with localization around both XM and MM stacking sites.
To quantify this qualitative observation, we calculate the layer polarization, defined as
\begin{equation}
    P_l=\frac{n_{bot}-n_{top}}{n_{bot}+n_{top}},
\end{equation}
where $n_{top/bot}=\sum_\mathbf{r}n_l (\mathbf{r})$ represents the total charge density on the top/bottom layer.
In Fig.~\ref{fig:LP}(d), we show $P_l$, averaged over the nearly degenerate members in the ground state, as a function of $\theta$, at a fixed value of the displacement field $V_D=4$ meV.  
Clearly, the layer polarization $P_l$ decreases as the twist angle $\theta$ increases, yet it exhibits a small but noticeable region around $\theta \sim 2.18^\circ$, where the slope of the curve changes significantly. 
Away from this point, $P_l$ for $24$ and $27$ unit cell clusters closely track each other; however, they deviate close to the transition point, showcasing stronger finite-size effects. 
The $N_{uc}=27$ cluster shows a pronounced effect in the $P_l$ variation compared to $N_{uc}=24$, although these two cluster sizes are comparable. 
To fully elucidate the true nature of phase transition, larger clusters containing tripled unit cells are necessary; however, the next available cluster size is $36$, with a  Hilbert space dimension of $\sim 35 \times 10^6$ for each momentum sector, which is computationally very expensive for calculations at multiple parameters.
Nevertheless, examining the variation in the slope of $P_l$ may provide insights less susceptible to finite size effects.
In the inset of Fig.~\ref{fig:LP}(d), we illustrate the slope of $P_l$ around the transition point as a function of $\theta$. 
This analysis clearly shows a divergence at the critical value of $\theta$, providing stronger evidence of the first-order transition between CDW-$1$ and CDW-$2$.
%
\begin{figure}
\includegraphics[width=0.5\linewidth]{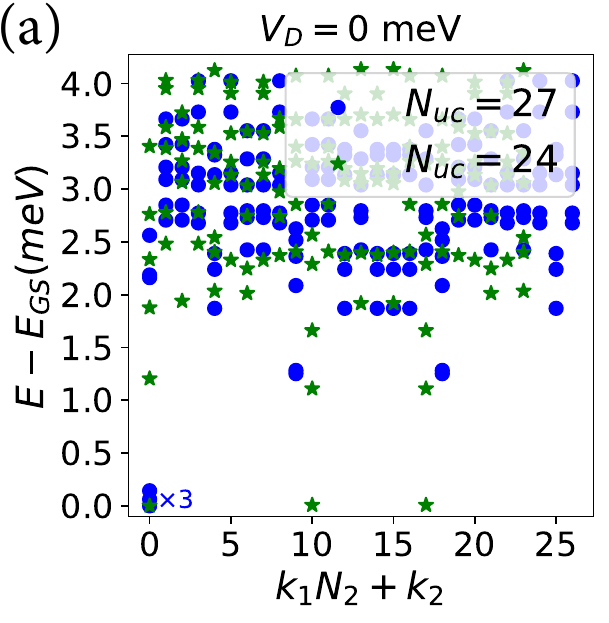} 
\includegraphics[width=0.48\linewidth]{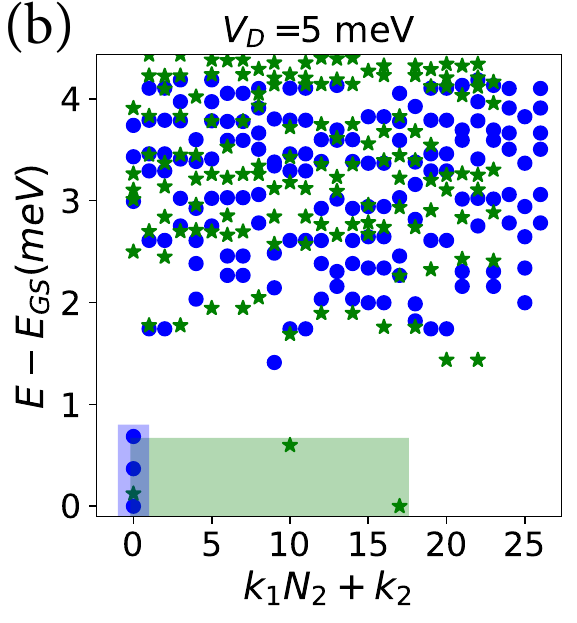} 
\\
\includegraphics[width=0.49\linewidth]{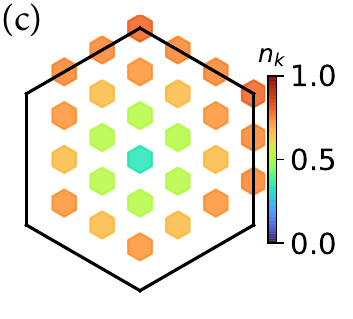}
\includegraphics[width=0.49\linewidth]{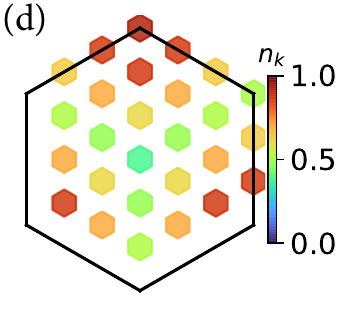}
\caption{The low lying many-body energy spectra at $\nu=\frac{2}{3}$ filling for $\theta=2.5^\circ$, and displacement fields: (a) $V_D=0$ and (b) $V_D=5$ meV. The light blue and green rectangular regions mark the nearly degenerate ground states in the corresponding clusters.
(c) and (d) are the corresponding Bloch state occupations averaged over the members of nearly degenerate ground states shown in the panel directly above them for $N_{uc}=27$. 
}
\label{fig:spectra_23}
\vspace{-0.15cm}
\end{figure}
%

\begin{figure*}
\includegraphics[width=0.44\linewidth]{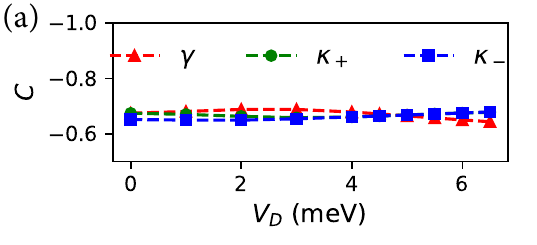} 
\includegraphics[width=0.55\linewidth]{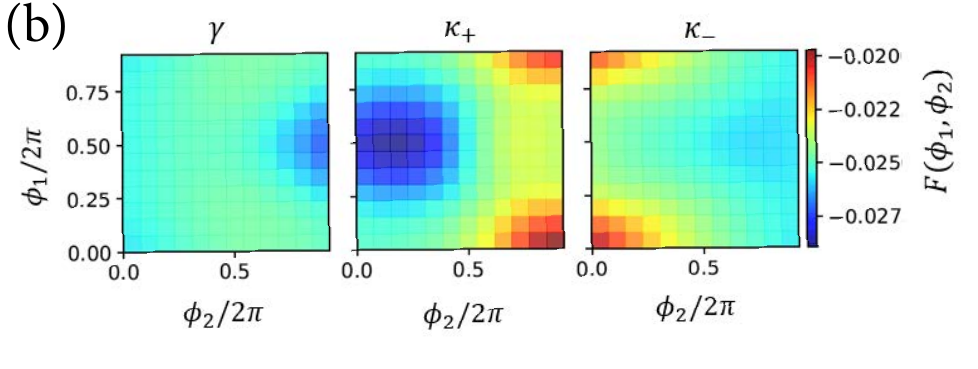} 
\caption{(a) The many-body Chern numbers of nearly degenerate ground states as a function of $V_D$ at three distinct COM momenta on a $24$-unit cell cluster for $\nu=\frac{2}{3}$ filling at $\theta=2.5^\circ$. (b) The many-body Berry curvature of the ground states as a function of discretized boundary phases at $\theta=2.5$ and $V_D=4$ meV.
}
\label{fig:chernnum_23}
\vspace{-0.15cm}
\end{figure*}
%

We further analyze ground state energies and their first derivatives with respect to parameter $\theta$ at COM momenta $\gamma$ and $\kappa_\pm$ separately in Fig.~\ref{fig:LP}(e) (where we take the average of the two energies at $\kappa_+ \text{ and } \kappa_-$).
The ground state energies at $\gamma$ and $\kappa_\pm$ are almost perfectly degenerate in the range of twist angle we study. However, there is a tiny splitting around the critical value of $\theta$ that lifts the energies at $\kappa_\pm$ to slightly higher values (see Appendix~\ref{sec:CDW1_CDW2}).
This ground state energy at $\kappa_\pm$ contains a small but noticeable kink, as shown in the inset of Fig.~\ref{fig:LP}(e), and its slope exhibits clear jumps at the transition point, providing consistent numerical evidence of first-order phase transition between $\text{CDW-}1$ and CDW-$2$.
Importantly, this discontinuity in the energy derivative manifests only in the $\kappa_\pm$ sector, whereas $\gamma$ is continuous for a given cluster with $N_{uc}=27$ unit cells.
The apparent two jumps in the slope at $\kappa_\pm$ are due to finite size effects, which can occur when there exists a broader region for level crossing, as shown in Fig.~\ref{fig:level_cross_Vd=4} in the Appendix~\ref{sec:CDW1_CDW2}.
\subsection{Robust FCI at $\nu=\frac{2}{3}$}
Given that the particle-hole symmetry is broken within the band-projected Hamiltonian~\cite{yu2024fractional,abouelkomsan2020particle}, unlike at $\nu=\frac{1}{3}$, we observe a robust FCI phase for $\nu=\frac{2}{3}$ filling, as depicted in the right panel of Fig.\ref{fig:phase_diagram}.
The resilience of the FCI phase at $\nu=\frac{2}{3}$ against twist angles has been previously reported in the literature~\cite{reddy2023fractional}, and here we further affirm its resilience against the displacement field.
Only at large displacement fields, very close to the transition of the single-particle band Chern number, does the FCI become unstable to the CDW at small twist angle $\theta\sim 1.9$ (see Appendix~\ref{app:twothird}), which we identify as a topologically trivial state via many-body Chern number calculations.
Our study supports the conclusion~\cite{wang2024fractional} that the disappearance of FCI occurs well before the ferromagnetism disappears.
For example, at $\theta=2^\circ$, FCI-CDW transition occurs around $V_D=5.8$ meV, while the full spin gap ($\Delta_S\sim 2$ meV) at this value of $V_D$ is still robust.

To show the robustness of FCI, we begin by analyzing the low-lying many-body energy spectrum, shown in Fig.~\ref{fig:spectra_23}(a)-(b), at a twist angle of $\theta=2.5^\circ$ and representative $V_D$.
These spectra reveal nearly three-fold degenerate ground states for both cluster sizes, clearly gapped from the excited states and positioned at the COM momenta permitted by the counting rule. 
However, at a large displacement field, $V_D=5$ meV, the ground states are more spread out but still remain gapped from the excited states. 
The bottom panels (c) and (d) in Fig.~\ref{fig:spectra_23} represent the Bloch state occupation numbers, which are almost uniformly distributed inside the mBZ, with some variation at the corners at finite $V_D$ due to valley polarization. 
All these evidences point towards the FCI phase being resilient against displacement field.

Now, we present stronger evidence for the robustness of the FQAH effect by calculating the many-body Berry curvature and Chern number defined in equation~\eqref{eq:mchernnum}. 
For representative parameters $\theta=2.5^\circ$ and $V_D=4$ meV, Fig.~\ref{fig:chernnum_23}(b) represents the Berry curvature of the nearly degenerate ground states as a function of discretized boundary phase (of $14\times 14$ mesh) in $24$ unit cell cluster (in which ground states occur at three distinct COM momenta $\gamma$ and $\kappa_\pm$).
We observe small fluctuations in Berry curvatures at each ground state.
Notably, along the $\phi_1$ direction, each graph exhibits perfect periodicity, while along $\phi_2$, the three panels form a continuous spectral flow-cycle $\gamma\rightarrow \kappa_+ \rightarrow \kappa_- \rightarrow \gamma$, confirming the existence of three ground state manifolds where the insertion of one flux quantum shifts the momentum sector from $\gamma \rightarrow \kappa_+$, $\kappa_+ \rightarrow \kappa_-$, and $\kappa_- \rightarrow \gamma$.
Adding up the discretized values of $F(\phi_1,\phi_2)$ separately in each sector yields the many-body body Chern number $\mathcal{C}$, which is presented in Fig.~\ref{fig:chernnum_23}(a) as a function of $V_D$, with the twist angle fixed at $\theta=2.5^\circ$.
In this plot, we observe that the total Chern number is almost uniformly distributed among three ground states, with each state exhibiting a Chern number close to $\mathcal{C}= \frac{2 }{3}$ 
(any small deviation from an ideal value $\mathcal{C}=\frac{2}{3}$ is due to the finite size effect; however, the sum of three Chern numbers is exactly equal to $2$ within numerical precision).
This is compelling evidence of charge fractionalization with a quantized fractional Hall response $\sigma_{xy}=\frac{2}{3}\frac{e^2}{h}$, where $e$ is the electron (hole) charge and $h$ is the Planck constant. 

Despite the broader spread in the ground state manifold and a small many-body gap within finite clusters studied 
, the persistence of a non-vanishing quantized many-body Chern number up to a large displacement field of $V_D=6.5$ meV convincingly demonstrates the robustness of the $\frac{2}{3}$ FQAH state against both the displacement field and twist angle variations.

\section{Conclusion}
In summary, we presented the topological quantum phase diagram for both $\frac{1}{3}$ and $\frac{2}{3}$ filling of the lowest moir\'e band of $t$MoTe$_2$ by varying both the twist angles and displacement fields. 
Our study demonstrates that out-of-plane displacement fields, in general, can have a non-trivial role in the band topology and many-body correlated states of bilayer moir\'e superlattices. 
While high displacement fields suppress the FM and drive the single-particle bands into topologically trivial states, the effects at intermediate values cannot be overlooked. 
For instance, we observe a transition from a fractional Chern insulator to a layer-polarized charge density wave (CDW-$1$) order at filling faction $\nu=\frac{1}{3}$ driven by the displacement field. 
Additionally, we identify a first-order phase transition from CDW-$1$ to a layer hybridized CDW-$2$ induced by twist angle at finite displacement fields. 
This first-order phase transition between two CDW orders results from the competition between the kinetic energy (controlled by twist angle) and layer potential energy (controlled by out-of-plane displacement field). 
These findings emphasize the need for further experimental and theoretical investigations in the future. 

Furthermore, determining the precise location of the phase boundary and further confirming the nature and order of the phase transitions may require studying larger system sizes, which could be explored using alternative techniques such as Density Matrix Renormalization Group (DMRG).
Recently, it has been claimed in ref.~\cite{yu2024fractional} that underlying magnetism is sensitive to band mixing, and ferromagnetism significantly weakens when two bands are considered. While their claim is based on large twist angle analysis ($\theta \geq 3.5^\circ$) where bands are dispersive and FCI vanish at $\frac{1}{3}$ filling (for parameters in ref.~\cite{reddy2023fractional}), a detailed study of magnetism for smaller twist angle under band mixing remains to be explored. Such an investigation would require finite-size analysis, which is challenging with small-size ED methods; however, DMRG could be useful in this context.
Nevertheless, our exact diagonalization study within the lowest band projected Coulomb interaction adds valuable insights into the interplay between strong correlation effects and band topology of TMD-based moir\'e bilayer semiconductors under the influence of displacement fields. 
Moreover, given that the displacement field is easily controlled in experiments, our study sheds light on how topological phases of bilayers can be manipulated to achieve desired outcomes.

Last but not least, our study further strengthens the previous findings regarding the resilience of the $\frac{2}{3}$ FQAH state against variations in the moir\'e twist angles. 
Additionally, we demonstrate its stability against electric fields, which could make it viable for applications in future quantum devices.
While our focus was on $t$MoTe$_2$, above findings are expected to have broader applicability to TMD-based homobilayers in general.

\begin{acknowledgements}
P. S. thanks Ken K. W. Ma for some initial conversations.
This work was supported by the US National Science Foundation (NSF)
 through the Partnership in Research and Education in Materials Grant
 No. DMR-1828019 (P.S, D.N.S) and NSF PREP Grant No. PHY-2216774 (Y.P.,D.N.S).
\end{acknowledgements}

\appendix

\begin{widetext}
\section{Single particle bandwidth, charge distribution, and geometric tensor}
\label{sec:single_particle}
In the main text, we show the effects of displacement fields in the band topology. Here, we further show its effect in the band dispersion, real-space local charge distribution, and quantum geometry within the lowest moir\'e band. Bandwidth is defined as the difference between maxima and minima of band energy, i.e., $\Delta E_w=max(\epsilon_\mathbf{k})-min(\epsilon_\mathbf{k})$.
\begin{figure}[H]
\includegraphics[width=0.35\linewidth]{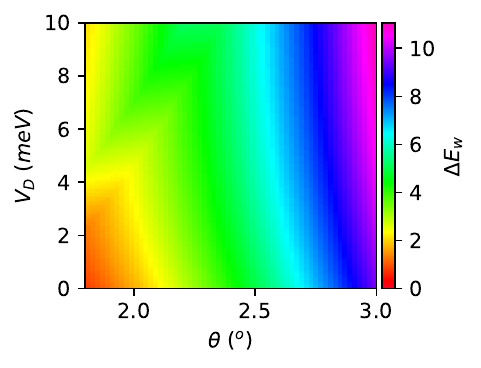} 
\includegraphics[width=0.33\linewidth]{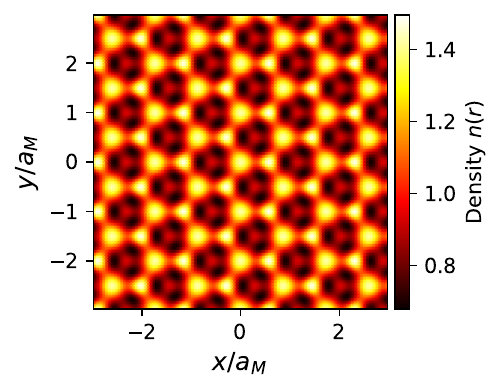}
\includegraphics[width=0.3\linewidth]{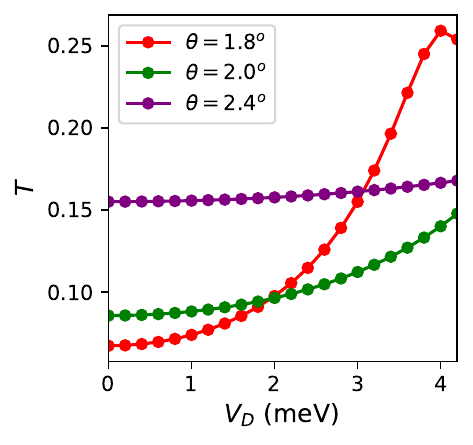}
\caption{(Left) Bandwidth of the lowest moir\'e band from the continuum model as a function of twist angle $\theta$ and displacement field $V_D$. 
(Middle) The real space charge distribution (corresponding to $\nu=1$ filling) of the lowest hole moir\'e band at $\theta=3^\circ$ and $V_D=4$ meV.
(Right) Trace condition deviation obtained from the quantum metric tensor as a function of $V_D$ for various twist angles.}
 \label{fig:band_width}
\end{figure}
In the left panel of Fig.~\ref{fig:band_width}, we present the bandwidth as a function of twist angles and displacement fields. Clearly, the bandwidth increases, consequently enhancing the effective kinetic energies of holes with increasing twist angle. However, it remains less sensitive to the displacement fields ($0-10$ meV), particularly at large twist angles. 

In the middle panel, we show the charge distribution $n(\mathbf{r})$  at large twist angle $\theta=3^\circ$ and $V_D=4$ meV. Unlike the case at $\theta=2^\circ$ discussed in the main text, here we observe an almost uniform distribution of charge density across the top and bottom layers, indicating layer hybridization tendency due to larger bandwidth. In the absence of $V_D$, the charge density is peaked at MM stacking sites, forming a triangular lattice at~\cite{reddy2023fractional}. This triangular pattern gradually changes into a kagome geometry as $V_D$ is turned on. This new kagome structure exhibits density concentration at both MM and XM stacking sites.

We also analyze the so-called trace condition of the lowest Chern band, which is encoded in the quantum geometric tensor:
\begin{equation}
    \mathcal{Q}^{\mu\nu}(\mathbf{k}) = \bra{\partial^\mu u_\mathbf{k}}(1-\ket{u_\mathbf{k}}\bra{u_\mathbf{k}})\ket{\partial^\nu u_\mathbf{k}}
\end{equation}
in the unit of mBZ area, where $\mu \nu=x,y$. The imaginary part of $\mathcal{Q}^{\mu \nu}$ defines the Berry curvature,
    $F^{\mu\nu}(\mathbf{k})=-2Im(Q^{\mu \nu})$,
which resembles the magnetic field in the reciprocal space. The integral Berry curvature over the mBZ provides a topological invariant quantity (which does not change with the band dispersion unless the band gap closes) called Chern number,
\begin{equation}
    \mathcal{C} = \frac{1}{2\pi}\int_{\mathbf{k}\in mBZ}F^{xy}(\mathbf{k})d^2\mathbf{k}
\end{equation}
which must be an integer number, as we have shown in the main text. On the other hand, the real part of $\mathcal{Q^{\mu\nu}}$ is the Fubini-Study (FS) metric, $g^{\mu\nu}(\mathbf{k})=Re(\mathcal{Q^{\mu\nu}})$, which is the measurement of distance between two Bloch states in the momentum space. The Berry curvature and FS metric are related by two important inequalities~\cite{roy2014band},
\begin{equation}
    tr[g(\mathbf{k})]\geq |F(\mathbf{k})| \quad\quad\text{ and }\quad \quad det[g(\mathbf{k})] \geq \frac{1}{4} |F(\mathbf{k})|^2.
\end{equation}
Since the determinant condition can be obtained from the trace condition~\cite{roy2014band}, the second is automatically satisfied when the first condition is fulfilled. Therefore, in Fig.~\ref{fig:band_width}(right panel), we show the deviation from the saturated trace condition
\begin{equation}
    T = \frac{1}{2\pi}\int_{\mathbf{k}\in mBZ} d^2\mathbf{k}\left( tr[g(\mathbf{k})]- |F(\mathbf{k})| \right)
    = \int_{\mathbf{k}\in mBZ} d^2\mathbf{k} \frac{tr[g(\mathbf{k})]}{2\pi}-|C|.
\end{equation}
The smaller the value of $T$ (i.e., $T\rightarrow 0$), the closer the Bloch bands mimic the lowest Landau level (LL) physics. 
The trance condition is minimum for small twist angles ($\theta=1.8^\circ \text{ and } 2^\circ$) and small values of $V_D$, while it deviates a lot from $0$ at large twist angle, as shown in the right panel of Fig~\ref{fig:band_width}.
This indicates that FCI is more favorable at smaller twist angles and displacement fields as we demonstrated in the main text. 
Most remarkably, $T$ increases rapidly for small angle $\theta=1.8^\circ$ as $V_D$ is increased and surpasses its value for $\theta=2^\circ$ at $V_D\geq 2$ meV. In other words, fluctuations in $T$ occur at a slower rate with larger twist angles, possibly supporting the persistence of FCI at higher $V_D$ values. This observation concurs with our main text findings, wherein we observe the FCI phase boundary at $\nu=\frac{1}{3}$ shifting towards larger $V_D$ as the twist angle increases from $\theta=1.8^\circ$. 
\section{Band projected Coulomb interaction}
\label{sec:coulomb}
We begin by diagonalizing the continuum model Hamiltonian, $h_\mathbf{k}$, defined in the main text in equation~\eqref{eqn:continuum_model}. To diagonalize $h_\mathbf{k}$, we denote $\ket{\mathbf{k}+\mathbf{g},l,\sigma}$ as the original (local) momentum basis with lattice vector $\mathbf{g}$, layer index $l=1,2$, and single particle (crystal) momenta $\mathbf{k}$. 
The original hole operators that create this state are 
\begin{equation}
    c_{\mathbf{k+g},l,\sigma}^\dagger = \frac{1}{\sqrt{N_{uc}}}\sum_\mathbf{r} e^{i(\mathbf{k+g})\cdot \mathbf{r}} c_{\mathbf{r},l,\sigma}^\dagger
    \label{eq:local_creation_operator}
\end{equation}
 where $c^\dagger_{\mathbf{r},l,\sigma}$ creates a hole with spin index $\sigma$ at position $\mathbf{r}$ in the layer $l$. Note that $\mathbf{k}$ is the unique point in the ``mesh grid" restricted to the first moir\'e Brillouin Zone (mBZ).

To add Coulomb interaction, we first construct the periodic part of the Bloch state which can be written as
\begin{equation}
    \ket{u_\mathbf{k},\sigma} = \sum_{\mathbf{g},l} z_{\mathbf{g},l,\sigma}(\mathbf{k}) \ket{\mathbf{k+g},l,\sigma}
    \label{blochstate}
\end{equation}
where $z_{\mathbf{g},l,\sigma}(\mathbf{k})$ is the eigenvector component of lowest flat band of non-interacting hamiltonian in equation~\eqref{eqn:continuum_model} at wavevector $\mathbf{k}\in$ mBZ. Because we are interested in projecting Coulomb interactions to the lowest flat band, we omitted the band index $n$ assuming the Bloch states \eqref{blochstate} are constructed from the lowest flat band.

Having defined band eigenvectors, one can now define a new set of fermionic operators $\gamma_\mathbf{k}^\dagger$ that creates a particle with crystal momentum $\mathbf{k}\in$ mBZ in the lowest Bloch state such that
\begin{equation}
    \gamma_{\mathbf{k},\sigma}^\dagger = \sum_{\mathbf{g},l} z_{\mathbf{g},l,\sigma}(\mathbf{k}) c_{\mathbf{k+g},l,\sigma}^\dagger
    \label{eq: block operator}
\end{equation}
which means $\ket{u_{\mathbf{k},\sigma}}= \gamma_{\mathbf{k},\sigma}^\dagger \ket{\text{vac}}$. 

The kinetic term of the Hamiltonian, with projection to the lowest flat band, reads
\begin{equation}
     H_{\text{kin}} = \sum_{\mathbf{k},\sigma}\epsilon_\mathbf{k,\sigma} \gamma_\mathbf{k,\sigma}^\dagger \gamma_\mathbf{k,\sigma}.
\end{equation}
where $\epsilon_\mathbf{k}$ are the lowest flat band energies at $\mathbf{k\in}\text{ mBZ}$.
\\
Inverting the relation~\eqref{eq: block operator}, each component of the original fermionic operator in the continuum model can be written as
\begin{equation}
    c_{\mathbf{k+g},l,\sigma} = \langle \mathbf{k+g},l,\sigma|u_\mathbf{k,\sigma} \rangle \gamma_\mathbf{k,\sigma}=
    z_{\mathbf{g},l,\sigma}(\mathbf{k}) \gamma_\mathbf{k,\sigma}
    \label{eq:operator transformation}
\end{equation}
For mathematical clarity, let us explicitly denote the momentum in the continuum model by $\Tilde{\mathbf{k}}$ which in the moir\'e band basis transforms to $\mathbf{k}$ by a reciprocal vector $\mathbf{g_k}$ as
\begin{equation}
\Tilde{\mathbf{k}} = \mathbf{k+g_k}.   
\label{eq:transformation}
\end{equation}
Here on, we will stick to this convention where momentum strictly in the mBZ will be denoted without a tilde, otherwise with a tilde sign (unless specified). 
\\
The generic Coulomb Hamiltonian can be expressed in terms of the normal ordering of density-density interactions,
\begin{equation}
    H_C = \frac{1}{2A} \sum_{\Tilde{\mathbf{q}}} V(\Tilde{q}) :\Bar{\rho}(\Tilde{\mathbf{q}}) \Bar{\rho}(-\Tilde{\mathbf{q}}):
    \label{eq:coulomb_interaction}
\end{equation}
where $V(\Tilde{q})=\frac{2\pi e^2}{\epsilon |\Tilde{\mathbf{q}}|}$ is the Fourier transform of the Coulomb interaction in 2D, and $A=\frac{\sqrt{3}}{2}a^2_M L_x\times L_y$ is the area of moir\'e Brillouin zone, $L_x$ and $L_y$ are the number of  moir\'e unit cells along two directions. Note that the density operator $\Bar{\rho}(\Tilde{\mathbf{q}})$ is projected to the lowest flat band. The projected density operator is defined in the original continuum basis with momentum index $\Tilde{\mathbf{k}}$ and layer index $l$ as
\begin{equation}
\begin{split}
    \Bar{\rho}(\Tilde{\mathbf{q}}) &= \sum_{\Tilde{\mathbf{k}},l,\sigma} c^\dagger_{\Tilde{\mathbf{k}}+\Tilde{\mathbf{q}},l,\sigma}
    c_{\Tilde{\mathbf{k}},l,\sigma}
    =
    \sum_{\mathbf{k,g_k},l,\sigma} c^\dagger_{\mathbf{(k+\Tilde{q})+g_k},l,\sigma} c_{\mathbf{k+g_k},l,\sigma}
\end{split}
\label{eq:density operator}
\end{equation}
Folding $(\mathbf{k}+\Tilde{\mathbf{q}})$ back to the mBZ using $(\mathbf{k}+\Tilde{\mathbf{q}})= \mathbf{k+q+g_{k+q}}$ and subsequently applying relation~\eqref{eq:operator transformation}, equation~\eqref{eq:density operator} takes the form
\begin{equation}
    \begin{split}
       \Bar{\rho}(\Tilde{\mathbf{q}}) &= \sum_{\mathbf{k,g_k},l,\sigma} c^\dagger_{\mathbf{k+q+g_k+g_{k+q}},l,\sigma} c_{\mathbf{k+g_k},l,\sigma}
        \\
        &=
         \sum_{\mathbf{k,g_k},l,\sigma} z^*_{\mathbf{g_k+g_{k+q}},l,\sigma}(\mathbf{k+q}) z_{\mathbf{g_k},l,\sigma}(\mathbf{k}) \gamma^\dagger_\mathbf{k+q,\sigma} \gamma_\mathbf{k,\sigma}
         \\
         &=
         \sum_{\mathbf{k},\sigma} F(\mathbf{k+q,k,g_{k+q},\sigma}) \gamma^\dagger_\mathbf{k+q,\sigma} \gamma_\mathbf{k,\sigma} 
    \end{split}
\end{equation}
where we define a $F$-function,
\begin{equation}  F(\mathbf{k}_1,\mathbf{k}_2,\mathbf{g_{k^\prime},\sigma}) = \sum_{\mathbf{g_k},l} z^*_{\mathbf{g_k+g_{k^\prime}},l,\sigma}(\mathbf{k}_1) z_{\mathbf{g_k},l,\sigma}(\mathbf{k}_2)
\end{equation}
Notice that all $\mathbf{k_1,k_2}\in \text{mBZ}$ in the final expression of the density operator.
Therefore, Coulomb interaction hamiltonian~\eqref{eq:coulomb interaction} takes the form,
\begin{equation}
\begin{split}
    H_C &= \frac{1}{2A}\sum_{\mathbf{q}\in R^2} V(\mathbf{q}) \sum_{\{\mathbf{k}\}, \sigma,\sigma^\prime} F(\mathbf{k_1,k_3, g_{k_3+q},\sigma}) F(\mathbf{k_2,k_4, g_{k_4-q},\sigma^\prime})
    \gamma^\dagger_{\mathbf{k_1},\sigma} \gamma^\dagger_{\mathbf{k_2},\sigma^\prime} \gamma_\mathbf{k_4,\sigma^\prime} \gamma_\mathbf{k_3,\sigma}       
\end{split}
    \label{eq:interaction}
\end{equation}
with the constraints $\mathbf{k_1}=\mathbf{(k_3+q)} \text{ modulo } \mathbf{g_{k_3+q}}$ and $\mathbf{k_2}=\mathbf{(k_4-q)} \text{ modulo } \mathbf{g_{k_4-q}}$. 
\\
Finally, the full many-body Hamiltonian with the lowest band projection takes the following form,
\begin{equation}
    H = \sum_\mathbf{k,\sigma} \epsilon_\mathbf{k,\sigma} \gamma_\mathbf{k,\sigma}^\dagger \gamma_\mathbf{k,\sigma} + \frac{1}{2} \sum_{\{\mathbf{k}\},\sigma,\sigma^\prime} V^{\sigma,\sigma^\prime}_{\mathbf{k_1, k_2, k_3, k_4}} \gamma^\dagger_\mathbf{k_1,\sigma} \gamma^\dagger_\mathbf{k_2,\sigma^\prime} \gamma_\mathbf{k_4,\sigma^\prime} \gamma_\mathbf{k_3,\sigma} 
    \label{eq:manybody_hamiltonian}
\end{equation}
where 
$$
V^{\sigma,\sigma^\prime}_{\mathbf{k_1, k_2, k_3, k_4}} = \sum_{\mathbf{q}\in R^2} \frac{2\pi e^2}{A\epsilon|\mathbf{q}|} 
F(\mathbf{k_1,k_3, g_{k_3+q},\sigma}) 
F(\mathbf{k_2,k_4, g_{k_4-q},\sigma^\prime})
$$
After this point, we will drop the spin indices $\sigma$ as all the observables are computed in the fully polarized ground states. 
\section{Charge Structure Factor}
\label{sec:CSF}
The connected charge structure factor is defined as,
\begin{equation}
\begin{split}
    S(\mathbf{q}) &= \frac{1}{N_{uc}} \left[  \langle \Bar{\rho}_\mathbf{q} \Bar{\rho}_\mathbf{-q} \rangle - \langle \Bar{\rho}_\textbf{q}\rangle \langle \Bar{\rho}_{-q}\rangle  \right]
    \\ &=
    \frac{1}{N_{uc}}\left[  
    \sum_{\mathbf{k_1,k_2}\in \text{mBZ}} 
    F(\mathbf{k_1+q,k_1, g_{k_1+q}}) F(\mathbf{k_2-q,k_2, g_{k_2-q}}) 
   \langle \gamma^\dagger_\mathbf{k_1+q} \gamma_\mathbf{k_1}
     \gamma^\dagger_\mathbf{k_2-q} \gamma_\mathbf{k_2} \rangle
     - N^2 \delta_{\mathbf{q,}0}\right]
\end{split}
\end{equation}
where $N$ is the total number of particles. Notice that $\mathbf{k_1+q}$ and $\mathbf{k_2-q}$ are already folded back to the mBZ using the transformation relation~\eqref{eq:transformation} where the corresponding reciprocal vectors are incorporated into the $F$ functions. In general, $\mathbf{q}$ are not restricted to mBZ, but it is sufficient for us to consider $\mathbf{q}$ in the mBZ to understand the underlying charge structures of material.  
The expectation value $\langle \gamma_\mathbf{k_1+q}^\dagger \gamma_\mathbf{k_1} \gamma_\mathbf{k_2-q}^\dagger \gamma_\mathbf{k_2} \rangle $ is calculated on the many-body (nearly) degenerate ground states and averaged out. 
\section{Many-body local density operator}
\label{sec:local density}
In the main text, we have computed the real-space distribution of charge density within the many-body ground state for a specified filling fraction. Here, we proceed to derive the many-body local density operator $n(\mathbf{r})$, which can be defined for each layer with index $l$ as,
\begin{equation}
    n_l(\mathbf{r}) = c^\dagger_{\mathbf{r},l} c_{\mathbf{r},l},
    \label{eq:manybody_density}
\end{equation}
where $c^\dagger_{\mathbf{r},l}$ and $c_{\mathbf{r},l}$ are local creation and annihilation operators in the original monolayer as per our definition in equation~\eqref{eq:local_creation_operator}. Substituting the Fourier transform of these operators,
\begin{equation}
\begin{split}
    n_l(\mathbf{r}) &= \frac{1}{N_{uc}} \sum_\mathbf{k,g} e^{-i(\mathbf{k+g})\cdot \mathbf{r}} c^\dagger_{\mathbf{k+g},l} \sum_{\mathbf{k^\prime, g^\prime}} 
    e^{i(\mathbf{k^\prime+g^\prime})\cdot \mathbf{r}} c_{\mathbf{k^\prime+g^\prime},l}
    \\ &=
    \frac{1}{N_{uc}} \sum_\mathbf{k} \psi^*_{\mathbf{k},l}(\mathbf{r})  \psi_{\mathbf{k},l}(\mathbf{r}) \gamma^\dagger_\mathbf{k} \gamma_\mathbf{k}
    \end{split} 
    \label{eq:nr}
\end{equation}
where, 
\begin{equation}
    \psi_{\mathbf{k},l}(\mathbf{r}) = \sum_{\mathbf{g}} 
    e^{i(\mathbf{k+g})\cdot \mathbf{r}} z_{\mathbf{g},l} (\mathbf{k})
\end{equation}
Note that the factor $\frac{1}{N_u}$ helps to make comparison of the expectation value $\langle n_l(\mathbf{r})\rangle$ between different system sizes. 
The total density operator is the sum of local density operators in each layer,
\begin{equation}
    n_{tot}(\mathbf{r}) = \sum_{l=1,2} n_l(\mathbf{r})
\end{equation}
While $\mathbf{r}$ is a continuous variable, in practice, we sample discrete values of $\mathbf{r}$ within the unit cell. If $N_\mu$ $(\mu=1,2)$ are the number of sampling points taken along the lattice vectors $\mathbf{a}_1$ and $\mathbf{a}_2$ within the unit cell, then in the limit of an infinite number of real space sample points $\mathbf{r}$ (i.e., $N_1,N_2 \rightarrow \infty$), the total charge density must satisfy the following sum rule
\begin{equation}
    \nu = \frac{1}{N_1 N_2}\sum_\mathbf{r} \langle n_{tot}(\mathbf{r}) \rangle
\end{equation}
where $\nu$ is the filling fraction.

\section{Full spin gap at large displacement field}
\label{APP:Ising_FM}
In the main text, we demonstrate the spin-1 gap and full spin gap for representative values of displacement fields. To address the issue at large displacement field, where the ferromagnetism is weaker, here we further analyse the full spin gap in the smallest $|Sz|$ sector at large displacement field $V_D=6$ meV for both angles $\theta=2^\circ$ and $\theta=3^\circ$ on three different system sizes. 

\begin{figure}[H]
\includegraphics[width=0.4\linewidth]{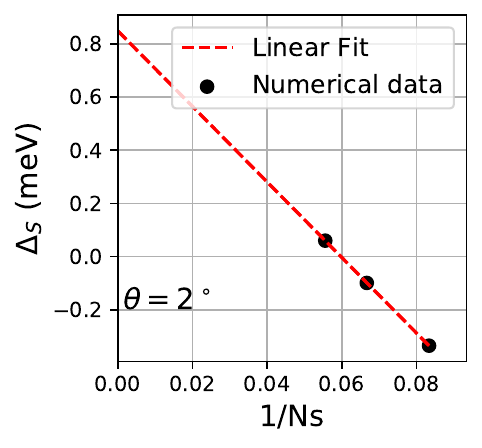} 
\includegraphics[width=0.4\linewidth]{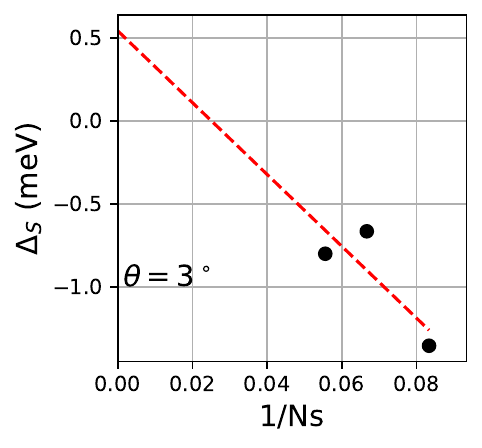}
\caption{Finite size scaling of the full spin gap, $\Delta_S $, at one third filling at large displacement field, $V_D=6$ meV. Numerical calculations are carried out in $Sz=0$ sector for $12$ and $18$ sites clusters, and in the $Sz=1/2$ sector for $15$ site clusters. The red dashed line represents linear extrapolation to the thermodynamic limit.}
 \label{fig:spinfull_gap_Vd6}
\end{figure}
As shown in the left panel of Fig.~\ref{fig:spinfull_gap_Vd6}, at $\theta=2^\circ$ the spin gap linearly increases with the system size and extrapolates to $\Delta_S=0.85$ meV in the thermodynamic limit, indicating ferromagnetic ground state. 
However, at large twist angle $\theta=3^\circ$, finite size effects are strong and there is no clear trend, posing challenge to draw a conclusion.

\section{Extended data for $\nu=\frac{1}{3}$ filling}
\label{sec:extended13}
In this section, we presented ED data to further characterize the phase transitions at $\nu=\frac{1}{3}$ filling of the lowest moir\'e band of $t$MoTe$_2$.
Throughout the calculations, we assumed full spin/valley polarization for the range of displacement fields between $V_D=0-6.5$ meV. 

\begin{figure}[H]
\includegraphics[width=0.247\linewidth]{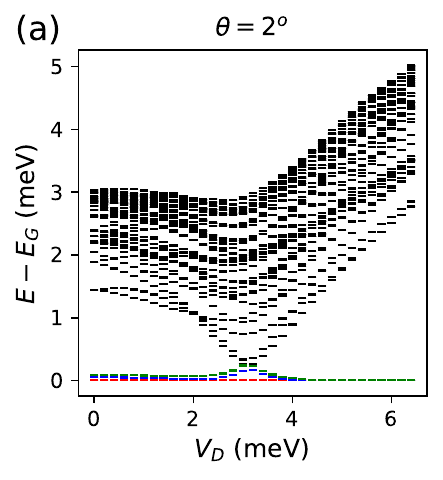} 
\includegraphics[width=0.247\linewidth]{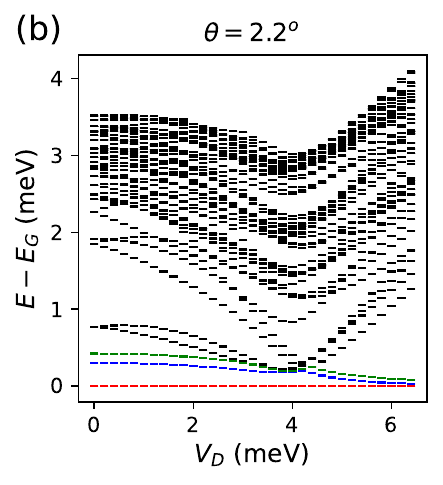}
\includegraphics[width=0.247\linewidth]{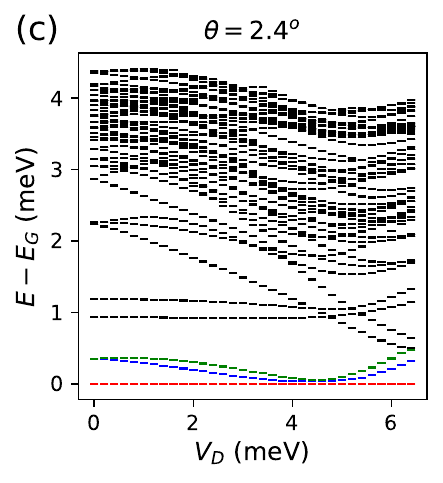}
\includegraphics[width=0.247\linewidth]{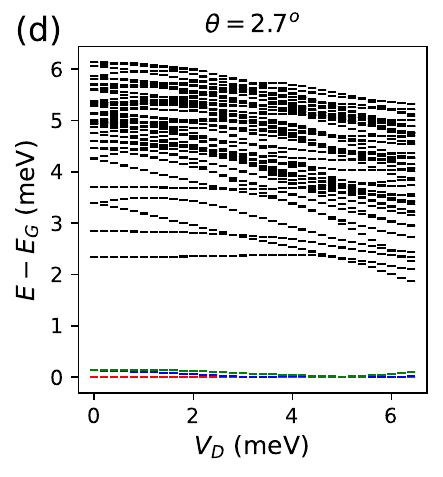}
\caption{The low-lying many-body energy spectrum (after subtraction from the corresponding ground state energies) as a function of displacement field $V_D$ for various twist angles.  Calculations are performed on the 27 unit cell clusters at $\nu=1/3$ filling. Red, blue, and green markers denote the lowest three states.}
 \label{fig:level_corssing13}
\end{figure}
In the main text, we demonstrated the phase transition from FCI to a layer-polarized CDW-$1$ at low twist angles. Here, we further illustrate the low-lying many-body energy spectrum, which clearly manifests level crossing at phase transition as shown in  Fig.~\ref{fig:level_corssing13}(a)-(b).
The ground state approximate degeneracy (denoted by red, blue, and green) at lower values of $V_D$ (i.e., before the level crossing) correspond to the FCI phase, while those at larger values of $V_D$ (i.e., after the level crossing) denote CDW-$1$ state.  
Additionally, we can see that the many-body gap after the phase transitions (level crossings) monotonically increases with the displacement field $V_D$ (for both twist angles $\theta=2^\circ$ and $2.2^\circ$), indicating a robust CDW-$1$ phase at larger $V_D$. 
On the other hand, at larger twist angles side ($\theta=2.4$ and $2.7^\circ$), shown in Fig.~\ref{fig:level_corssing13}(c)-(d) for example, many-body gap remains robust against the range of displacement fields we considered. This indicates the robustness of the CDW-$2$ phase against the displacement fields.

\begin{figure}[H]
\includegraphics[width=0.43\linewidth]{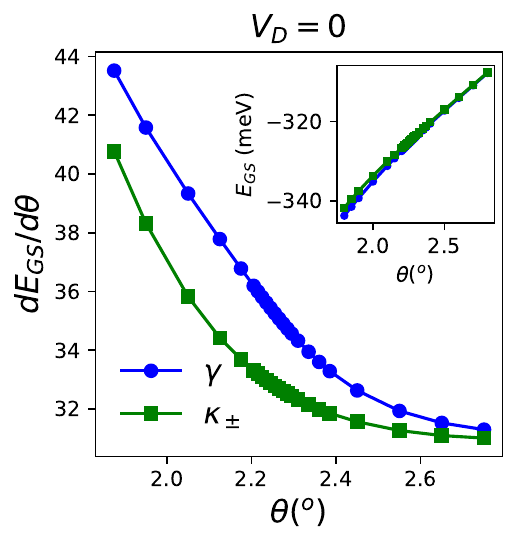} 
\includegraphics[width=0.44\linewidth]{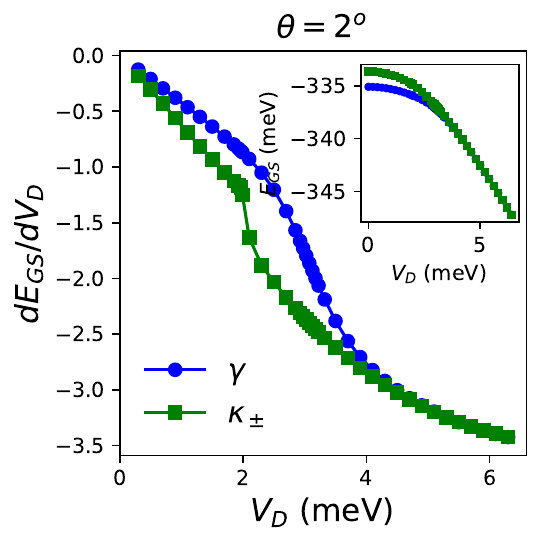}
\caption{The first derivative of ground state energy at $\gamma$ and $\kappa_\pm$ (averaged) for FCI to CDW-$2$ transition at fixed $V_D=0$ (left panel), and for FCI to CDW-$1$ transition at fixed $\theta=2^\circ$ (right panel). The inset on each plot represents energy at the corresponding COM momenta as a function of corresponding parameters. Calculations are performed on $27$ unit cell clusters.}
 \label{fig:FCI_CDWs}
\end{figure}
 \label{fig:FCI_CDW2}
\subsection{Phase transition between FCI and CDWs}
\label{sec:FCI_CDWs}
In the main text, we demonstrated the phase transitions from FCI to two CDW states: CDW-$1$ driven by displacement field and CDW-$2$ driven by twist angle. We also commented on the nature of these phase transitions. Here, we further discuss this based on our finite system size ED results on $27$ site clusters. 

\begin{figure}[H]
\includegraphics[width=0.247\linewidth]{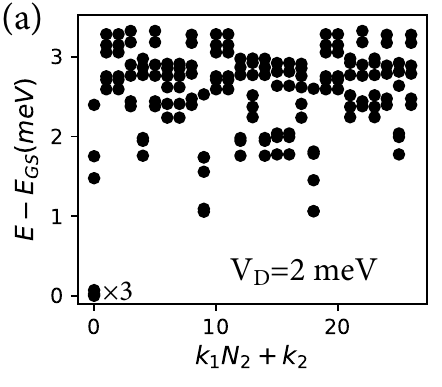} 
\includegraphics[width=0.247\linewidth]{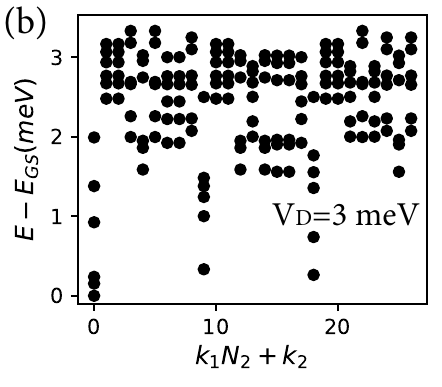}
\includegraphics[width=0.247\linewidth]{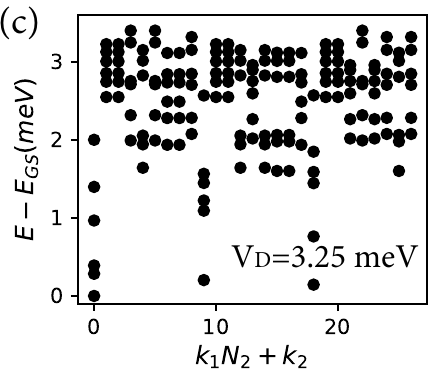}
\includegraphics[width=0.247\linewidth]{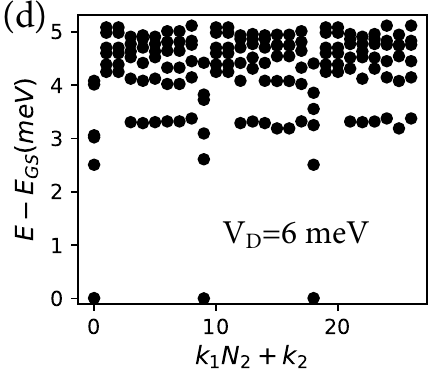}
\caption{Low lying energy spectrum on $N_{uc}=27$ for representative values of $V_D$ at fixed value of twist angle $\theta=2^\circ$. The evolution of low-lying many-body eigen states indicate FCI to CDW-1 transition.}
 \label{fig:FCI_CDW1_spectra}
\end{figure}

In Fig.~\ref{fig:FCI_CDWs}, we show the first derivative of the lowest energies at COM momentum $\gamma$ and $\kappa_\pm$ (average) for fixed $V_D=0$ (left panel) and for fixed $\theta=2^\circ$ (right panel). The former corresponds to the FCI-CDW-$2$ transition, while the latter represents the FCI-CDW-$1$ transition. The corresponding energies at $\gamma$ and $\kappa_\pm$, depicted in the insets of each plot, exhibit deviations at smaller $\theta$ ($V_D$). These deviations arise because the ground state degeneracy of FCI occurs at $\gamma$ on $27$ site clusters, however, these two energies closely track each other in the CDW phases. The momentum resolved low-lying many-body spectra for FCI to CDW-$1$ transition at fixed twist angle $\theta=2^\circ$ is demonstrated in Fig.~\ref{fig:FCI_CDW1_spectra}. 

For both phase transitions in Fig.~\ref{fig:FCI_CDWs}, energies and their corresponding first derivatives are continuous functions of parameter $\theta$ (and $V_D$). These distinctive traits of the low-lying states suggest a continuous phase transition. Notably, there is a substantial change in the slope within the FCI and CDW regions, yet there is no indication of non-analytic behavior in the energy function or its derivative.
However, it's crucial to validate these findings further, particularly in larger system sizes, which can be achieved using alternative methods such as the Density Matrix Renormalization Group (DMRG).   

\subsection{CDW-1 to CDW-2 transition}
\label{sec:CDW1_CDW2}
\begin{figure}[H]
    \centering
    \begin{minipage}[b]{0.5\linewidth}
        \centering
        \includegraphics[width=0.49\linewidth]{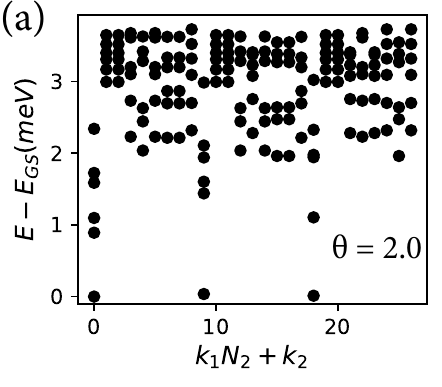}
        \includegraphics[width=0.49\linewidth]{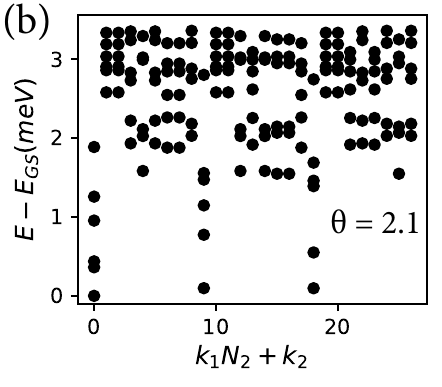}
        \includegraphics[width=0.49\linewidth]{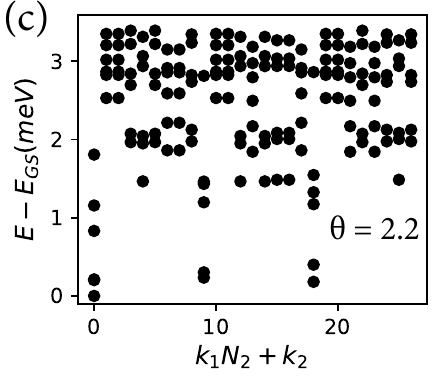}
        \includegraphics[width=0.49\linewidth]{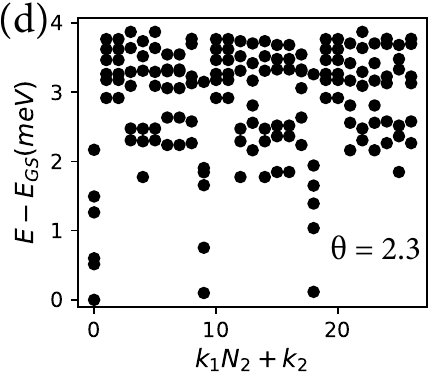}
    \end{minipage}
    \begin{minipage}[b]{0.49\linewidth}
        \centering
        \includegraphics[width=0.95\linewidth]{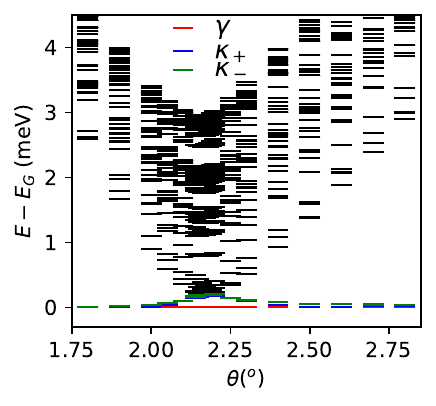}
    \end{minipage}
    \caption{(a)-(d) Momentum resolved low-lying many-body energy spectrum at various twist angles $\theta$ near the phase transition and (e)
    the many-body energy spectrum (after subtraction from the corresponding ground state energies) as a function of twist angle $\theta$ for fixed $V_D=4$ meV obtained by ED calculations on the 27 unit cell clusters at $\nu=1/3$ filling. Red, blue, and green markers in (e) denote the lowest three states.
    }
    \label{fig:level_cross_Vd=4}
\end{figure}
In Fig.~\ref{fig:level_cross_Vd=4}(a)-(d), we show momentum resolved low-lying many-body energy eigenvalues at various $\theta$ near the phase transition for fixed $V_D=4$ meV. 
At $\theta=2^\circ$, three-fold degenerate ground states are situated at COM momenta $\gamma$, $\kappa_+$, and $\kappa_-$, which is a CDW-$1$ state. The many-body gap becomes smaller as we increase the twist angle and closes at around $\theta=2.18^\circ$ and reopens upon further increasing $\theta$. However, momentum dependence of ground state degeneracy doesn't change after level crossing, which is a CDW-$2$ phase.  
In Fig.~\ref{fig:level_cross_Vd=4}(e), we show the evolution of low-lying many-body spectrum (subtracted from the corresponding ground states) as a function of twist angle for fixed displacement field value $V_D=4$ meV. First, we observe a level crossing around $\theta\approx 2.18^\circ$, as discussed in the main text. Second, we notice that the phase boundary around the transition point is broader due to finite-size effects. This broadening gives rise to two apparent jumps in the energy slope at $\kappa_\pm$, as discussed in Fig.~\ref{fig:LP}(e).

\section{Extended results for $\nu=\frac{2}{3}$} 
\label{app:twothird}
\begin{figure}[H]
\includegraphics[width=0.247\linewidth]{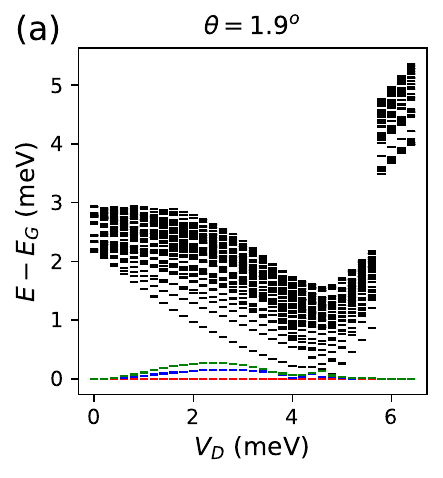} 
\includegraphics[width=0.248\linewidth]{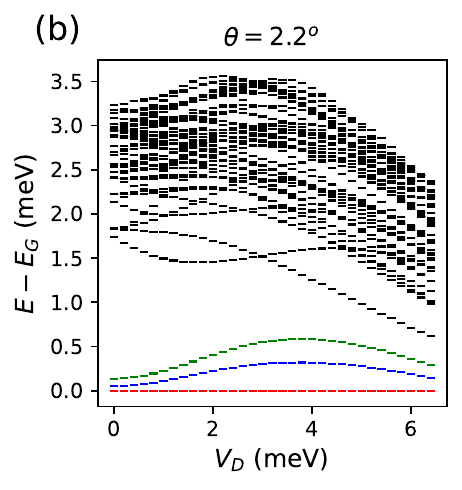}
\includegraphics[width=0.247\linewidth]{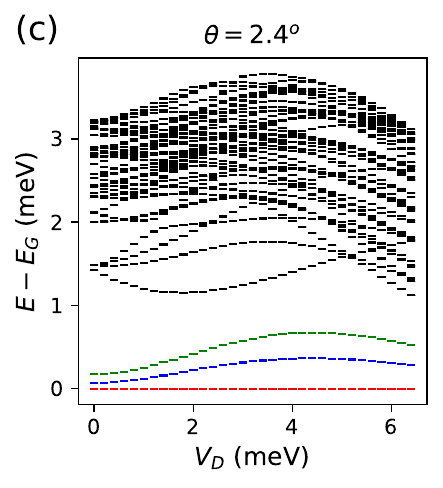}
\includegraphics[width=0.247\linewidth]{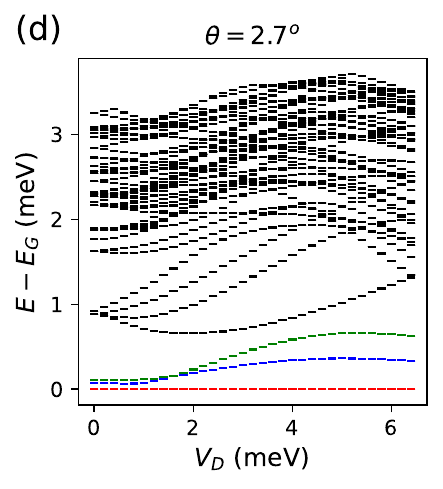}
\caption{The low-lying many-body energy spectrum (after subtraction from the corresponding ground state energies) as a function of displacement field $V_D$ for fixed at various twist angles obtained by ED calculations on the 27 unit cell clusters at $\nu=\frac{2}{3}$ filling. Red, blue, and green markers denote the lowest three states.}
 \label{fig:level_corssing23}
\end{figure}
 In the main text, we discussed the robustness of the FCI phase at $\nu=\frac{2}{3}$ filling against the displacement field. Here, we further demonstrate the low-lying energies as a function of $V_D$ for representative twist angles, as depicted in Fig.~\ref{fig:level_corssing23}. At a small twist angle of $\theta=1.9^\circ$, a phase transition from FCI to CDW is observed, as indicated by the level crossing. This many-body phase transition occurs close to the single-particle topological phase transition, as evidenced by a sudden jump in the many-body spectra at $V_D\approx 6$ meV. 

 At larger twist angles, while the three-fold ground state degeneracy is not perfect, there is no phase transition observed, and the system remains within the FCI phase. This conclusion was confirmed through the computation of the many-body Chern number for each ground state, as discussed in the main text.

\section{Comparison with Wang et al.'s parameters}
\label{APP:wang}

We have extensively discussed our results by adapting single-particle parameters from ref.~\cite{reddy2023fractional}.
This section will briefly discuss how our results (single-particle and many-body quantum phase transitions) are affected when using different sets of parameters from ref.~\cite{wang2024fractional}. 
The large scale DFT parameters provided in ref.~\cite{wang2024fractional} are $(v,w,\phi)= (20.8 \text{ meV},-23.8 \text{ meV}, 107.7^\circ)$. 
Since parameters are obtained at commensurate twist angle $\theta=3.89^\circ$, which is also close to the experimental value in ref.~\cite{cai2023signatures}, we fix the twist angle at this value for accuracy and vary the displacement field.

In Fig.~\ref{fig:band_wang}, we present the lowest single-particle bands with their corresponding Chern numbers at displacement fields $V_D=5,10,$ and $15$ meV. The lowest two bands remain well separated up to $V_D=15$ meV, and the band gap closes at a large displacement field $V_D\sim 20$ meV. This indicates band topology is robust under an external electric field.

For many-body simulations, we consider only the lowest hole band projecting the Coulomb interaction, just like before. The Coulomb interaction strength is consistently fixed at $\epsilon=10$.

\begin{figure}[H]
\includegraphics[width=0.32\linewidth]{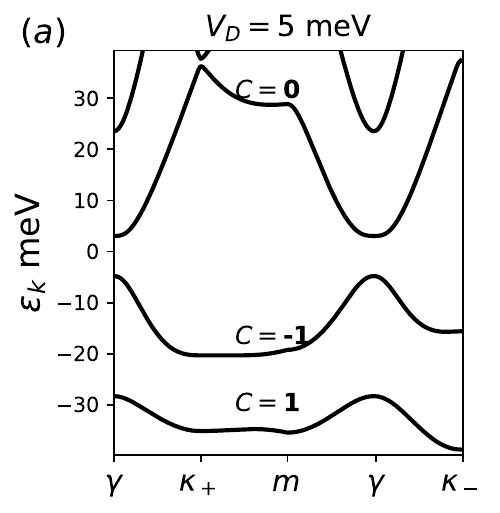} 
\includegraphics[width=0.32\linewidth]{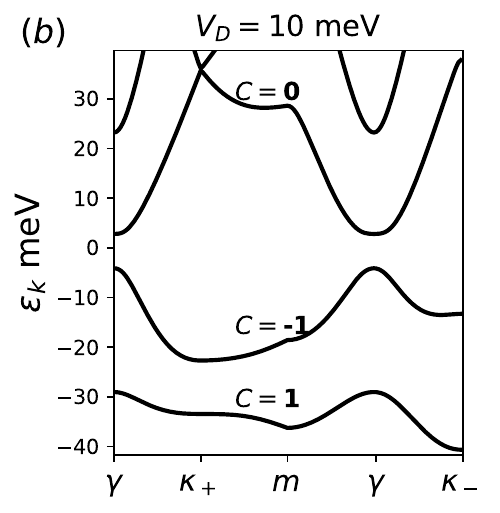}
\includegraphics[width=0.32\linewidth]{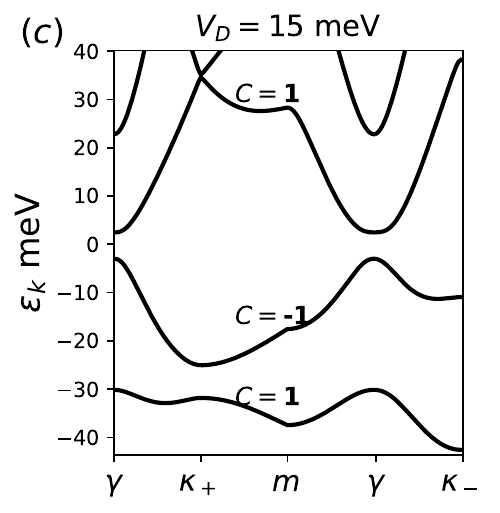}

\caption{Band structure of the lowest three bands with their chern numbers at representative values of the displacement fields at fixed twist angle $\theta=3.89^\circ$.}
 \label{fig:band_wang}
\end{figure}

\begin{figure}[H]
\includegraphics[width=0.247\linewidth]{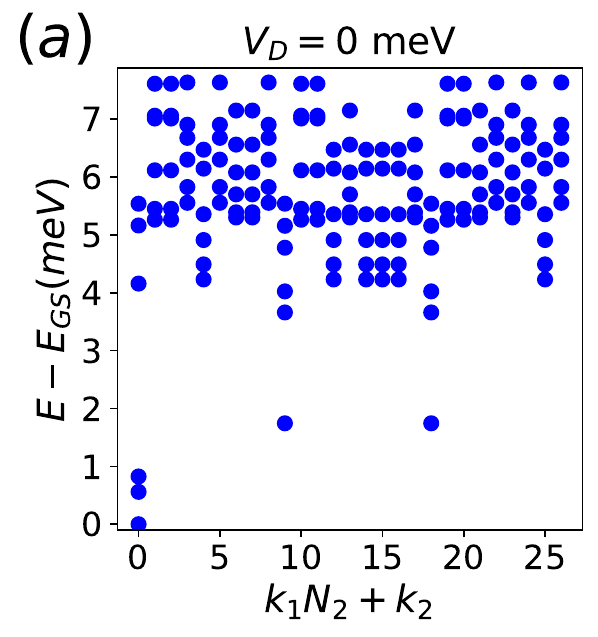} 
\includegraphics[width=0.248\linewidth]{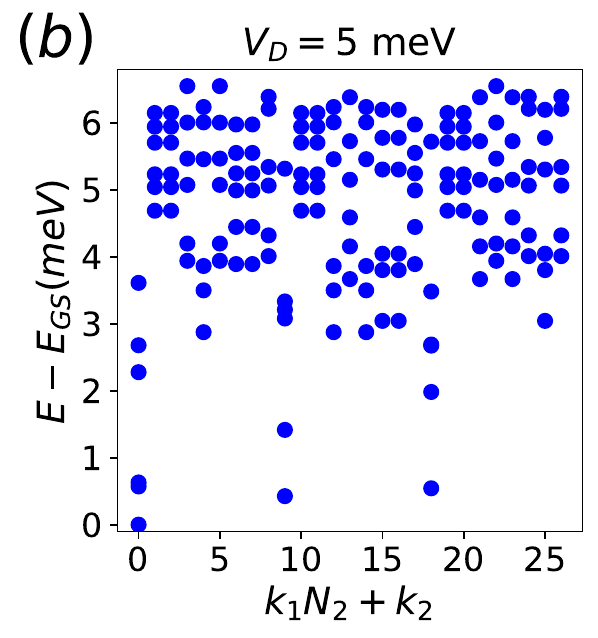}
\includegraphics[width=0.247\linewidth]{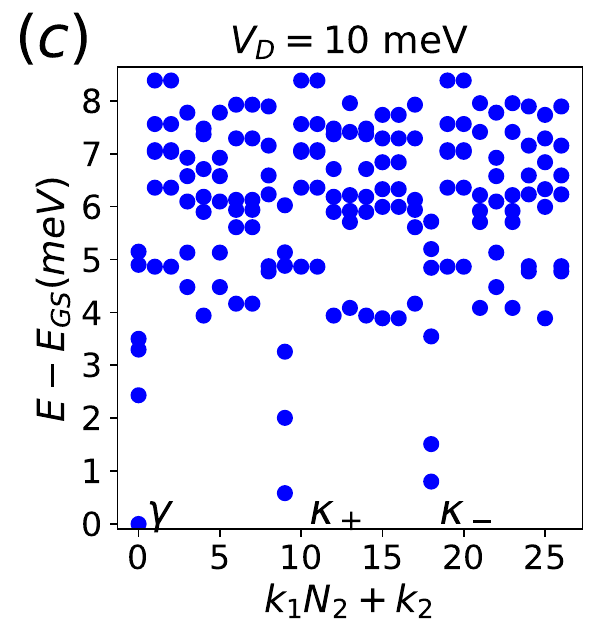}
\includegraphics[width=0.247\linewidth]{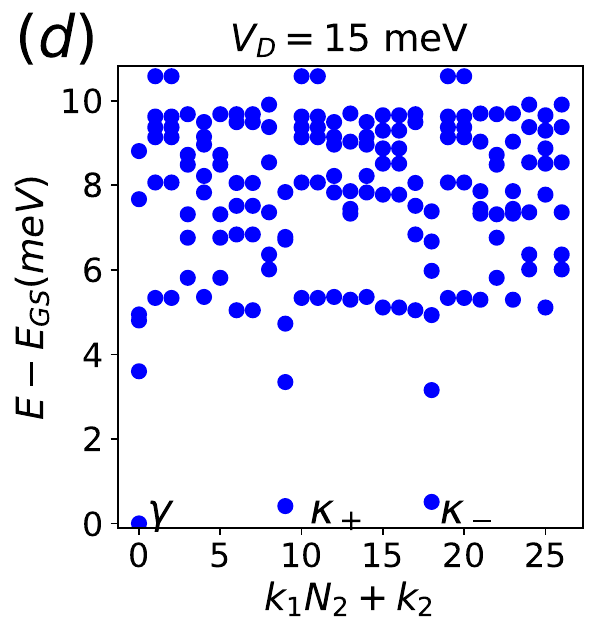}
\caption{The momentum resolved low-lying many-body energy spectrum at $\nu=\frac{1}{3}$ filling on $N_{uc}=27$ unit cell clusters at various displacement fields for fixed twist angle $\theta=3.89^\circ$. }
 \label{fig:ED_wang}
\end{figure}

Unlike in ref.~\cite{reddy2023fractional}, $\frac{1}{3}$ FQAH state survives at a larger twist angle in ref.~\cite{wang2024fractional}. In Fig.~\ref{fig:ED_wang}, we show the lowest six many-body eigen energies at each momentum sector for representative values of the displacement field. 
At zero field, there are nearly three-fold degenerate ground states at $\gamma$, separated by a many-body gap from the rest of the states, indicating the FQAH state as before.
As we increase the displacement field, this many-body gap closes at around $V_D= 5$ meV and reopens again with further increase in $V_D$, but with three-fold degeneracy migrating to $\gamma, \kappa_+,$ and $\kappa_-$.
This observation is consistent with our observation of the FCI to CDW-1 transition in the main text. The only difference is that this transition occurs at a relatively higher displacement field than previously observed, indicating FCI is more robust in this case.  

\end{widetext}
\bibliography{bibfile}

\begin{thebibliography}{77}%
\makeatletter
\providecommand \@ifxundefined [1]{%
 \@ifx{#1\undefined}
}%
\providecommand \@ifnum [1]{%
 \ifnum #1\expandafter \@firstoftwo
 \else \expandafter \@secondoftwo
 \fi
}%
\providecommand \@ifx [1]{%
 \ifx #1\expandafter \@firstoftwo
 \else \expandafter \@secondoftwo
 \fi
}%
\providecommand \natexlab [1]{#1}%
\providecommand \enquote  [1]{``#1''}%
\providecommand \bibnamefont  [1]{#1}%
\providecommand \bibfnamefont [1]{#1}%
\providecommand \citenamefont [1]{#1}%
\providecommand \href@noop [0]{\@secondoftwo}%
\providecommand \href [0]{\begingroup \@sanitize@url \@href}%
\providecommand \@href[1]{\@@startlink{#1}\@@href}%
\providecommand \@@href[1]{\endgroup#1\@@endlink}%
\providecommand \@sanitize@url [0]{\catcode `\\12\catcode `\$12\catcode `\&12\catcode `\#12\catcode `\^12\catcode `\_12\catcode `\%12\relax}%
\providecommand \@@startlink[1]{}%
\providecommand \@@endlink[0]{}%
\providecommand \url  [0]{\begingroup\@sanitize@url \@url }%
\providecommand \@url [1]{\endgroup\@href {#1}{\urlprefix }}%
\providecommand \urlprefix  [0]{URL }%
\providecommand \Eprint [0]{\href }%
\providecommand \doibase [0]{http://dx.doi.org/}%
\providecommand \selectlanguage [0]{\@gobble}%
\providecommand \bibinfo  [0]{\@secondoftwo}%
\providecommand \bibfield  [0]{\@secondoftwo}%
\providecommand \translation [1]{[#1]}%
\providecommand \BibitemOpen [0]{}%
\providecommand \bibitemStop [0]{}%
\providecommand \bibitemNoStop [0]{.\EOS\space}%
\providecommand \EOS [0]{\spacefactor3000\relax}%
\providecommand \BibitemShut  [1]{\csname bibitem#1\endcsname}%
\let\auto@bib@innerbib\@empty
\bibitem [{\citenamefont {Cao}\ \emph {et~al.}(2018{\natexlab{a}})\citenamefont {Cao}, \citenamefont {Fatemi}, \citenamefont {Demir}, \citenamefont {Fang}, \citenamefont {Tomarken}, \citenamefont {Luo}, \citenamefont {Sanchez-Yamagishi}, \citenamefont {Watanabe}, \citenamefont {Taniguchi}, \citenamefont {Kaxiras} \emph {et~al.}}]{cao2018correlated}%
  \BibitemOpen
  \bibfield  {author} {\bibinfo {author} {\bibfnamefont {Yuan}\ \bibnamefont {Cao}}, \bibinfo {author} {\bibfnamefont {Valla}\ \bibnamefont {Fatemi}}, \bibinfo {author} {\bibfnamefont {Ahmet}\ \bibnamefont {Demir}}, \bibinfo {author} {\bibfnamefont {Shiang}\ \bibnamefont {Fang}}, \bibinfo {author} {\bibfnamefont {Spencer~L}\ \bibnamefont {Tomarken}}, \bibinfo {author} {\bibfnamefont {Jason~Y}\ \bibnamefont {Luo}}, \bibinfo {author} {\bibfnamefont {Javier~D}\ \bibnamefont {Sanchez-Yamagishi}}, \bibinfo {author} {\bibfnamefont {Kenji}\ \bibnamefont {Watanabe}}, \bibinfo {author} {\bibfnamefont {Takashi}\ \bibnamefont {Taniguchi}}, \bibinfo {author} {\bibfnamefont {Efthimios}\ \bibnamefont {Kaxiras}},  \emph {et~al.},\ }\bibfield  {title} {\enquote {\bibinfo {title} {Correlated insulator behaviour at half-filling in magic-angle graphene superlattices},}\ }\href@noop {} {\bibfield  {journal} {\bibinfo  {journal} {Nature}\ }\textbf {\bibinfo {volume} {556}},\ \bibinfo {pages} {80--84} (\bibinfo {year}
  {2018}{\natexlab{a}})}\BibitemShut {NoStop}%
\bibitem [{\citenamefont {Tang}\ \emph {et~al.}(2020)\citenamefont {Tang}, \citenamefont {Li}, \citenamefont {Li}, \citenamefont {Xu}, \citenamefont {Liu}, \citenamefont {Barmak}, \citenamefont {Watanabe}, \citenamefont {Taniguchi}, \citenamefont {MacDonald}, \citenamefont {Shan} \emph {et~al.}}]{tang2020simulation}%
  \BibitemOpen
  \bibfield  {author} {\bibinfo {author} {\bibfnamefont {Yanhao}\ \bibnamefont {Tang}}, \bibinfo {author} {\bibfnamefont {Lizhong}\ \bibnamefont {Li}}, \bibinfo {author} {\bibfnamefont {Tingxin}\ \bibnamefont {Li}}, \bibinfo {author} {\bibfnamefont {Yang}\ \bibnamefont {Xu}}, \bibinfo {author} {\bibfnamefont {Song}\ \bibnamefont {Liu}}, \bibinfo {author} {\bibfnamefont {Katayun}\ \bibnamefont {Barmak}}, \bibinfo {author} {\bibfnamefont {Kenji}\ \bibnamefont {Watanabe}}, \bibinfo {author} {\bibfnamefont {Takashi}\ \bibnamefont {Taniguchi}}, \bibinfo {author} {\bibfnamefont {Allan~H}\ \bibnamefont {MacDonald}}, \bibinfo {author} {\bibfnamefont {Jie}\ \bibnamefont {Shan}},  \emph {et~al.},\ }\bibfield  {title} {\enquote {\bibinfo {title} {Simulation of hubbard model physics in wse2/ws2 moir{\'e} superlattices},}\ }\href@noop {} {\bibfield  {journal} {\bibinfo  {journal} {Nature}\ }\textbf {\bibinfo {volume} {579}},\ \bibinfo {pages} {353--358} (\bibinfo {year} {2020})}\BibitemShut {NoStop}%
\bibitem [{\citenamefont {Regan}\ \emph {et~al.}(2020)\citenamefont {Regan}, \citenamefont {Wang}, \citenamefont {Jin}, \citenamefont {Bakti~Utama}, \citenamefont {Gao}, \citenamefont {Wei}, \citenamefont {Zhao}, \citenamefont {Zhao}, \citenamefont {Zhang}, \citenamefont {Yumigeta} \emph {et~al.}}]{regan2020mott}%
  \BibitemOpen
  \bibfield  {author} {\bibinfo {author} {\bibfnamefont {Emma~C}\ \bibnamefont {Regan}}, \bibinfo {author} {\bibfnamefont {Danqing}\ \bibnamefont {Wang}}, \bibinfo {author} {\bibfnamefont {Chenhao}\ \bibnamefont {Jin}}, \bibinfo {author} {\bibfnamefont {M~Iqbal}\ \bibnamefont {Bakti~Utama}}, \bibinfo {author} {\bibfnamefont {Beini}\ \bibnamefont {Gao}}, \bibinfo {author} {\bibfnamefont {Xin}\ \bibnamefont {Wei}}, \bibinfo {author} {\bibfnamefont {Sihan}\ \bibnamefont {Zhao}}, \bibinfo {author} {\bibfnamefont {Wenyu}\ \bibnamefont {Zhao}}, \bibinfo {author} {\bibfnamefont {Zuocheng}\ \bibnamefont {Zhang}}, \bibinfo {author} {\bibfnamefont {Kentaro}\ \bibnamefont {Yumigeta}},  \emph {et~al.},\ }\bibfield  {title} {\enquote {\bibinfo {title} {Mott and generalized wigner crystal states in wse2/ws2 moir{\'e} superlattices},}\ }\href@noop {} {\bibfield  {journal} {\bibinfo  {journal} {Nature}\ }\textbf {\bibinfo {volume} {579}},\ \bibinfo {pages} {359--363} (\bibinfo {year} {2020})}\BibitemShut {NoStop}%
\bibitem [{\citenamefont {Li}\ \emph {et~al.}(2021{\natexlab{a}})\citenamefont {Li}, \citenamefont {Li}, \citenamefont {Regan}, \citenamefont {Wang}, \citenamefont {Zhao}, \citenamefont {Kahn}, \citenamefont {Yumigeta}, \citenamefont {Blei}, \citenamefont {Taniguchi}, \citenamefont {Watanabe} \emph {et~al.}}]{li2021imaging}%
  \BibitemOpen
  \bibfield  {author} {\bibinfo {author} {\bibfnamefont {Hongyuan}\ \bibnamefont {Li}}, \bibinfo {author} {\bibfnamefont {Shaowei}\ \bibnamefont {Li}}, \bibinfo {author} {\bibfnamefont {Emma~C}\ \bibnamefont {Regan}}, \bibinfo {author} {\bibfnamefont {Danqing}\ \bibnamefont {Wang}}, \bibinfo {author} {\bibfnamefont {Wenyu}\ \bibnamefont {Zhao}}, \bibinfo {author} {\bibfnamefont {Salman}\ \bibnamefont {Kahn}}, \bibinfo {author} {\bibfnamefont {Kentaro}\ \bibnamefont {Yumigeta}}, \bibinfo {author} {\bibfnamefont {Mark}\ \bibnamefont {Blei}}, \bibinfo {author} {\bibfnamefont {Takashi}\ \bibnamefont {Taniguchi}}, \bibinfo {author} {\bibfnamefont {Kenji}\ \bibnamefont {Watanabe}},  \emph {et~al.},\ }\bibfield  {title} {\enquote {\bibinfo {title} {Imaging two-dimensional generalized wigner crystals},}\ }\href@noop {} {\bibfield  {journal} {\bibinfo  {journal} {Nature}\ }\textbf {\bibinfo {volume} {597}},\ \bibinfo {pages} {650--654} (\bibinfo {year} {2021}{\natexlab{a}})}\BibitemShut {NoStop}%
\bibitem [{\citenamefont {Huang}\ \emph {et~al.}(2021)\citenamefont {Huang}, \citenamefont {Wang}, \citenamefont {Miao}, \citenamefont {Wang}, \citenamefont {Li}, \citenamefont {Lian}, \citenamefont {Taniguchi}, \citenamefont {Watanabe}, \citenamefont {Okamoto}, \citenamefont {Xiao} \emph {et~al.}}]{huang2021correlated}%
  \BibitemOpen
  \bibfield  {author} {\bibinfo {author} {\bibfnamefont {Xiong}\ \bibnamefont {Huang}}, \bibinfo {author} {\bibfnamefont {Tianmeng}\ \bibnamefont {Wang}}, \bibinfo {author} {\bibfnamefont {Shengnan}\ \bibnamefont {Miao}}, \bibinfo {author} {\bibfnamefont {Chong}\ \bibnamefont {Wang}}, \bibinfo {author} {\bibfnamefont {Zhipeng}\ \bibnamefont {Li}}, \bibinfo {author} {\bibfnamefont {Zhen}\ \bibnamefont {Lian}}, \bibinfo {author} {\bibfnamefont {Takashi}\ \bibnamefont {Taniguchi}}, \bibinfo {author} {\bibfnamefont {Kenji}\ \bibnamefont {Watanabe}}, \bibinfo {author} {\bibfnamefont {Satoshi}\ \bibnamefont {Okamoto}}, \bibinfo {author} {\bibfnamefont {Di}~\bibnamefont {Xiao}},  \emph {et~al.},\ }\bibfield  {title} {\enquote {\bibinfo {title} {Correlated insulating states at fractional fillings of the ws2/wse2 moir{\'e} lattice},}\ }\href@noop {} {\bibfield  {journal} {\bibinfo  {journal} {Nature Physics}\ }\textbf {\bibinfo {volume} {17}},\ \bibinfo {pages} {715--719} (\bibinfo {year} {2021})}\BibitemShut
  {NoStop}%
\bibitem [{\citenamefont {Zhou}\ \emph {et~al.}(2021)\citenamefont {Zhou}, \citenamefont {Sung}, \citenamefont {Brutschea}, \citenamefont {Esterlis}, \citenamefont {Wang}, \citenamefont {Scuri}, \citenamefont {Gelly}, \citenamefont {Heo}, \citenamefont {Taniguchi}, \citenamefont {Watanabe} \emph {et~al.}}]{zhou2021bilayer}%
  \BibitemOpen
  \bibfield  {author} {\bibinfo {author} {\bibfnamefont {You}\ \bibnamefont {Zhou}}, \bibinfo {author} {\bibfnamefont {Jiho}\ \bibnamefont {Sung}}, \bibinfo {author} {\bibfnamefont {Elise}\ \bibnamefont {Brutschea}}, \bibinfo {author} {\bibfnamefont {Ilya}\ \bibnamefont {Esterlis}}, \bibinfo {author} {\bibfnamefont {Yao}\ \bibnamefont {Wang}}, \bibinfo {author} {\bibfnamefont {Giovanni}\ \bibnamefont {Scuri}}, \bibinfo {author} {\bibfnamefont {Ryan~J}\ \bibnamefont {Gelly}}, \bibinfo {author} {\bibfnamefont {Hoseok}\ \bibnamefont {Heo}}, \bibinfo {author} {\bibfnamefont {Takashi}\ \bibnamefont {Taniguchi}}, \bibinfo {author} {\bibfnamefont {Kenji}\ \bibnamefont {Watanabe}},  \emph {et~al.},\ }\bibfield  {title} {\enquote {\bibinfo {title} {Bilayer wigner crystals in a transition metal dichalcogenide heterostructure},}\ }\href@noop {} {\bibfield  {journal} {\bibinfo  {journal} {Nature}\ }\textbf {\bibinfo {volume} {595}},\ \bibinfo {pages} {48--52} (\bibinfo {year} {2021})}\BibitemShut {NoStop}%
\bibitem [{\citenamefont {Ki}\ \emph {et~al.}(2014)\citenamefont {Ki}, \citenamefont {Fal’ko}, \citenamefont {Abanin},\ and\ \citenamefont {Morpurgo}}]{ki2014observation}%
  \BibitemOpen
  \bibfield  {author} {\bibinfo {author} {\bibfnamefont {Dong-Keun}\ \bibnamefont {Ki}}, \bibinfo {author} {\bibfnamefont {Vladimir~I}\ \bibnamefont {Fal’ko}}, \bibinfo {author} {\bibfnamefont {Dmitry~A}\ \bibnamefont {Abanin}}, \ and\ \bibinfo {author} {\bibfnamefont {Alberto~F}\ \bibnamefont {Morpurgo}},\ }\bibfield  {title} {\enquote {\bibinfo {title} {Observation of even denominator fractional quantum hall effect in suspended bilayer graphene},}\ }\href@noop {} {\bibfield  {journal} {\bibinfo  {journal} {Nano letters}\ }\textbf {\bibinfo {volume} {14}},\ \bibinfo {pages} {2135--2139} (\bibinfo {year} {2014})}\BibitemShut {NoStop}%
\bibitem [{\citenamefont {Dean}\ \emph {et~al.}(2013)\citenamefont {Dean}, \citenamefont {Wang}, \citenamefont {Maher}, \citenamefont {Forsythe}, \citenamefont {Ghahari}, \citenamefont {Gao}, \citenamefont {Katoch}, \citenamefont {Ishigami}, \citenamefont {Moon}, \citenamefont {Koshino} \emph {et~al.}}]{dean2013hofstadter}%
  \BibitemOpen
  \bibfield  {author} {\bibinfo {author} {\bibfnamefont {Cory~R}\ \bibnamefont {Dean}}, \bibinfo {author} {\bibfnamefont {L}~\bibnamefont {Wang}}, \bibinfo {author} {\bibfnamefont {P}~\bibnamefont {Maher}}, \bibinfo {author} {\bibfnamefont {C}~\bibnamefont {Forsythe}}, \bibinfo {author} {\bibfnamefont {Fereshte}\ \bibnamefont {Ghahari}}, \bibinfo {author} {\bibfnamefont {Y}~\bibnamefont {Gao}}, \bibinfo {author} {\bibfnamefont {Jyoti}\ \bibnamefont {Katoch}}, \bibinfo {author} {\bibfnamefont {M}~\bibnamefont {Ishigami}}, \bibinfo {author} {\bibfnamefont {P}~\bibnamefont {Moon}}, \bibinfo {author} {\bibfnamefont {M}~\bibnamefont {Koshino}},  \emph {et~al.},\ }\bibfield  {title} {\enquote {\bibinfo {title} {Hofstadter’s butterfly and the fractal quantum hall effect in moir{\'e} superlattices},}\ }\href@noop {} {\bibfield  {journal} {\bibinfo  {journal} {Nature}\ }\textbf {\bibinfo {volume} {497}},\ \bibinfo {pages} {598--602} (\bibinfo {year} {2013})}\BibitemShut {NoStop}%
\bibitem [{\citenamefont {Kou}\ \emph {et~al.}(2014)\citenamefont {Kou}, \citenamefont {Feldman}, \citenamefont {Levin}, \citenamefont {Halperin}, \citenamefont {Watanabe}, \citenamefont {Taniguchi},\ and\ \citenamefont {Yacoby}}]{kou2014electron}%
  \BibitemOpen
  \bibfield  {author} {\bibinfo {author} {\bibfnamefont {Angela}\ \bibnamefont {Kou}}, \bibinfo {author} {\bibfnamefont {Benjamin~E}\ \bibnamefont {Feldman}}, \bibinfo {author} {\bibfnamefont {Andrei~J}\ \bibnamefont {Levin}}, \bibinfo {author} {\bibfnamefont {Bertrand~I}\ \bibnamefont {Halperin}}, \bibinfo {author} {\bibfnamefont {Kenji}\ \bibnamefont {Watanabe}}, \bibinfo {author} {\bibfnamefont {Takashi}\ \bibnamefont {Taniguchi}}, \ and\ \bibinfo {author} {\bibfnamefont {Amir}\ \bibnamefont {Yacoby}},\ }\bibfield  {title} {\enquote {\bibinfo {title} {Electron-hole asymmetric integer and fractional quantum hall effect in bilayer graphene},}\ }\href@noop {} {\bibfield  {journal} {\bibinfo  {journal} {Science}\ }\textbf {\bibinfo {volume} {345}},\ \bibinfo {pages} {55--57} (\bibinfo {year} {2014})}\BibitemShut {NoStop}%
\bibitem [{\citenamefont {Maher}\ \emph {et~al.}(2014)\citenamefont {Maher}, \citenamefont {Wang}, \citenamefont {Gao}, \citenamefont {Forsythe}, \citenamefont {Taniguchi}, \citenamefont {Watanabe}, \citenamefont {Abanin}, \citenamefont {Papi{\'c}}, \citenamefont {Cadden-Zimansky}, \citenamefont {Hone} \emph {et~al.}}]{maher2014tunable}%
  \BibitemOpen
  \bibfield  {author} {\bibinfo {author} {\bibfnamefont {Patrick}\ \bibnamefont {Maher}}, \bibinfo {author} {\bibfnamefont {Lei}\ \bibnamefont {Wang}}, \bibinfo {author} {\bibfnamefont {Yuanda}\ \bibnamefont {Gao}}, \bibinfo {author} {\bibfnamefont {Carlos}\ \bibnamefont {Forsythe}}, \bibinfo {author} {\bibfnamefont {Takashi}\ \bibnamefont {Taniguchi}}, \bibinfo {author} {\bibfnamefont {Kenji}\ \bibnamefont {Watanabe}}, \bibinfo {author} {\bibfnamefont {Dmitry}\ \bibnamefont {Abanin}}, \bibinfo {author} {\bibfnamefont {Zlatko}\ \bibnamefont {Papi{\'c}}}, \bibinfo {author} {\bibfnamefont {Paul}\ \bibnamefont {Cadden-Zimansky}}, \bibinfo {author} {\bibfnamefont {James}\ \bibnamefont {Hone}},  \emph {et~al.},\ }\bibfield  {title} {\enquote {\bibinfo {title} {Tunable fractional quantum hall phases in bilayer graphene},}\ }\href@noop {} {\bibfield  {journal} {\bibinfo  {journal} {Science}\ }\textbf {\bibinfo {volume} {345}},\ \bibinfo {pages} {61--64} (\bibinfo {year} {2014})}\BibitemShut {NoStop}%
\bibitem [{\citenamefont {Cao}\ \emph {et~al.}(2018{\natexlab{b}})\citenamefont {Cao}, \citenamefont {Fatemi}, \citenamefont {Fang}, \citenamefont {Watanabe}, \citenamefont {Taniguchi}, \citenamefont {Kaxiras},\ and\ \citenamefont {Jarillo-Herrero}}]{cao2018unconventional}%
  \BibitemOpen
  \bibfield  {author} {\bibinfo {author} {\bibfnamefont {Yuan}\ \bibnamefont {Cao}}, \bibinfo {author} {\bibfnamefont {Valla}\ \bibnamefont {Fatemi}}, \bibinfo {author} {\bibfnamefont {Shiang}\ \bibnamefont {Fang}}, \bibinfo {author} {\bibfnamefont {Kenji}\ \bibnamefont {Watanabe}}, \bibinfo {author} {\bibfnamefont {Takashi}\ \bibnamefont {Taniguchi}}, \bibinfo {author} {\bibfnamefont {Efthimios}\ \bibnamefont {Kaxiras}}, \ and\ \bibinfo {author} {\bibfnamefont {Pablo}\ \bibnamefont {Jarillo-Herrero}},\ }\bibfield  {title} {\enquote {\bibinfo {title} {Unconventional superconductivity in magic-angle graphene superlattices},}\ }\href@noop {} {\bibfield  {journal} {\bibinfo  {journal} {Nature}\ }\textbf {\bibinfo {volume} {556}},\ \bibinfo {pages} {43--50} (\bibinfo {year} {2018}{\natexlab{b}})}\BibitemShut {NoStop}%
\bibitem [{\citenamefont {Yankowitz}\ \emph {et~al.}(2019)\citenamefont {Yankowitz}, \citenamefont {Chen}, \citenamefont {Polshyn}, \citenamefont {Zhang}, \citenamefont {Watanabe}, \citenamefont {Taniguchi}, \citenamefont {Graf}, \citenamefont {Young},\ and\ \citenamefont {Dean}}]{yankowitz2019tuning}%
  \BibitemOpen
  \bibfield  {author} {\bibinfo {author} {\bibfnamefont {Matthew}\ \bibnamefont {Yankowitz}}, \bibinfo {author} {\bibfnamefont {Shaowen}\ \bibnamefont {Chen}}, \bibinfo {author} {\bibfnamefont {Hryhoriy}\ \bibnamefont {Polshyn}}, \bibinfo {author} {\bibfnamefont {Yuxuan}\ \bibnamefont {Zhang}}, \bibinfo {author} {\bibfnamefont {K}~\bibnamefont {Watanabe}}, \bibinfo {author} {\bibfnamefont {T}~\bibnamefont {Taniguchi}}, \bibinfo {author} {\bibfnamefont {David}\ \bibnamefont {Graf}}, \bibinfo {author} {\bibfnamefont {Andrea~F}\ \bibnamefont {Young}}, \ and\ \bibinfo {author} {\bibfnamefont {Cory~R}\ \bibnamefont {Dean}},\ }\bibfield  {title} {\enquote {\bibinfo {title} {Tuning superconductivity in twisted bilayer graphene},}\ }\href@noop {} {\bibfield  {journal} {\bibinfo  {journal} {Science}\ }\textbf {\bibinfo {volume} {363}},\ \bibinfo {pages} {1059--1064} (\bibinfo {year} {2019})}\BibitemShut {NoStop}%
\bibitem [{\citenamefont {Stepanov}\ \emph {et~al.}(2020)\citenamefont {Stepanov}, \citenamefont {Das}, \citenamefont {Lu}, \citenamefont {Fahimniya}, \citenamefont {Watanabe}, \citenamefont {Taniguchi}, \citenamefont {Koppens}, \citenamefont {Lischner}, \citenamefont {Levitov},\ and\ \citenamefont {Efetov}}]{stepanov2020untying}%
  \BibitemOpen
  \bibfield  {author} {\bibinfo {author} {\bibfnamefont {Petr}\ \bibnamefont {Stepanov}}, \bibinfo {author} {\bibfnamefont {Ipsita}\ \bibnamefont {Das}}, \bibinfo {author} {\bibfnamefont {Xiaobo}\ \bibnamefont {Lu}}, \bibinfo {author} {\bibfnamefont {Ali}\ \bibnamefont {Fahimniya}}, \bibinfo {author} {\bibfnamefont {Kenji}\ \bibnamefont {Watanabe}}, \bibinfo {author} {\bibfnamefont {Takashi}\ \bibnamefont {Taniguchi}}, \bibinfo {author} {\bibfnamefont {Frank~HL}\ \bibnamefont {Koppens}}, \bibinfo {author} {\bibfnamefont {Johannes}\ \bibnamefont {Lischner}}, \bibinfo {author} {\bibfnamefont {Leonid}\ \bibnamefont {Levitov}}, \ and\ \bibinfo {author} {\bibfnamefont {Dmitri~K}\ \bibnamefont {Efetov}},\ }\bibfield  {title} {\enquote {\bibinfo {title} {Untying the insulating and superconducting orders in magic-angle graphene},}\ }\href@noop {} {\bibfield  {journal} {\bibinfo  {journal} {Nature}\ }\textbf {\bibinfo {volume} {583}},\ \bibinfo {pages} {375--378} (\bibinfo {year} {2020})}\BibitemShut {NoStop}%
\bibitem [{\citenamefont {Oh}\ \emph {et~al.}(2021)\citenamefont {Oh}, \citenamefont {Nuckolls}, \citenamefont {Wong}, \citenamefont {Lee}, \citenamefont {Liu}, \citenamefont {Watanabe}, \citenamefont {Taniguchi},\ and\ \citenamefont {Yazdani}}]{oh2021evidence}%
  \BibitemOpen
  \bibfield  {author} {\bibinfo {author} {\bibfnamefont {Myungchul}\ \bibnamefont {Oh}}, \bibinfo {author} {\bibfnamefont {Kevin~P}\ \bibnamefont {Nuckolls}}, \bibinfo {author} {\bibfnamefont {Dillon}\ \bibnamefont {Wong}}, \bibinfo {author} {\bibfnamefont {Ryan~L}\ \bibnamefont {Lee}}, \bibinfo {author} {\bibfnamefont {Xiaomeng}\ \bibnamefont {Liu}}, \bibinfo {author} {\bibfnamefont {Kenji}\ \bibnamefont {Watanabe}}, \bibinfo {author} {\bibfnamefont {Takashi}\ \bibnamefont {Taniguchi}}, \ and\ \bibinfo {author} {\bibfnamefont {Ali}\ \bibnamefont {Yazdani}},\ }\bibfield  {title} {\enquote {\bibinfo {title} {Evidence for unconventional superconductivity in twisted bilayer graphene},}\ }\href@noop {} {\bibfield  {journal} {\bibinfo  {journal} {Nature}\ }\textbf {\bibinfo {volume} {600}},\ \bibinfo {pages} {240--245} (\bibinfo {year} {2021})}\BibitemShut {NoStop}%
\bibitem [{\citenamefont {Po}\ \emph {et~al.}(2018)\citenamefont {Po}, \citenamefont {Zou}, \citenamefont {Vishwanath},\ and\ \citenamefont {Senthil}}]{po2018origin}%
  \BibitemOpen
  \bibfield  {author} {\bibinfo {author} {\bibfnamefont {Hoi~Chun}\ \bibnamefont {Po}}, \bibinfo {author} {\bibfnamefont {Liujun}\ \bibnamefont {Zou}}, \bibinfo {author} {\bibfnamefont {Ashvin}\ \bibnamefont {Vishwanath}}, \ and\ \bibinfo {author} {\bibfnamefont {T}~\bibnamefont {Senthil}},\ }\bibfield  {title} {\enquote {\bibinfo {title} {Origin of mott insulating behavior and superconductivity in twisted bilayer graphene},}\ }\href@noop {} {\bibfield  {journal} {\bibinfo  {journal} {Physical Review X}\ }\textbf {\bibinfo {volume} {8}},\ \bibinfo {pages} {031089} (\bibinfo {year} {2018})}\BibitemShut {NoStop}%
\bibitem [{\citenamefont {Chen}\ \emph {et~al.}(2021)\citenamefont {Chen}, \citenamefont {Liao}, \citenamefont {Chen}, \citenamefont {Vafek}, \citenamefont {Kang}, \citenamefont {Li},\ and\ \citenamefont {Meng}}]{chen2021realization}%
  \BibitemOpen
  \bibfield  {author} {\bibinfo {author} {\bibfnamefont {Bin-Bin}\ \bibnamefont {Chen}}, \bibinfo {author} {\bibfnamefont {Yuan~Da}\ \bibnamefont {Liao}}, \bibinfo {author} {\bibfnamefont {Ziyu}\ \bibnamefont {Chen}}, \bibinfo {author} {\bibfnamefont {Oskar}\ \bibnamefont {Vafek}}, \bibinfo {author} {\bibfnamefont {Jian}\ \bibnamefont {Kang}}, \bibinfo {author} {\bibfnamefont {Wei}\ \bibnamefont {Li}}, \ and\ \bibinfo {author} {\bibfnamefont {Zi~Yang}\ \bibnamefont {Meng}},\ }\bibfield  {title} {\enquote {\bibinfo {title} {Realization of topological mott insulator in a twisted bilayer graphene lattice model},}\ }\href@noop {} {\bibfield  {journal} {\bibinfo  {journal} {Nature communications}\ }\textbf {\bibinfo {volume} {12}},\ \bibinfo {pages} {5480} (\bibinfo {year} {2021})}\BibitemShut {NoStop}%
\bibitem [{\citenamefont {Ledwith}\ \emph {et~al.}(2020)\citenamefont {Ledwith}, \citenamefont {Tarnopolsky}, \citenamefont {Khalaf},\ and\ \citenamefont {Vishwanath}}]{ledwith2020fractional}%
  \BibitemOpen
  \bibfield  {author} {\bibinfo {author} {\bibfnamefont {Patrick~J}\ \bibnamefont {Ledwith}}, \bibinfo {author} {\bibfnamefont {Grigory}\ \bibnamefont {Tarnopolsky}}, \bibinfo {author} {\bibfnamefont {Eslam}\ \bibnamefont {Khalaf}}, \ and\ \bibinfo {author} {\bibfnamefont {Ashvin}\ \bibnamefont {Vishwanath}},\ }\bibfield  {title} {\enquote {\bibinfo {title} {Fractional chern insulator states in twisted bilayer graphene: An analytical approach},}\ }\href@noop {} {\bibfield  {journal} {\bibinfo  {journal} {Physical Review Research}\ }\textbf {\bibinfo {volume} {2}},\ \bibinfo {pages} {023237} (\bibinfo {year} {2020})}\BibitemShut {NoStop}%
\bibitem [{\citenamefont {Abouelkomsan}\ \emph {et~al.}(2020)\citenamefont {Abouelkomsan}, \citenamefont {Liu},\ and\ \citenamefont {Bergholtz}}]{abouelkomsan2020particle}%
  \BibitemOpen
  \bibfield  {author} {\bibinfo {author} {\bibfnamefont {Ahmed}\ \bibnamefont {Abouelkomsan}}, \bibinfo {author} {\bibfnamefont {Zhao}\ \bibnamefont {Liu}}, \ and\ \bibinfo {author} {\bibfnamefont {Emil~J}\ \bibnamefont {Bergholtz}},\ }\bibfield  {title} {\enquote {\bibinfo {title} {Particle-hole duality, emergent fermi liquids, and fractional chern insulators in moir{\'e} flatbands},}\ }\href@noop {} {\bibfield  {journal} {\bibinfo  {journal} {Physical review letters}\ }\textbf {\bibinfo {volume} {124}},\ \bibinfo {pages} {106803} (\bibinfo {year} {2020})}\BibitemShut {NoStop}%
\bibitem [{\citenamefont {Matty}\ and\ \citenamefont {Kim}(2022)}]{matty2022melting}%
  \BibitemOpen
  \bibfield  {author} {\bibinfo {author} {\bibfnamefont {Michael}\ \bibnamefont {Matty}}\ and\ \bibinfo {author} {\bibfnamefont {Eun-Ah}\ \bibnamefont {Kim}},\ }\bibfield  {title} {\enquote {\bibinfo {title} {Melting of generalized wigner crystals in transition metal dichalcogenide heterobilayer moir{\'e} systems},}\ }\href@noop {} {\bibfield  {journal} {\bibinfo  {journal} {Nature Communications}\ }\textbf {\bibinfo {volume} {13}},\ \bibinfo {pages} {7098} (\bibinfo {year} {2022})}\BibitemShut {NoStop}%
\bibitem [{\citenamefont {Tao}\ \emph {et~al.}(2024)\citenamefont {Tao}, \citenamefont {Shen}, \citenamefont {Jiang}, \citenamefont {Li}, \citenamefont {Li}, \citenamefont {Ma}, \citenamefont {Zhao}, \citenamefont {Hu}, \citenamefont {Pistunova}, \citenamefont {Watanabe} \emph {et~al.}}]{tao2024valley}%
  \BibitemOpen
  \bibfield  {author} {\bibinfo {author} {\bibfnamefont {Zui}\ \bibnamefont {Tao}}, \bibinfo {author} {\bibfnamefont {Bowen}\ \bibnamefont {Shen}}, \bibinfo {author} {\bibfnamefont {Shengwei}\ \bibnamefont {Jiang}}, \bibinfo {author} {\bibfnamefont {Tingxin}\ \bibnamefont {Li}}, \bibinfo {author} {\bibfnamefont {Lizhong}\ \bibnamefont {Li}}, \bibinfo {author} {\bibfnamefont {Liguo}\ \bibnamefont {Ma}}, \bibinfo {author} {\bibfnamefont {Wenjin}\ \bibnamefont {Zhao}}, \bibinfo {author} {\bibfnamefont {Jenny}\ \bibnamefont {Hu}}, \bibinfo {author} {\bibfnamefont {Kateryna}\ \bibnamefont {Pistunova}}, \bibinfo {author} {\bibfnamefont {Kenji}\ \bibnamefont {Watanabe}},  \emph {et~al.},\ }\bibfield  {title} {\enquote {\bibinfo {title} {Valley-coherent quantum anomalous hall state in ab-stacked mote 2/w s e 2 bilayers},}\ }\href@noop {} {\bibfield  {journal} {\bibinfo  {journal} {Physical Review X}\ }\textbf {\bibinfo {volume} {14}},\ \bibinfo {pages} {011004} (\bibinfo {year} {2024})}\BibitemShut {NoStop}%
\bibitem [{\citenamefont {Balents}\ \emph {et~al.}(2020)\citenamefont {Balents}, \citenamefont {Dean}, \citenamefont {Efetov},\ and\ \citenamefont {Young}}]{balents2020superconductivity}%
  \BibitemOpen
  \bibfield  {author} {\bibinfo {author} {\bibfnamefont {Leon}\ \bibnamefont {Balents}}, \bibinfo {author} {\bibfnamefont {Cory~R}\ \bibnamefont {Dean}}, \bibinfo {author} {\bibfnamefont {Dmitri~K}\ \bibnamefont {Efetov}}, \ and\ \bibinfo {author} {\bibfnamefont {Andrea~F}\ \bibnamefont {Young}},\ }\bibfield  {title} {\enquote {\bibinfo {title} {Superconductivity and strong correlations in moir{\'e} flat bands},}\ }\href@noop {} {\bibfield  {journal} {\bibinfo  {journal} {Nature Physics}\ }\textbf {\bibinfo {volume} {16}},\ \bibinfo {pages} {725--733} (\bibinfo {year} {2020})}\BibitemShut {NoStop}%
\bibitem [{\citenamefont {Xu}\ and\ \citenamefont {Balents}(2018)}]{xu2018topological}%
  \BibitemOpen
  \bibfield  {author} {\bibinfo {author} {\bibfnamefont {Cenke}\ \bibnamefont {Xu}}\ and\ \bibinfo {author} {\bibfnamefont {Leon}\ \bibnamefont {Balents}},\ }\bibfield  {title} {\enquote {\bibinfo {title} {Topological superconductivity in twisted multilayer graphene},}\ }\href@noop {} {\bibfield  {journal} {\bibinfo  {journal} {Physical review letters}\ }\textbf {\bibinfo {volume} {121}},\ \bibinfo {pages} {087001} (\bibinfo {year} {2018})}\BibitemShut {NoStop}%
\bibitem [{\citenamefont {Kezilebieke}\ \emph {et~al.}(2022)\citenamefont {Kezilebieke}, \citenamefont {Vano}, \citenamefont {Huda}, \citenamefont {Aapro}, \citenamefont {Ganguli}, \citenamefont {Liljeroth},\ and\ \citenamefont {Lado}}]{kezilebieke2022moire}%
  \BibitemOpen
  \bibfield  {author} {\bibinfo {author} {\bibfnamefont {Shawulienu}\ \bibnamefont {Kezilebieke}}, \bibinfo {author} {\bibfnamefont {Viliam}\ \bibnamefont {Vano}}, \bibinfo {author} {\bibfnamefont {Md~N}\ \bibnamefont {Huda}}, \bibinfo {author} {\bibfnamefont {Markus}\ \bibnamefont {Aapro}}, \bibinfo {author} {\bibfnamefont {Somesh~C}\ \bibnamefont {Ganguli}}, \bibinfo {author} {\bibfnamefont {Peter}\ \bibnamefont {Liljeroth}}, \ and\ \bibinfo {author} {\bibfnamefont {Jose~L}\ \bibnamefont {Lado}},\ }\bibfield  {title} {\enquote {\bibinfo {title} {Moir{\'e}-enabled topological superconductivity},}\ }\href@noop {} {\bibfield  {journal} {\bibinfo  {journal} {Nano Letters}\ }\textbf {\bibinfo {volume} {22}},\ \bibinfo {pages} {328--333} (\bibinfo {year} {2022})}\BibitemShut {NoStop}%
\bibitem [{\citenamefont {Cr{\'e}pel}\ \emph {et~al.}(2023)\citenamefont {Cr{\'e}pel}, \citenamefont {Guerci}, \citenamefont {Cano}, \citenamefont {Pixley},\ and\ \citenamefont {Millis}}]{crepel2023topological}%
  \BibitemOpen
  \bibfield  {author} {\bibinfo {author} {\bibfnamefont {Valentin}\ \bibnamefont {Cr{\'e}pel}}, \bibinfo {author} {\bibfnamefont {Daniele}\ \bibnamefont {Guerci}}, \bibinfo {author} {\bibfnamefont {Jennifer}\ \bibnamefont {Cano}}, \bibinfo {author} {\bibfnamefont {JH}~\bibnamefont {Pixley}}, \ and\ \bibinfo {author} {\bibfnamefont {Andrew}\ \bibnamefont {Millis}},\ }\bibfield  {title} {\enquote {\bibinfo {title} {Topological superconductivity in doped magnetic moir{\'e} semiconductors},}\ }\href@noop {} {\bibfield  {journal} {\bibinfo  {journal} {Physical review letters}\ }\textbf {\bibinfo {volume} {131}},\ \bibinfo {pages} {056001} (\bibinfo {year} {2023})}\BibitemShut {NoStop}%
\bibitem [{\citenamefont {Lee}\ \emph {et~al.}(2023)\citenamefont {Lee}, \citenamefont {Sharma}, \citenamefont {Vafek},\ and\ \citenamefont {Changlani}}]{lee2023triangular}%
  \BibitemOpen
  \bibfield  {author} {\bibinfo {author} {\bibfnamefont {Kyungmin}\ \bibnamefont {Lee}}, \bibinfo {author} {\bibfnamefont {Prakash}\ \bibnamefont {Sharma}}, \bibinfo {author} {\bibfnamefont {Oskar}\ \bibnamefont {Vafek}}, \ and\ \bibinfo {author} {\bibfnamefont {Hitesh~J}\ \bibnamefont {Changlani}},\ }\bibfield  {title} {\enquote {\bibinfo {title} {Triangular lattice hubbard model physics at intermediate temperatures},}\ }\href@noop {} {\bibfield  {journal} {\bibinfo  {journal} {Physical Review B}\ }\textbf {\bibinfo {volume} {107}},\ \bibinfo {pages} {235105} (\bibinfo {year} {2023})}\BibitemShut {NoStop}%
\bibitem [{\citenamefont {Cai}\ \emph {et~al.}(2023)\citenamefont {Cai}, \citenamefont {Anderson}, \citenamefont {Wang}, \citenamefont {Zhang}, \citenamefont {Liu}, \citenamefont {Holtzmann}, \citenamefont {Zhang}, \citenamefont {Fan}, \citenamefont {Taniguchi}, \citenamefont {Watanabe} \emph {et~al.}}]{cai2023signatures}%
  \BibitemOpen
  \bibfield  {author} {\bibinfo {author} {\bibfnamefont {Jiaqi}\ \bibnamefont {Cai}}, \bibinfo {author} {\bibfnamefont {Eric}\ \bibnamefont {Anderson}}, \bibinfo {author} {\bibfnamefont {Chong}\ \bibnamefont {Wang}}, \bibinfo {author} {\bibfnamefont {Xiaowei}\ \bibnamefont {Zhang}}, \bibinfo {author} {\bibfnamefont {Xiaoyu}\ \bibnamefont {Liu}}, \bibinfo {author} {\bibfnamefont {William}\ \bibnamefont {Holtzmann}}, \bibinfo {author} {\bibfnamefont {Yinong}\ \bibnamefont {Zhang}}, \bibinfo {author} {\bibfnamefont {Fengren}\ \bibnamefont {Fan}}, \bibinfo {author} {\bibfnamefont {Takashi}\ \bibnamefont {Taniguchi}}, \bibinfo {author} {\bibfnamefont {Kenji}\ \bibnamefont {Watanabe}},  \emph {et~al.},\ }\bibfield  {title} {\enquote {\bibinfo {title} {Signatures of fractional quantum anomalous hall states in twisted mote2},}\ }\href@noop {} {\bibfield  {journal} {\bibinfo  {journal} {Nature}\ }\textbf {\bibinfo {volume} {622}},\ \bibinfo {pages} {63--68} (\bibinfo {year} {2023})}\BibitemShut {NoStop}%
\bibitem [{\citenamefont {Park}\ \emph {et~al.}(2023)\citenamefont {Park}, \citenamefont {Cai}, \citenamefont {Anderson}, \citenamefont {Zhang}, \citenamefont {Zhu}, \citenamefont {Liu}, \citenamefont {Wang}, \citenamefont {Holtzmann}, \citenamefont {Hu}, \citenamefont {Liu} \emph {et~al.}}]{park2023observation}%
  \BibitemOpen
  \bibfield  {author} {\bibinfo {author} {\bibfnamefont {Heonjoon}\ \bibnamefont {Park}}, \bibinfo {author} {\bibfnamefont {Jiaqi}\ \bibnamefont {Cai}}, \bibinfo {author} {\bibfnamefont {Eric}\ \bibnamefont {Anderson}}, \bibinfo {author} {\bibfnamefont {Yinong}\ \bibnamefont {Zhang}}, \bibinfo {author} {\bibfnamefont {Jiayi}\ \bibnamefont {Zhu}}, \bibinfo {author} {\bibfnamefont {Xiaoyu}\ \bibnamefont {Liu}}, \bibinfo {author} {\bibfnamefont {Chong}\ \bibnamefont {Wang}}, \bibinfo {author} {\bibfnamefont {William}\ \bibnamefont {Holtzmann}}, \bibinfo {author} {\bibfnamefont {Chaowei}\ \bibnamefont {Hu}}, \bibinfo {author} {\bibfnamefont {Zhaoyu}\ \bibnamefont {Liu}},  \emph {et~al.},\ }\bibfield  {title} {\enquote {\bibinfo {title} {Observation of fractionally quantized anomalous hall effect},}\ }\href@noop {} {\bibfield  {journal} {\bibinfo  {journal} {Nature}\ }\textbf {\bibinfo {volume} {622}},\ \bibinfo {pages} {74--79} (\bibinfo {year} {2023})}\BibitemShut {NoStop}%
\bibitem [{\citenamefont {Xu}\ \emph {et~al.}(2023)\citenamefont {Xu}, \citenamefont {Sun}, \citenamefont {Jia}, \citenamefont {Liu}, \citenamefont {Xu}, \citenamefont {Li}, \citenamefont {Gu}, \citenamefont {Watanabe}, \citenamefont {Taniguchi}, \citenamefont {Tong} \emph {et~al.}}]{xu2023observation}%
  \BibitemOpen
  \bibfield  {author} {\bibinfo {author} {\bibfnamefont {Fan}\ \bibnamefont {Xu}}, \bibinfo {author} {\bibfnamefont {Zheng}\ \bibnamefont {Sun}}, \bibinfo {author} {\bibfnamefont {Tongtong}\ \bibnamefont {Jia}}, \bibinfo {author} {\bibfnamefont {Chang}\ \bibnamefont {Liu}}, \bibinfo {author} {\bibfnamefont {Cheng}\ \bibnamefont {Xu}}, \bibinfo {author} {\bibfnamefont {Chushan}\ \bibnamefont {Li}}, \bibinfo {author} {\bibfnamefont {Yu}~\bibnamefont {Gu}}, \bibinfo {author} {\bibfnamefont {Kenji}\ \bibnamefont {Watanabe}}, \bibinfo {author} {\bibfnamefont {Takashi}\ \bibnamefont {Taniguchi}}, \bibinfo {author} {\bibfnamefont {Bingbing}\ \bibnamefont {Tong}},  \emph {et~al.},\ }\bibfield  {title} {\enquote {\bibinfo {title} {Observation of integer and fractional quantum anomalous hall effects in twisted bilayer mote 2},}\ }\href@noop {} {\bibfield  {journal} {\bibinfo  {journal} {Physical Review X}\ }\textbf {\bibinfo {volume} {13}},\ \bibinfo {pages} {031037} (\bibinfo {year} {2023})}\BibitemShut {NoStop}%
\bibitem [{\citenamefont {Zeng}\ \emph {et~al.}(2023)\citenamefont {Zeng}, \citenamefont {Xia}, \citenamefont {Kang}, \citenamefont {Zhu}, \citenamefont {Kn{\"u}ppel}, \citenamefont {Vaswani}, \citenamefont {Watanabe}, \citenamefont {Taniguchi}, \citenamefont {Mak},\ and\ \citenamefont {Shan}}]{zeng2023thermodynamic}%
  \BibitemOpen
  \bibfield  {author} {\bibinfo {author} {\bibfnamefont {Yihang}\ \bibnamefont {Zeng}}, \bibinfo {author} {\bibfnamefont {Zhengchao}\ \bibnamefont {Xia}}, \bibinfo {author} {\bibfnamefont {Kaifei}\ \bibnamefont {Kang}}, \bibinfo {author} {\bibfnamefont {Jiacheng}\ \bibnamefont {Zhu}}, \bibinfo {author} {\bibfnamefont {Patrick}\ \bibnamefont {Kn{\"u}ppel}}, \bibinfo {author} {\bibfnamefont {Chirag}\ \bibnamefont {Vaswani}}, \bibinfo {author} {\bibfnamefont {Kenji}\ \bibnamefont {Watanabe}}, \bibinfo {author} {\bibfnamefont {Takashi}\ \bibnamefont {Taniguchi}}, \bibinfo {author} {\bibfnamefont {Kin~Fai}\ \bibnamefont {Mak}}, \ and\ \bibinfo {author} {\bibfnamefont {Jie}\ \bibnamefont {Shan}},\ }\bibfield  {title} {\enquote {\bibinfo {title} {Thermodynamic evidence of fractional chern insulator in moir{\'e} mote2},}\ }\href@noop {} {\bibfield  {journal} {\bibinfo  {journal} {Nature}\ }\textbf {\bibinfo {volume} {622}},\ \bibinfo {pages} {69--73} (\bibinfo {year} {2023})}\BibitemShut {NoStop}%
\bibitem [{\citenamefont {Kang}\ \emph {et~al.}(2024)\citenamefont {Kang}, \citenamefont {Shen}, \citenamefont {Qiu}, \citenamefont {Zeng}, \citenamefont {Xia}, \citenamefont {Watanabe}, \citenamefont {Taniguchi}, \citenamefont {Shan},\ and\ \citenamefont {Mak}}]{kang2024evidence}%
  \BibitemOpen
  \bibfield  {author} {\bibinfo {author} {\bibfnamefont {Kaifei}\ \bibnamefont {Kang}}, \bibinfo {author} {\bibfnamefont {Bowen}\ \bibnamefont {Shen}}, \bibinfo {author} {\bibfnamefont {Yichen}\ \bibnamefont {Qiu}}, \bibinfo {author} {\bibfnamefont {Yihang}\ \bibnamefont {Zeng}}, \bibinfo {author} {\bibfnamefont {Zhengchao}\ \bibnamefont {Xia}}, \bibinfo {author} {\bibfnamefont {Kenji}\ \bibnamefont {Watanabe}}, \bibinfo {author} {\bibfnamefont {Takashi}\ \bibnamefont {Taniguchi}}, \bibinfo {author} {\bibfnamefont {Jie}\ \bibnamefont {Shan}}, \ and\ \bibinfo {author} {\bibfnamefont {Kin~Fai}\ \bibnamefont {Mak}},\ }\bibfield  {title} {\enquote {\bibinfo {title} {Evidence of the fractional quantum spin hall effect in moir{\'e} mote2},}\ }\href@noop {} {\bibfield  {journal} {\bibinfo  {journal} {Nature}\ ,\ \bibinfo {pages} {1--5}} (\bibinfo {year} {2024})}\BibitemShut {NoStop}%
\bibitem [{\citenamefont {Lu}\ \emph {et~al.}(2024)\citenamefont {Lu}, \citenamefont {Han}, \citenamefont {Yao}, \citenamefont {Reddy}, \citenamefont {Yang}, \citenamefont {Seo}, \citenamefont {Watanabe}, \citenamefont {Taniguchi}, \citenamefont {Fu},\ and\ \citenamefont {Ju}}]{lu2024fractional}%
  \BibitemOpen
  \bibfield  {author} {\bibinfo {author} {\bibfnamefont {Zhengguang}\ \bibnamefont {Lu}}, \bibinfo {author} {\bibfnamefont {Tonghang}\ \bibnamefont {Han}}, \bibinfo {author} {\bibfnamefont {Yuxuan}\ \bibnamefont {Yao}}, \bibinfo {author} {\bibfnamefont {Aidan~P}\ \bibnamefont {Reddy}}, \bibinfo {author} {\bibfnamefont {Jixiang}\ \bibnamefont {Yang}}, \bibinfo {author} {\bibfnamefont {Junseok}\ \bibnamefont {Seo}}, \bibinfo {author} {\bibfnamefont {Kenji}\ \bibnamefont {Watanabe}}, \bibinfo {author} {\bibfnamefont {Takashi}\ \bibnamefont {Taniguchi}}, \bibinfo {author} {\bibfnamefont {Liang}\ \bibnamefont {Fu}}, \ and\ \bibinfo {author} {\bibfnamefont {Long}\ \bibnamefont {Ju}},\ }\bibfield  {title} {\enquote {\bibinfo {title} {Fractional quantum anomalous hall effect in multilayer graphene},}\ }\href@noop {} {\bibfield  {journal} {\bibinfo  {journal} {Nature}\ }\textbf {\bibinfo {volume} {626}},\ \bibinfo {pages} {759--764} (\bibinfo {year} {2024})}\BibitemShut {NoStop}%
\bibitem [{\citenamefont {Wu}\ \emph {et~al.}(2019)\citenamefont {Wu}, \citenamefont {Lovorn}, \citenamefont {Tutuc}, \citenamefont {Martin},\ and\ \citenamefont {MacDonald}}]{wu2019topological}%
  \BibitemOpen
  \bibfield  {author} {\bibinfo {author} {\bibfnamefont {Fengcheng}\ \bibnamefont {Wu}}, \bibinfo {author} {\bibfnamefont {Timothy}\ \bibnamefont {Lovorn}}, \bibinfo {author} {\bibfnamefont {Emanuel}\ \bibnamefont {Tutuc}}, \bibinfo {author} {\bibfnamefont {Ivar}\ \bibnamefont {Martin}}, \ and\ \bibinfo {author} {\bibfnamefont {AH}~\bibnamefont {MacDonald}},\ }\bibfield  {title} {\enquote {\bibinfo {title} {Topological insulators in twisted transition metal dichalcogenide homobilayers},}\ }\href@noop {} {\bibfield  {journal} {\bibinfo  {journal} {Physical review letters}\ }\textbf {\bibinfo {volume} {122}},\ \bibinfo {pages} {086402} (\bibinfo {year} {2019})}\BibitemShut {NoStop}%
\bibitem [{\citenamefont {Devakul}\ \emph {et~al.}(2021)\citenamefont {Devakul}, \citenamefont {Cr{\'e}pel}, \citenamefont {Zhang},\ and\ \citenamefont {Fu}}]{devakul2021magic}%
  \BibitemOpen
  \bibfield  {author} {\bibinfo {author} {\bibfnamefont {Trithep}\ \bibnamefont {Devakul}}, \bibinfo {author} {\bibfnamefont {Valentin}\ \bibnamefont {Cr{\'e}pel}}, \bibinfo {author} {\bibfnamefont {Yang}\ \bibnamefont {Zhang}}, \ and\ \bibinfo {author} {\bibfnamefont {Liang}\ \bibnamefont {Fu}},\ }\bibfield  {title} {\enquote {\bibinfo {title} {Magic in twisted transition metal dichalcogenide bilayers},}\ }\href@noop {} {\bibfield  {journal} {\bibinfo  {journal} {Nature communications}\ }\textbf {\bibinfo {volume} {12}},\ \bibinfo {pages} {6730} (\bibinfo {year} {2021})}\BibitemShut {NoStop}%
\bibitem [{\citenamefont {Li}\ \emph {et~al.}(2021{\natexlab{b}})\citenamefont {Li}, \citenamefont {Kumar}, \citenamefont {Sun},\ and\ \citenamefont {Lin}}]{li2021spontaneous}%
  \BibitemOpen
  \bibfield  {author} {\bibinfo {author} {\bibfnamefont {Heqiu}\ \bibnamefont {Li}}, \bibinfo {author} {\bibfnamefont {Umesh}\ \bibnamefont {Kumar}}, \bibinfo {author} {\bibfnamefont {Kai}\ \bibnamefont {Sun}}, \ and\ \bibinfo {author} {\bibfnamefont {Shi-Zeng}\ \bibnamefont {Lin}},\ }\bibfield  {title} {\enquote {\bibinfo {title} {Spontaneous fractional chern insulators in transition metal dichalcogenide moir{\'e} superlattices},}\ }\href@noop {} {\bibfield  {journal} {\bibinfo  {journal} {Physical Review Research}\ }\textbf {\bibinfo {volume} {3}},\ \bibinfo {pages} {L032070} (\bibinfo {year} {2021}{\natexlab{b}})}\BibitemShut {NoStop}%
\bibitem [{\citenamefont {Cr{\'e}pel}\ and\ \citenamefont {Fu}(2023)}]{crepel2023anomalous}%
  \BibitemOpen
  \bibfield  {author} {\bibinfo {author} {\bibfnamefont {Valentin}\ \bibnamefont {Cr{\'e}pel}}\ and\ \bibinfo {author} {\bibfnamefont {Liang}\ \bibnamefont {Fu}},\ }\bibfield  {title} {\enquote {\bibinfo {title} {Anomalous hall metal and fractional chern insulator in twisted transition metal dichalcogenides},}\ }\href@noop {} {\bibfield  {journal} {\bibinfo  {journal} {Physical Review B}\ }\textbf {\bibinfo {volume} {107}},\ \bibinfo {pages} {L201109} (\bibinfo {year} {2023})}\BibitemShut {NoStop}%
\bibitem [{\citenamefont {Neupert}\ \emph {et~al.}(2011)\citenamefont {Neupert}, \citenamefont {Santos}, \citenamefont {Chamon},\ and\ \citenamefont {Mudry}}]{neupert2011fractional}%
  \BibitemOpen
  \bibfield  {author} {\bibinfo {author} {\bibfnamefont {Titus}\ \bibnamefont {Neupert}}, \bibinfo {author} {\bibfnamefont {Luiz}\ \bibnamefont {Santos}}, \bibinfo {author} {\bibfnamefont {Claudio}\ \bibnamefont {Chamon}}, \ and\ \bibinfo {author} {\bibfnamefont {Christopher}\ \bibnamefont {Mudry}},\ }\bibfield  {title} {\enquote {\bibinfo {title} {Fractional quantum hall states at zero magnetic field},}\ }\href@noop {} {\bibfield  {journal} {\bibinfo  {journal} {Physical review letters}\ }\textbf {\bibinfo {volume} {106}},\ \bibinfo {pages} {236804} (\bibinfo {year} {2011})}\BibitemShut {NoStop}%
\bibitem [{\citenamefont {Sheng}\ \emph {et~al.}(2011)\citenamefont {Sheng}, \citenamefont {Gu}, \citenamefont {Sun},\ and\ \citenamefont {Sheng}}]{sheng2011fractional}%
  \BibitemOpen
  \bibfield  {author} {\bibinfo {author} {\bibfnamefont {DN}~\bibnamefont {Sheng}}, \bibinfo {author} {\bibfnamefont {Zheng-Cheng}\ \bibnamefont {Gu}}, \bibinfo {author} {\bibfnamefont {Kai}\ \bibnamefont {Sun}}, \ and\ \bibinfo {author} {\bibfnamefont {L}~\bibnamefont {Sheng}},\ }\bibfield  {title} {\enquote {\bibinfo {title} {Fractional quantum hall effect in the absence of landau levels},}\ }\href@noop {} {\bibfield  {journal} {\bibinfo  {journal} {Nature communications}\ }\textbf {\bibinfo {volume} {2}},\ \bibinfo {pages} {389} (\bibinfo {year} {2011})}\BibitemShut {NoStop}%
\bibitem [{\citenamefont {Regnault}\ and\ \citenamefont {Bernevig}(2011)}]{regnault2011fractional}%
  \BibitemOpen
  \bibfield  {author} {\bibinfo {author} {\bibfnamefont {Nicolas}\ \bibnamefont {Regnault}}\ and\ \bibinfo {author} {\bibfnamefont {B~Andrei}\ \bibnamefont {Bernevig}},\ }\bibfield  {title} {\enquote {\bibinfo {title} {Fractional chern insulator},}\ }\href@noop {} {\bibfield  {journal} {\bibinfo  {journal} {Physical Review X}\ }\textbf {\bibinfo {volume} {1}},\ \bibinfo {pages} {021014} (\bibinfo {year} {2011})}\BibitemShut {NoStop}%
\bibitem [{\citenamefont {Tang}\ \emph {et~al.}(2011)\citenamefont {Tang}, \citenamefont {Mei},\ and\ \citenamefont {Wen}}]{tang2011high}%
  \BibitemOpen
  \bibfield  {author} {\bibinfo {author} {\bibfnamefont {Evelyn}\ \bibnamefont {Tang}}, \bibinfo {author} {\bibfnamefont {Jia-Wei}\ \bibnamefont {Mei}}, \ and\ \bibinfo {author} {\bibfnamefont {Xiao-Gang}\ \bibnamefont {Wen}},\ }\bibfield  {title} {\enquote {\bibinfo {title} {High-temperature fractional quantum hall states},}\ }\href@noop {} {\bibfield  {journal} {\bibinfo  {journal} {Physical review letters}\ }\textbf {\bibinfo {volume} {106}},\ \bibinfo {pages} {236802} (\bibinfo {year} {2011})}\BibitemShut {NoStop}%
\bibitem [{\citenamefont {Sun}\ \emph {et~al.}(2011)\citenamefont {Sun}, \citenamefont {Gu}, \citenamefont {Katsura},\ and\ \citenamefont {Sarma}}]{sun2011nearly}%
  \BibitemOpen
  \bibfield  {author} {\bibinfo {author} {\bibfnamefont {Kai}\ \bibnamefont {Sun}}, \bibinfo {author} {\bibfnamefont {Zhengcheng}\ \bibnamefont {Gu}}, \bibinfo {author} {\bibfnamefont {Hosho}\ \bibnamefont {Katsura}}, \ and\ \bibinfo {author} {\bibfnamefont {S~Das}\ \bibnamefont {Sarma}},\ }\bibfield  {title} {\enquote {\bibinfo {title} {Nearly flatbands with nontrivial topology},}\ }\href@noop {} {\bibfield  {journal} {\bibinfo  {journal} {Physical review letters}\ }\textbf {\bibinfo {volume} {106}},\ \bibinfo {pages} {236803} (\bibinfo {year} {2011})}\BibitemShut {NoStop}%
\bibitem [{\citenamefont {Reddy}\ \emph {et~al.}(2023)\citenamefont {Reddy}, \citenamefont {Alsallom}, \citenamefont {Zhang}, \citenamefont {Devakul},\ and\ \citenamefont {Fu}}]{reddy2023fractional}%
  \BibitemOpen
  \bibfield  {author} {\bibinfo {author} {\bibfnamefont {Aidan~P}\ \bibnamefont {Reddy}}, \bibinfo {author} {\bibfnamefont {Faisal}\ \bibnamefont {Alsallom}}, \bibinfo {author} {\bibfnamefont {Yang}\ \bibnamefont {Zhang}}, \bibinfo {author} {\bibfnamefont {Trithep}\ \bibnamefont {Devakul}}, \ and\ \bibinfo {author} {\bibfnamefont {Liang}\ \bibnamefont {Fu}},\ }\bibfield  {title} {\enquote {\bibinfo {title} {Fractional quantum anomalous hall states in twisted bilayer mote 2 and wse 2},}\ }\href@noop {} {\bibfield  {journal} {\bibinfo  {journal} {Physical Review B}\ }\textbf {\bibinfo {volume} {108}},\ \bibinfo {pages} {085117} (\bibinfo {year} {2023})}\BibitemShut {NoStop}%
\bibitem [{\citenamefont {Reddy}\ and\ \citenamefont {Fu}(2023)}]{reddy2023toward}%
  \BibitemOpen
  \bibfield  {author} {\bibinfo {author} {\bibfnamefont {Aidan~P}\ \bibnamefont {Reddy}}\ and\ \bibinfo {author} {\bibfnamefont {Liang}\ \bibnamefont {Fu}},\ }\bibfield  {title} {\enquote {\bibinfo {title} {Toward a global phase diagram of the fractional quantum anomalous hall effect},}\ }\href@noop {} {\bibfield  {journal} {\bibinfo  {journal} {Physical Review B}\ }\textbf {\bibinfo {volume} {108}},\ \bibinfo {pages} {245159} (\bibinfo {year} {2023})}\BibitemShut {NoStop}%
\bibitem [{\citenamefont {Dong}\ \emph {et~al.}(2023{\natexlab{a}})\citenamefont {Dong}, \citenamefont {Patri},\ and\ \citenamefont {Senthil}}]{dong2023theory}%
  \BibitemOpen
  \bibfield  {author} {\bibinfo {author} {\bibfnamefont {Zhihuan}\ \bibnamefont {Dong}}, \bibinfo {author} {\bibfnamefont {Adarsh~S}\ \bibnamefont {Patri}}, \ and\ \bibinfo {author} {\bibfnamefont {T}~\bibnamefont {Senthil}},\ }\bibfield  {title} {\enquote {\bibinfo {title} {Theory of fractional quantum anomalous hall phases in pentalayer rhombohedral graphene moir$\backslash$'e structures},}\ }\href@noop {} {\bibfield  {journal} {\bibinfo  {journal} {arXiv preprint arXiv:2311.03445}\ } (\bibinfo {year} {2023}{\natexlab{a}})}\BibitemShut {NoStop}%
\bibitem [{\citenamefont {Abouelkomsan}\ \emph {et~al.}(2023)\citenamefont {Abouelkomsan}, \citenamefont {Reddy}, \citenamefont {Fu},\ and\ \citenamefont {Bergholtz}}]{abouelkomsan_band_2023}%
  \BibitemOpen
  \bibfield  {author} {\bibinfo {author} {\bibfnamefont {Ahmed}\ \bibnamefont {Abouelkomsan}}, \bibinfo {author} {\bibfnamefont {Aidan~P.}\ \bibnamefont {Reddy}}, \bibinfo {author} {\bibfnamefont {Liang}\ \bibnamefont {Fu}}, \ and\ \bibinfo {author} {\bibfnamefont {Emil~J.}\ \bibnamefont {Bergholtz}},\ }\bibfield  {title} {\enquote {\bibinfo {title} {Band mixing in the quantum anomalous {Hall} regime of twisted semiconductor bilayers},}\ }\href {https://arxiv.org/abs/2309.16548} {\bibfield  {journal} {\bibinfo  {journal} {arXiv preprint arXiv:2309.16548}\ } (\bibinfo {year} {2023})}\BibitemShut {NoStop}%
\bibitem [{\citenamefont {Wang}\ \emph {et~al.}(2024{\natexlab{a}})\citenamefont {Wang}, \citenamefont {Zhang}, \citenamefont {Liu}, \citenamefont {He}, \citenamefont {Xu}, \citenamefont {Ran}, \citenamefont {Cao},\ and\ \citenamefont {Xiao}}]{wang2024fractional}%
  \BibitemOpen
  \bibfield  {author} {\bibinfo {author} {\bibfnamefont {Chong}\ \bibnamefont {Wang}}, \bibinfo {author} {\bibfnamefont {Xiao-Wei}\ \bibnamefont {Zhang}}, \bibinfo {author} {\bibfnamefont {Xiaoyu}\ \bibnamefont {Liu}}, \bibinfo {author} {\bibfnamefont {Yuchi}\ \bibnamefont {He}}, \bibinfo {author} {\bibfnamefont {Xiaodong}\ \bibnamefont {Xu}}, \bibinfo {author} {\bibfnamefont {Ying}\ \bibnamefont {Ran}}, \bibinfo {author} {\bibfnamefont {Ting}\ \bibnamefont {Cao}}, \ and\ \bibinfo {author} {\bibfnamefont {Di}~\bibnamefont {Xiao}},\ }\bibfield  {title} {\enquote {\bibinfo {title} {Fractional chern insulator in twisted bilayer mote 2},}\ }\href@noop {} {\bibfield  {journal} {\bibinfo  {journal} {Physical Review Letters}\ }\textbf {\bibinfo {volume} {132}},\ \bibinfo {pages} {036501} (\bibinfo {year} {2024}{\natexlab{a}})}\BibitemShut {NoStop}%
\bibitem [{\citenamefont {Yu}\ \emph {et~al.}(2024)\citenamefont {Yu}, \citenamefont {Herzog-Arbeitman}, \citenamefont {Wang}, \citenamefont {Vafek}, \citenamefont {Bernevig},\ and\ \citenamefont {Regnault}}]{yu2024fractional}%
  \BibitemOpen
  \bibfield  {author} {\bibinfo {author} {\bibfnamefont {Jiabin}\ \bibnamefont {Yu}}, \bibinfo {author} {\bibfnamefont {Jonah}\ \bibnamefont {Herzog-Arbeitman}}, \bibinfo {author} {\bibfnamefont {Minxuan}\ \bibnamefont {Wang}}, \bibinfo {author} {\bibfnamefont {Oskar}\ \bibnamefont {Vafek}}, \bibinfo {author} {\bibfnamefont {B~Andrei}\ \bibnamefont {Bernevig}}, \ and\ \bibinfo {author} {\bibfnamefont {Nicolas}\ \bibnamefont {Regnault}},\ }\bibfield  {title} {\enquote {\bibinfo {title} {Fractional chern insulators versus nonmagnetic states in twisted bilayer mote 2},}\ }\href@noop {} {\bibfield  {journal} {\bibinfo  {journal} {Physical Review B}\ }\textbf {\bibinfo {volume} {109}},\ \bibinfo {pages} {045147} (\bibinfo {year} {2024})}\BibitemShut {NoStop}%
\bibitem [{\citenamefont {Xu}\ \emph {et~al.}(2024{\natexlab{a}})\citenamefont {Xu}, \citenamefont {Li}, \citenamefont {Xu}, \citenamefont {Bi},\ and\ \citenamefont {Zhang}}]{xu2024maximally}%
  \BibitemOpen
  \bibfield  {author} {\bibinfo {author} {\bibfnamefont {Cheng}\ \bibnamefont {Xu}}, \bibinfo {author} {\bibfnamefont {Jiangxu}\ \bibnamefont {Li}}, \bibinfo {author} {\bibfnamefont {Yong}\ \bibnamefont {Xu}}, \bibinfo {author} {\bibfnamefont {Zhen}\ \bibnamefont {Bi}}, \ and\ \bibinfo {author} {\bibfnamefont {Yang}\ \bibnamefont {Zhang}},\ }\bibfield  {title} {\enquote {\bibinfo {title} {Maximally localized wannier functions, interaction models, and fractional quantum anomalous hall effect in twisted bilayer mote2},}\ }\href@noop {} {\bibfield  {journal} {\bibinfo  {journal} {Proceedings of the National Academy of Sciences}\ }\textbf {\bibinfo {volume} {121}},\ \bibinfo {pages} {e2316749121} (\bibinfo {year} {2024}{\natexlab{a}})}\BibitemShut {NoStop}%
\bibitem [{\citenamefont {Song}\ \emph {et~al.}(2024)\citenamefont {Song}, \citenamefont {Jian}, \citenamefont {Fu},\ and\ \citenamefont {Xu}}]{song2024intertwined}%
  \BibitemOpen
  \bibfield  {author} {\bibinfo {author} {\bibfnamefont {Xue-Yang}\ \bibnamefont {Song}}, \bibinfo {author} {\bibfnamefont {Chao-Ming}\ \bibnamefont {Jian}}, \bibinfo {author} {\bibfnamefont {Liang}\ \bibnamefont {Fu}}, \ and\ \bibinfo {author} {\bibfnamefont {Cenke}\ \bibnamefont {Xu}},\ }\bibfield  {title} {\enquote {\bibinfo {title} {Intertwined fractional quantum anomalous hall states and charge density waves},}\ }\href@noop {} {\bibfield  {journal} {\bibinfo  {journal} {Physical Review B}\ }\textbf {\bibinfo {volume} {109}},\ \bibinfo {pages} {115116} (\bibinfo {year} {2024})}\BibitemShut {NoStop}%
\bibitem [{\citenamefont {Reddy}\ \emph {et~al.}(2024)\citenamefont {Reddy}, \citenamefont {Paul}, \citenamefont {Abouelkomsan},\ and\ \citenamefont {Fu}}]{reddy2024non}%
  \BibitemOpen
  \bibfield  {author} {\bibinfo {author} {\bibfnamefont {Aidan~P}\ \bibnamefont {Reddy}}, \bibinfo {author} {\bibfnamefont {Nisarga}\ \bibnamefont {Paul}}, \bibinfo {author} {\bibfnamefont {Ahmed}\ \bibnamefont {Abouelkomsan}}, \ and\ \bibinfo {author} {\bibfnamefont {Liang}\ \bibnamefont {Fu}},\ }\bibfield  {title} {\enquote {\bibinfo {title} {Non-abelian fractionalization in topological minibands},}\ }\href@noop {} {\bibfield  {journal} {\bibinfo  {journal} {arXiv preprint arXiv:2403.00059}\ } (\bibinfo {year} {2024})}\BibitemShut {NoStop}%
\bibitem [{\citenamefont {Xu}\ \emph {et~al.}(2024{\natexlab{b}})\citenamefont {Xu}, \citenamefont {Mao}, \citenamefont {Zeng},\ and\ \citenamefont {Zhang}}]{xu2024multiple}%
  \BibitemOpen
  \bibfield  {author} {\bibinfo {author} {\bibfnamefont {Cheng}\ \bibnamefont {Xu}}, \bibinfo {author} {\bibfnamefont {Ning}\ \bibnamefont {Mao}}, \bibinfo {author} {\bibfnamefont {Tiansheng}\ \bibnamefont {Zeng}}, \ and\ \bibinfo {author} {\bibfnamefont {Yang}\ \bibnamefont {Zhang}},\ }\href@noop {} {\enquote {\bibinfo {title} {Multiple chern bands in twisted mote$_2$ and possible non-abelian states},}\ } (\bibinfo {year} {2024}{\natexlab{b}}),\ \Eprint {http://arxiv.org/abs/2403.17003} {arXiv:2403.17003 [cond-mat.str-el]} \BibitemShut {NoStop}%
\bibitem [{\citenamefont {Ahn}\ \emph {et~al.}(2024)\citenamefont {Ahn}, \citenamefont {Lee}, \citenamefont {Yananose}, \citenamefont {Kim},\ and\ \citenamefont {Cho}}]{ahn2024landau}%
  \BibitemOpen
  \bibfield  {author} {\bibinfo {author} {\bibfnamefont {Cheong-Eung}\ \bibnamefont {Ahn}}, \bibinfo {author} {\bibfnamefont {Wonjun}\ \bibnamefont {Lee}}, \bibinfo {author} {\bibfnamefont {Kunihiro}\ \bibnamefont {Yananose}}, \bibinfo {author} {\bibfnamefont {Youngwook}\ \bibnamefont {Kim}}, \ and\ \bibinfo {author} {\bibfnamefont {Gil~Young}\ \bibnamefont {Cho}},\ }\href@noop {} {\enquote {\bibinfo {title} {First landau level physics in second moir\'e band of $2.1^\circ$ twisted bilayer mote${}_2$},}\ } (\bibinfo {year} {2024}),\ \Eprint {http://arxiv.org/abs/2403.19155} {arXiv:2403.19155 [cond-mat.str-el]} \BibitemShut {NoStop}%
\bibitem [{\citenamefont {Wang}\ \emph {et~al.}(2024{\natexlab{b}})\citenamefont {Wang}, \citenamefont {Zhang}, \citenamefont {Liu}, \citenamefont {Wang}, \citenamefont {Cao},\ and\ \citenamefont {Xiao}}]{wang2024higher}%
  \BibitemOpen
  \bibfield  {author} {\bibinfo {author} {\bibfnamefont {Chong}\ \bibnamefont {Wang}}, \bibinfo {author} {\bibfnamefont {Xiao-Wei}\ \bibnamefont {Zhang}}, \bibinfo {author} {\bibfnamefont {Xiaoyu}\ \bibnamefont {Liu}}, \bibinfo {author} {\bibfnamefont {Jie}\ \bibnamefont {Wang}}, \bibinfo {author} {\bibfnamefont {Ting}\ \bibnamefont {Cao}}, \ and\ \bibinfo {author} {\bibfnamefont {Di}~\bibnamefont {Xiao}},\ }\href@noop {} {\enquote {\bibinfo {title} {Higher landau-level analogues and signatures of non-abelian states in twisted bilayer mote$_2$},}\ } (\bibinfo {year} {2024}{\natexlab{b}}),\ \Eprint {http://arxiv.org/abs/2404.05697} {arXiv:2404.05697 [cond-mat.str-el]} \BibitemShut {NoStop}%
\bibitem [{\citenamefont {{Jian}}\ \emph {et~al.}(2024)\citenamefont {{Jian}}, \citenamefont {{Cheng}},\ and\ \citenamefont {{Xu}}}]{ChaomingJian2024}%
  \BibitemOpen
  \bibfield  {author} {\bibinfo {author} {\bibfnamefont {Chao-Ming}\ \bibnamefont {{Jian}}}, \bibinfo {author} {\bibfnamefont {Meng}\ \bibnamefont {{Cheng}}}, \ and\ \bibinfo {author} {\bibfnamefont {Cenke}\ \bibnamefont {{Xu}}},\ }\bibfield  {title} {\enquote {\bibinfo {title} {{Minimal Fractional Topological Insulator in half-filled conjugate moir{\'e} Chern bands}},}\ }\href {\doibase 10.48550/arXiv.2403.07054} {\bibfield  {journal} {\bibinfo  {journal} {arXiv e-prints}\ ,\ \bibinfo {eid} {arXiv:2403.07054}} (\bibinfo {year} {2024})},\ \Eprint {http://arxiv.org/abs/2403.07054} {arXiv:2403.07054 [cond-mat.str-el]} \BibitemShut {NoStop}%
\bibitem [{\citenamefont {Goldman}\ \emph {et~al.}(2023)\citenamefont {Goldman}, \citenamefont {Reddy}, \citenamefont {Paul},\ and\ \citenamefont {Fu}}]{goldman2023zero}%
  \BibitemOpen
  \bibfield  {author} {\bibinfo {author} {\bibfnamefont {Hart}\ \bibnamefont {Goldman}}, \bibinfo {author} {\bibfnamefont {Aidan~P}\ \bibnamefont {Reddy}}, \bibinfo {author} {\bibfnamefont {Nisarga}\ \bibnamefont {Paul}}, \ and\ \bibinfo {author} {\bibfnamefont {Liang}\ \bibnamefont {Fu}},\ }\bibfield  {title} {\enquote {\bibinfo {title} {Zero-field composite fermi liquid in twisted semiconductor bilayers},}\ }\href@noop {} {\bibfield  {journal} {\bibinfo  {journal} {Physical Review Letters}\ }\textbf {\bibinfo {volume} {131}},\ \bibinfo {pages} {136501} (\bibinfo {year} {2023})}\BibitemShut {NoStop}%
\bibitem [{\citenamefont {Dong}\ \emph {et~al.}(2023{\natexlab{b}})\citenamefont {Dong}, \citenamefont {Wang}, \citenamefont {Ledwith}, \citenamefont {Vishwanath},\ and\ \citenamefont {Parker}}]{dong2023composite}%
  \BibitemOpen
  \bibfield  {author} {\bibinfo {author} {\bibfnamefont {Junkai}\ \bibnamefont {Dong}}, \bibinfo {author} {\bibfnamefont {Jie}\ \bibnamefont {Wang}}, \bibinfo {author} {\bibfnamefont {Patrick~J}\ \bibnamefont {Ledwith}}, \bibinfo {author} {\bibfnamefont {Ashvin}\ \bibnamefont {Vishwanath}}, \ and\ \bibinfo {author} {\bibfnamefont {Daniel~E}\ \bibnamefont {Parker}},\ }\bibfield  {title} {\enquote {\bibinfo {title} {Composite fermi liquid at zero magnetic field in twisted mote $ \_2$},}\ }\href@noop {} {\bibfield  {journal} {\bibinfo  {journal} {arXiv preprint arXiv:2306.01719}\ } (\bibinfo {year} {2023}{\natexlab{b}})}\BibitemShut {NoStop}%
\bibitem [{\citenamefont {Parameswaran}\ \emph {et~al.}(2013)\citenamefont {Parameswaran}, \citenamefont {Roy},\ and\ \citenamefont {Sondhi}}]{parameswaran2013fractional}%
  \BibitemOpen
  \bibfield  {author} {\bibinfo {author} {\bibfnamefont {Siddharth~A}\ \bibnamefont {Parameswaran}}, \bibinfo {author} {\bibfnamefont {Rahul}\ \bibnamefont {Roy}}, \ and\ \bibinfo {author} {\bibfnamefont {Shivaji~L}\ \bibnamefont {Sondhi}},\ }\bibfield  {title} {\enquote {\bibinfo {title} {Fractional quantum hall physics in topological flat bands},}\ }\href@noop {} {\bibfield  {journal} {\bibinfo  {journal} {Comptes Rendus Physique}\ }\textbf {\bibinfo {volume} {14}},\ \bibinfo {pages} {816--839} (\bibinfo {year} {2013})}\BibitemShut {NoStop}%
\bibitem [{\citenamefont {Liu}\ and\ \citenamefont {Bergholtz}(2022)}]{liu2022recent}%
  \BibitemOpen
  \bibfield  {author} {\bibinfo {author} {\bibfnamefont {Zhao}\ \bibnamefont {Liu}}\ and\ \bibinfo {author} {\bibfnamefont {Emil~J}\ \bibnamefont {Bergholtz}},\ }\bibfield  {title} {\enquote {\bibinfo {title} {Recent developments in fractional chern insulators},}\ }\href@noop {} {\bibfield  {journal} {\bibinfo  {journal} {arXiv preprint arXiv:2208.08449}\ } (\bibinfo {year} {2022})}\BibitemShut {NoStop}%
\bibitem [{\citenamefont {Roy}(2014)}]{roy2014band}%
  \BibitemOpen
  \bibfield  {author} {\bibinfo {author} {\bibfnamefont {Rahul}\ \bibnamefont {Roy}},\ }\bibfield  {title} {\enquote {\bibinfo {title} {Band geometry of fractional topological insulators},}\ }\href@noop {} {\bibfield  {journal} {\bibinfo  {journal} {Physical Review B}\ }\textbf {\bibinfo {volume} {90}},\ \bibinfo {pages} {165139} (\bibinfo {year} {2014})}\BibitemShut {NoStop}%
\bibitem [{\citenamefont {Bistritzer}\ and\ \citenamefont {MacDonald}(2011)}]{bistritzer2011moire}%
  \BibitemOpen
  \bibfield  {author} {\bibinfo {author} {\bibfnamefont {Rafi}\ \bibnamefont {Bistritzer}}\ and\ \bibinfo {author} {\bibfnamefont {Allan~H}\ \bibnamefont {MacDonald}},\ }\bibfield  {title} {\enquote {\bibinfo {title} {Moir{\'e} bands in twisted double-layer graphene},}\ }\href@noop {} {\bibfield  {journal} {\bibinfo  {journal} {Proceedings of the National Academy of Sciences}\ }\textbf {\bibinfo {volume} {108}},\ \bibinfo {pages} {12233--12237} (\bibinfo {year} {2011})}\BibitemShut {NoStop}%
\bibitem [{\citenamefont {Sheng}\ \emph {et~al.}(2024)\citenamefont {Sheng}, \citenamefont {Reddy}, \citenamefont {Abouelkomsan}, \citenamefont {Bergholtz},\ and\ \citenamefont {Fu}}]{sheng2024quantum}%
  \BibitemOpen
  \bibfield  {author} {\bibinfo {author} {\bibfnamefont {DN}~\bibnamefont {Sheng}}, \bibinfo {author} {\bibfnamefont {AP}~\bibnamefont {Reddy}}, \bibinfo {author} {\bibfnamefont {A}~\bibnamefont {Abouelkomsan}}, \bibinfo {author} {\bibfnamefont {EJ}~\bibnamefont {Bergholtz}}, \ and\ \bibinfo {author} {\bibfnamefont {L}~\bibnamefont {Fu}},\ }\bibfield  {title} {\enquote {\bibinfo {title} {Quantum anomalous hall crystal at fractional filling of moir$\backslash$'e superlattices},}\ }\href@noop {} {\bibfield  {journal} {\bibinfo  {journal} {arXiv preprint arXiv:2402.17832}\ } (\bibinfo {year} {2024})}\BibitemShut {NoStop}%
\bibitem [{\citenamefont {Wang}\ \emph {et~al.}(2024{\natexlab{c}})\citenamefont {Wang}, \citenamefont {Vila}, \citenamefont {Zaletel},\ and\ \citenamefont {Chatterjee}}]{wang2024electrical}%
  \BibitemOpen
  \bibfield  {author} {\bibinfo {author} {\bibfnamefont {Taige}\ \bibnamefont {Wang}}, \bibinfo {author} {\bibfnamefont {Marc}\ \bibnamefont {Vila}}, \bibinfo {author} {\bibfnamefont {Michael~P}\ \bibnamefont {Zaletel}}, \ and\ \bibinfo {author} {\bibfnamefont {Shubhayu}\ \bibnamefont {Chatterjee}},\ }\bibfield  {title} {\enquote {\bibinfo {title} {Electrical control of spin and valley in spin-orbit coupled graphene multilayers},}\ }\href@noop {} {\bibfield  {journal} {\bibinfo  {journal} {Physical Review Letters}\ }\textbf {\bibinfo {volume} {132}},\ \bibinfo {pages} {116504} (\bibinfo {year} {2024}{\natexlab{c}})}\BibitemShut {NoStop}%
\bibitem [{\citenamefont {Adak}\ \emph {et~al.}(2022)\citenamefont {Adak}, \citenamefont {Sinha}, \citenamefont {Giri}, \citenamefont {Mukherjee}, \citenamefont {Chandan}, \citenamefont {Sangani}, \citenamefont {Layek}, \citenamefont {Mukherjee}, \citenamefont {Watanabe}, \citenamefont {Taniguchi} \emph {et~al.}}]{adak2022perpendicular}%
  \BibitemOpen
  \bibfield  {author} {\bibinfo {author} {\bibfnamefont {Pratap~Chandra}\ \bibnamefont {Adak}}, \bibinfo {author} {\bibfnamefont {Subhajit}\ \bibnamefont {Sinha}}, \bibinfo {author} {\bibfnamefont {Debasmita}\ \bibnamefont {Giri}}, \bibinfo {author} {\bibfnamefont {Dibya~Kanti}\ \bibnamefont {Mukherjee}}, \bibinfo {author} {\bibnamefont {Chandan}}, \bibinfo {author} {\bibfnamefont {LD~Varma}\ \bibnamefont {Sangani}}, \bibinfo {author} {\bibfnamefont {Surat}\ \bibnamefont {Layek}}, \bibinfo {author} {\bibfnamefont {Ayshi}\ \bibnamefont {Mukherjee}}, \bibinfo {author} {\bibfnamefont {Kenji}\ \bibnamefont {Watanabe}}, \bibinfo {author} {\bibfnamefont {Takashi}\ \bibnamefont {Taniguchi}},  \emph {et~al.},\ }\bibfield  {title} {\enquote {\bibinfo {title} {Perpendicular electric field drives chern transitions and layer polarization changes in hofstadter bands},}\ }\href@noop {} {\bibfield  {journal} {\bibinfo  {journal} {Nature Communications}\ }\textbf {\bibinfo {volume} {13}},\ \bibinfo {pages} {7781} (\bibinfo
  {year} {2022})}\BibitemShut {NoStop}%
\bibitem [{\citenamefont {Cao}\ \emph {et~al.}(2020)\citenamefont {Cao}, \citenamefont {Rodan-Legrain}, \citenamefont {Rubies-Bigorda}, \citenamefont {Park}, \citenamefont {Watanabe}, \citenamefont {Taniguchi},\ and\ \citenamefont {Jarillo-Herrero}}]{cao2020tunable}%
  \BibitemOpen
  \bibfield  {author} {\bibinfo {author} {\bibfnamefont {Yuan}\ \bibnamefont {Cao}}, \bibinfo {author} {\bibfnamefont {Daniel}\ \bibnamefont {Rodan-Legrain}}, \bibinfo {author} {\bibfnamefont {Oriol}\ \bibnamefont {Rubies-Bigorda}}, \bibinfo {author} {\bibfnamefont {Jeong~Min}\ \bibnamefont {Park}}, \bibinfo {author} {\bibfnamefont {Kenji}\ \bibnamefont {Watanabe}}, \bibinfo {author} {\bibfnamefont {Takashi}\ \bibnamefont {Taniguchi}}, \ and\ \bibinfo {author} {\bibfnamefont {Pablo}\ \bibnamefont {Jarillo-Herrero}},\ }\bibfield  {title} {\enquote {\bibinfo {title} {Tunable correlated states and spin-polarized phases in twisted bilayer--bilayer graphene},}\ }\href@noop {} {\bibfield  {journal} {\bibinfo  {journal} {Nature}\ }\textbf {\bibinfo {volume} {583}},\ \bibinfo {pages} {215--220} (\bibinfo {year} {2020})}\BibitemShut {NoStop}%
\bibitem [{\citenamefont {Xiao}\ \emph {et~al.}(2012)\citenamefont {Xiao}, \citenamefont {Liu}, \citenamefont {Feng}, \citenamefont {Xu},\ and\ \citenamefont {Yao}}]{xiao2012coupled}%
  \BibitemOpen
  \bibfield  {author} {\bibinfo {author} {\bibfnamefont {Di}~\bibnamefont {Xiao}}, \bibinfo {author} {\bibfnamefont {Gui-Bin}\ \bibnamefont {Liu}}, \bibinfo {author} {\bibfnamefont {Wanxiang}\ \bibnamefont {Feng}}, \bibinfo {author} {\bibfnamefont {Xiaodong}\ \bibnamefont {Xu}}, \ and\ \bibinfo {author} {\bibfnamefont {Wang}\ \bibnamefont {Yao}},\ }\bibfield  {title} {\enquote {\bibinfo {title} {Coupled spin and valley physics in monolayers of mos 2 and other group-vi dichalcogenides},}\ }\href@noop {} {\bibfield  {journal} {\bibinfo  {journal} {Physical review letters}\ }\textbf {\bibinfo {volume} {108}},\ \bibinfo {pages} {196802} (\bibinfo {year} {2012})}\BibitemShut {NoStop}%
\bibitem [{\citenamefont {Yu}\ \emph {et~al.}(2020)\citenamefont {Yu}, \citenamefont {Chen},\ and\ \citenamefont {Yao}}]{yu2020giant}%
  \BibitemOpen
  \bibfield  {author} {\bibinfo {author} {\bibfnamefont {Hongyi}\ \bibnamefont {Yu}}, \bibinfo {author} {\bibfnamefont {Mingxing}\ \bibnamefont {Chen}}, \ and\ \bibinfo {author} {\bibfnamefont {Wang}\ \bibnamefont {Yao}},\ }\bibfield  {title} {\enquote {\bibinfo {title} {Giant magnetic field from moir{\'e} induced berry phase in homobilayer semiconductors},}\ }\href@noop {} {\bibfield  {journal} {\bibinfo  {journal} {National Science Review}\ }\textbf {\bibinfo {volume} {7}},\ \bibinfo {pages} {12--20} (\bibinfo {year} {2020})}\BibitemShut {NoStop}%
\bibitem [{\citenamefont {Jia}\ \emph {et~al.}(2024)\citenamefont {Jia}, \citenamefont {Yu}, \citenamefont {Liu}, \citenamefont {Herzog-Arbeitman}, \citenamefont {Qi}, \citenamefont {Pi}, \citenamefont {Regnault}, \citenamefont {Weng}, \citenamefont {Bernevig},\ and\ \citenamefont {Wu}}]{jia2024moire}%
  \BibitemOpen
  \bibfield  {author} {\bibinfo {author} {\bibfnamefont {Yujin}\ \bibnamefont {Jia}}, \bibinfo {author} {\bibfnamefont {Jiabin}\ \bibnamefont {Yu}}, \bibinfo {author} {\bibfnamefont {Jiaxuan}\ \bibnamefont {Liu}}, \bibinfo {author} {\bibfnamefont {Jonah}\ \bibnamefont {Herzog-Arbeitman}}, \bibinfo {author} {\bibfnamefont {Ziyue}\ \bibnamefont {Qi}}, \bibinfo {author} {\bibfnamefont {Hanqi}\ \bibnamefont {Pi}}, \bibinfo {author} {\bibfnamefont {Nicolas}\ \bibnamefont {Regnault}}, \bibinfo {author} {\bibfnamefont {Hongming}\ \bibnamefont {Weng}}, \bibinfo {author} {\bibfnamefont {B~Andrei}\ \bibnamefont {Bernevig}}, \ and\ \bibinfo {author} {\bibfnamefont {Quansheng}\ \bibnamefont {Wu}},\ }\bibfield  {title} {\enquote {\bibinfo {title} {Moir{\'e} fractional chern insulators. i. first-principles calculations and continuum models of twisted bilayer mote 2},}\ }\href@noop {} {\bibfield  {journal} {\bibinfo  {journal} {Physical Review B}\ }\textbf {\bibinfo {volume} {109}},\ \bibinfo {pages} {205121} (\bibinfo
  {year} {2024})}\BibitemShut {NoStop}%
\bibitem [{\citenamefont {Wang}\ \emph {et~al.}(2023)\citenamefont {Wang}, \citenamefont {Devakul}, \citenamefont {Zaletel},\ and\ \citenamefont {Fu}}]{wang2023topological}%
  \BibitemOpen
  \bibfield  {author} {\bibinfo {author} {\bibfnamefont {Taige}\ \bibnamefont {Wang}}, \bibinfo {author} {\bibfnamefont {Trithep}\ \bibnamefont {Devakul}}, \bibinfo {author} {\bibfnamefont {Michael~P}\ \bibnamefont {Zaletel}}, \ and\ \bibinfo {author} {\bibfnamefont {Liang}\ \bibnamefont {Fu}},\ }\bibfield  {title} {\enquote {\bibinfo {title} {Topological magnets and magnons in twisted bilayer mote $ \_2 $ and wse $ \_2$},}\ }\href@noop {} {\bibfield  {journal} {\bibinfo  {journal} {arXiv preprint arXiv:2306.02501}\ } (\bibinfo {year} {2023})}\BibitemShut {NoStop}%
\bibitem [{\citenamefont {Thouless}(1983)}]{thouless1983quantization}%
  \BibitemOpen
  \bibfield  {author} {\bibinfo {author} {\bibfnamefont {DJ}~\bibnamefont {Thouless}},\ }\bibfield  {title} {\enquote {\bibinfo {title} {Quantization of particle transport},}\ }\href@noop {} {\bibfield  {journal} {\bibinfo  {journal} {Physical Review B}\ }\textbf {\bibinfo {volume} {27}},\ \bibinfo {pages} {6083} (\bibinfo {year} {1983})}\BibitemShut {NoStop}%
\bibitem [{\citenamefont {Xiao}\ \emph {et~al.}(2010)\citenamefont {Xiao}, \citenamefont {Chang},\ and\ \citenamefont {Niu}}]{xiao2010berry}%
  \BibitemOpen
  \bibfield  {author} {\bibinfo {author} {\bibfnamefont {Di}~\bibnamefont {Xiao}}, \bibinfo {author} {\bibfnamefont {Ming-Che}\ \bibnamefont {Chang}}, \ and\ \bibinfo {author} {\bibfnamefont {Qian}\ \bibnamefont {Niu}},\ }\bibfield  {title} {\enquote {\bibinfo {title} {Berry phase effects on electronic properties},}\ }\href@noop {} {\bibfield  {journal} {\bibinfo  {journal} {Reviews of modern physics}\ }\textbf {\bibinfo {volume} {82}},\ \bibinfo {pages} {1959} (\bibinfo {year} {2010})}\BibitemShut {NoStop}%
\bibitem [{\citenamefont {Wimmer}\ \emph {et~al.}(2017)\citenamefont {Wimmer}, \citenamefont {Price}, \citenamefont {Carusotto},\ and\ \citenamefont {Peschel}}]{wimmer2017experimental}%
  \BibitemOpen
  \bibfield  {author} {\bibinfo {author} {\bibfnamefont {Martin}\ \bibnamefont {Wimmer}}, \bibinfo {author} {\bibfnamefont {Hannah~M}\ \bibnamefont {Price}}, \bibinfo {author} {\bibfnamefont {Iacopo}\ \bibnamefont {Carusotto}}, \ and\ \bibinfo {author} {\bibfnamefont {Ulf}\ \bibnamefont {Peschel}},\ }\bibfield  {title} {\enquote {\bibinfo {title} {Experimental measurement of the berry curvature from anomalous transport},}\ }\href@noop {} {\bibfield  {journal} {\bibinfo  {journal} {Nature Physics}\ }\textbf {\bibinfo {volume} {13}},\ \bibinfo {pages} {545--550} (\bibinfo {year} {2017})}\BibitemShut {NoStop}%
\bibitem [{\citenamefont {Kourtis}\ \emph {et~al.}(2014)\citenamefont {Kourtis}, \citenamefont {Neupert}, \citenamefont {Chamon},\ and\ \citenamefont {Mudry}}]{kourtis2014fractional}%
  \BibitemOpen
  \bibfield  {author} {\bibinfo {author} {\bibfnamefont {Stefanos}\ \bibnamefont {Kourtis}}, \bibinfo {author} {\bibfnamefont {Titus}\ \bibnamefont {Neupert}}, \bibinfo {author} {\bibfnamefont {Claudio}\ \bibnamefont {Chamon}}, \ and\ \bibinfo {author} {\bibfnamefont {Christopher}\ \bibnamefont {Mudry}},\ }\bibfield  {title} {\enquote {\bibinfo {title} {Fractional chern insulators with strong interactions that far exceed band gaps},}\ }\href@noop {} {\bibfield  {journal} {\bibinfo  {journal} {Physical review letters}\ }\textbf {\bibinfo {volume} {112}},\ \bibinfo {pages} {126806} (\bibinfo {year} {2014})}\BibitemShut {NoStop}%
\bibitem [{\citenamefont {Grushin}\ \emph {et~al.}(2015)\citenamefont {Grushin}, \citenamefont {Motruk}, \citenamefont {Zaletel},\ and\ \citenamefont {Pollmann}}]{grushin2015characterization}%
  \BibitemOpen
  \bibfield  {author} {\bibinfo {author} {\bibfnamefont {Adolfo~G}\ \bibnamefont {Grushin}}, \bibinfo {author} {\bibfnamefont {Johannes}\ \bibnamefont {Motruk}}, \bibinfo {author} {\bibfnamefont {Michael~P}\ \bibnamefont {Zaletel}}, \ and\ \bibinfo {author} {\bibfnamefont {Frank}\ \bibnamefont {Pollmann}},\ }\bibfield  {title} {\enquote {\bibinfo {title} {Characterization and stability of a fermionic $\nu$= 1/3 fractional chern insulator},}\ }\href@noop {} {\bibfield  {journal} {\bibinfo  {journal} {Physical Review B}\ }\textbf {\bibinfo {volume} {91}},\ \bibinfo {pages} {035136} (\bibinfo {year} {2015})}\BibitemShut {NoStop}%
\bibitem [{\citenamefont {Anderson}\ \emph {et~al.}(2023)\citenamefont {Anderson}, \citenamefont {Fan}, \citenamefont {Cai}, \citenamefont {Holtzmann}, \citenamefont {Taniguchi}, \citenamefont {Watanabe}, \citenamefont {Xiao}, \citenamefont {Yao},\ and\ \citenamefont {Xu}}]{anderson2023programming}%
  \BibitemOpen
  \bibfield  {author} {\bibinfo {author} {\bibfnamefont {Eric}\ \bibnamefont {Anderson}}, \bibinfo {author} {\bibfnamefont {Feng-Ren}\ \bibnamefont {Fan}}, \bibinfo {author} {\bibfnamefont {Jiaqi}\ \bibnamefont {Cai}}, \bibinfo {author} {\bibfnamefont {William}\ \bibnamefont {Holtzmann}}, \bibinfo {author} {\bibfnamefont {Takashi}\ \bibnamefont {Taniguchi}}, \bibinfo {author} {\bibfnamefont {Kenji}\ \bibnamefont {Watanabe}}, \bibinfo {author} {\bibfnamefont {Di}~\bibnamefont {Xiao}}, \bibinfo {author} {\bibfnamefont {Wang}\ \bibnamefont {Yao}}, \ and\ \bibinfo {author} {\bibfnamefont {Xiaodong}\ \bibnamefont {Xu}},\ }\bibfield  {title} {\enquote {\bibinfo {title} {Programming correlated magnetic states with gate-controlled moir{\'e} geometry},}\ }\href@noop {} {\bibfield  {journal} {\bibinfo  {journal} {Science}\ }\textbf {\bibinfo {volume} {381}},\ \bibinfo {pages} {325--330} (\bibinfo {year} {2023})}\BibitemShut {NoStop}%
\bibitem [{\citenamefont {Haldane}(1991)}]{haldane1991fractional}%
  \BibitemOpen
  \bibfield  {author} {\bibinfo {author} {\bibfnamefont {F~Duncan~M}\ \bibnamefont {Haldane}},\ }\bibfield  {title} {\enquote {\bibinfo {title} {‘‘fractional statistics’’in arbitrary dimensions: A generalization of the pauli principle},}\ }\href@noop {} {\bibfield  {journal} {\bibinfo  {journal} {Physical review letters}\ }\textbf {\bibinfo {volume} {67}},\ \bibinfo {pages} {937} (\bibinfo {year} {1991})}\BibitemShut {NoStop}%
\bibitem [{\citenamefont {Bernevig}\ and\ \citenamefont {Haldane}(2008)}]{bernevig2008model}%
  \BibitemOpen
  \bibfield  {author} {\bibinfo {author} {\bibfnamefont {B~Andrei}\ \bibnamefont {Bernevig}}\ and\ \bibinfo {author} {\bibfnamefont {FDM}\ \bibnamefont {Haldane}},\ }\bibfield  {title} {\enquote {\bibinfo {title} {Model fractional quantum hall states and jack polynomials},}\ }\href@noop {} {\bibfield  {journal} {\bibinfo  {journal} {Physical review letters}\ }\textbf {\bibinfo {volume} {100}},\ \bibinfo {pages} {246802} (\bibinfo {year} {2008})}\BibitemShut {NoStop}%
\bibitem [{\citenamefont {Wilhelm}\ \emph {et~al.}(2021)\citenamefont {Wilhelm}, \citenamefont {Lang},\ and\ \citenamefont {L{\"a}uchli}}]{wilhelm2021interplay}%
  \BibitemOpen
  \bibfield  {author} {\bibinfo {author} {\bibfnamefont {Patrick}\ \bibnamefont {Wilhelm}}, \bibinfo {author} {\bibfnamefont {Thomas~C}\ \bibnamefont {Lang}}, \ and\ \bibinfo {author} {\bibfnamefont {Andreas~M}\ \bibnamefont {L{\"a}uchli}},\ }\bibfield  {title} {\enquote {\bibinfo {title} {Interplay of fractional chern insulator and charge density wave phases in twisted bilayer graphene},}\ }\href@noop {} {\bibfield  {journal} {\bibinfo  {journal} {Physical Review B}\ }\textbf {\bibinfo {volume} {103}},\ \bibinfo {pages} {125406} (\bibinfo {year} {2021})}\BibitemShut {NoStop}%
\bibitem [{\citenamefont {Okamoto}\ \emph {et~al.}(2022)\citenamefont {Okamoto}, \citenamefont {Mohanta}, \citenamefont {Dagotto},\ and\ \citenamefont {Sheng}}]{okamoto2022topological}%
  \BibitemOpen
  \bibfield  {author} {\bibinfo {author} {\bibfnamefont {Satoshi}\ \bibnamefont {Okamoto}}, \bibinfo {author} {\bibfnamefont {Narayan}\ \bibnamefont {Mohanta}}, \bibinfo {author} {\bibfnamefont {Elbio}\ \bibnamefont {Dagotto}}, \ and\ \bibinfo {author} {\bibfnamefont {DN}~\bibnamefont {Sheng}},\ }\bibfield  {title} {\enquote {\bibinfo {title} {Topological flat bands in a kagome lattice multiorbital system},}\ }\href@noop {} {\bibfield  {journal} {\bibinfo  {journal} {Communications Physics}\ }\textbf {\bibinfo {volume} {5}},\ \bibinfo {pages} {198} (\bibinfo {year} {2022})}\BibitemShut {NoStop}%
\end{thebibliography}%

\end{document}